\documentclass[%
superscriptaddress,
groupedaddress,
preprint,
prfluids,
notitlepage,
onecolumn, 
preprintnumbers,
amsmath,amssymb,
aps,
floatfix,
]{revtex4-2}

\usepackage{dcolumn}
\usepackage{bm}
\usepackage[utf8]{inputenc}
\usepackage{textcomp}
\usepackage{hyperref}
\usepackage{cleveref}
\usepackage{graphicx,epstopdf}
\usepackage{amsmath}
\usepackage[version=4]{mhchem}
\usepackage{siunitx}
\usepackage{natbib}
\usepackage{subcaption}
\usepackage{xcolor}
\usepackage{amsmath}
\usepackage{mathtools,tikz,caption}

\begin{document}

\preprint{This draft was prepared using the LaTeX style file belonging to \textit{Physical Review Fluids}}

\title{Manuscript Title:\\Pressure-Informed Velocity Estimation in a Subsonic Jet}

\author{Songqi Li}
	\email{songzi32@ufl.edu}
	\affiliation{University of Florida, Gainesville, Florida 32611, USA}
\author{Lawrence Ukeiley}%
 \affiliation{University of Florida, Gainesville, Florida 32611, USA}
 
\date{\today}

\begin{abstract}
This work aims to estimate time-resolved velocity field that is directly associated with pressure fluctuations in a subsonic round jet.
To achieve this goal, synchronous measurements of the velocity field and in-flow pressure fluctuations were performed at Mach number 0.3.
Two different experiment campaigns were conducted, the first experimental campaign aims to explore the time-resolved dynamics of the axisymmetric velocity components, and second experiment focuses on the time-resolved, 2D velocity estimates on a streamwise plane.
Two different methods were utilized to estimate the input-output relation between velocity and in-flow pressure measurements.
A hybrid approach based on the spectral linear stochastic estimation and the proper orthogonal decomposition was applied to setup the model in a linear manner, and a wavelet-based filter was implemented to attenuate the noise level in the cross-correlation functions.
In addition, the pressure-velocity relationship was also described by neural network architectures based on the multi-layer perceptron (MLP) and bidirectional long-short-term-memory (LSTM).
In both experimental sets, pressure fluctuations inside the flow are found to be connected to the streamwise convection of large-scale coherent structures in the flow.
A unique advantage of the bidirectional LSTM method was found among all estimation schemes is also reported in this work.
The estimation result represents the space-time dynamics of the acoustic sources in the jet flow field, and it is of great importance to understand the noise generation mechanism.
\end{abstract}

\maketitle


\section{\label{sec:intro}Introduction}

The accurate depiction of spatial-temporal activities associated with acoustic sources in subsonic free jet flows remains a challenging engineering problem in the fluid dynamics and aeroacoustics communities. 
Although beamforming experiments \cite{dougherty2002,papamoschou2011,breen2015} have shown that the region near the end of the potential core is of great importance with respect to the noise generation, the real-time dynamics of the source activities in this region cannot be directly obtained from this approach. 
On the other hand, time-resolved (TR) measurements from PIV \cite{semeraro2012,berger2014} and hot-wire rakes \cite{citriniti2000jfm,iqbal2007} are capable to reveal the temporal dynamics of coherent structures in velocity field of the flow. 
However, the distillation of the portion of the velocity that directly contributes to the noise generation in the far field is still a prerequisite to the evaluation of the noise source mechanisms and their control.

Stochastic estimation (SE), first proposed by \citet{adrian1977}, is one of the most widely applied techniques to refine "conditional eddies" in turbulent flows from some dependent and correlated inputs.
As elucidated in \cite{bonnet1994}, the goal of this technique is to utilize conditional information about the flow in an attempt to estimate the correlated portion of the flow at other locations.
Pioneering work to use this techniques to estimate the time dependence of the conditional structures can be found in \cite{guezennec1989,cole1992,cole1998,murray2003} among many others. 
On the basis of the traditional SE, the combination of SE with the Proper Orthogonal Decomposition (POD) has also aroused great interest among researchers. 
In this modified approach the POD is utilized to provide reduced-order representations of the complex turbulent flows, and the original problem becomes the estimation of time-varying POD expansion coefficients (\cite{arunajatesan2007,tinney2008,durgesh2010,nickels2020,zhang2020}).
In the practical problem of noise generated from turbulent jets, various types of inputs have been applied in SE to better highlight the source activities inside the jet shear layer, including the turbulent velocities \cite{ewing1999}, far-field acoustics \cite{magstadt2016}, and near-field pressure fluctuations \cite{picard2000,tinney2008}. 
Apart from the above-mentioned attempts, direct measurements of in-flow pressure fluctuations \cite{fuchs1972a,fuchs1972b,jones1979,george1984} provide a novel opportunity to evaluate the source mechanism directly related to the pressure fluctuations in the convective turbulent flow. 
Pressure fluctuations in a subsonic jet has been experimentally measured and analyzed in \citet{li2021}, in which time-localized imprints of the wave-packets (\cite{jordan2013,cavalieri2012,cavalieri2019}) were effectively extracted from pressure fluctuations in the jet shear layer.

With the booming development of computational power, modern machine-leaning (ML) methods have come into the researcher’s field of vision for a wide range of applications. 
The potential of ML-based techniques to solve complex fluid dynamical problems have been expounded in \cite{brunton2020} and \cite{kutz2017}.
For the practical problem associated with this work, i.e., the estimation of real-time source activities from sequential inputs at discrete locations, pioneering attempts have been made and yield encouraging results.
In a relevant application to the work reported here \citet{tenney2020} trained a multi-layer perceptron (MLP) to estimate pressure fluctuations in the near field of a subsonic jet and the MLP outperforms the stochastic estimation with an improvement of accuracy.
In addition, architectures based on the recurrent neural networks (RNNs) \cite{mohan2018,jin2020,deng2019} have also been proposed to estimate or predict real-time flow dynamics with sequential inputs in time. 
The RNNs are specialized to reveal the temporal dynamics of the sequential inputs, which includes simple RNN, long-short-time-memory (LSTM), and gated recurrent unit (GRU), etc. 
These architectures are highlighted by their time-dependent parameter transmission mechanism and have shown huge success in relevant tasks involving time-dependent data such as music genre classification \cite{choi2017} and stock price prediction \cite{selvin2017}.

This study focuses on the application of experimental measurements of in-flow pressure fluctuations to elucidate the relationship between pressure and velocity in a subsonic axisymmetric jet. 
Synchronized measurements of in-flow pressure with low-frame-rate PIV were performed to reveal the connections between velocity and pressure at the sound-generating region.
In addition, time-resolved estimates of turbulent velocity associated to the in-flow pressure was realized via SE and ML approaches.
A hybrid approach combining SE and POD was adopted in this work, and a wavelet-based filter was implemented to denoise the cross-correlation functions between pressure and POD expansion coefficients that were directly calculated from experiments in an attempt to better highlight the wave-packet structures \cite{jordan2013} in the flow.
Moreover, two neural network architectures were also utilized in this work to provide time-resolved estimates of the velocity field as an alternative approach.
Detailed experimental procedure is described in \Cref{sec:setup}, followed by the introduction of the estimation techniques in \Cref{sec:analysis} which includes the wavelet-filtered stochastic estimation and two neural network architectures. 
\Cref{sec:results} reports the main results and a brief summary in \Cref{sec:conclusions} concludes the manuscript.
\section{\label{sec:setup}Experimental Setup}
All experiments were carried out in the Anechoic Jet Test Facility at the University of Florida. Details of this facility in terms of the anechoic room and the recently installed jet facility can be found in \cite{mathew2005} and \cite{li2020}. The installed subsonic jet  has a convergent nozzle with an exit diameter of $D=5.08$ cm and an area contraction ratio of 9:1. For all experiments in this work, the facility was operated under the blow-down mode at $M=0.3$, which corresponds to a jet exit Reynolds numbers of $Re_D=3.8\times10^5$. The air supply for the jet was controlled with a Fisher regulating valve coupled to a LabVIEW PID controller that allowed the jet exit Mach number to be maintained within $1\%$ of the desired value throughout a test.   

The fluctuating static pressure inside the flow was measured using both a B\&K 4138 1/8’’ microphone and a GRAS 46DD 1/8” microphone. 
Each microphone was equipped with an aerodynamically shaped nosecone to avoid any contamination from total pressure fluctuations \cite{dassen1996}. 
Although different vendor supplied nosecone configurations were adopted for the two microphones, in which the B\&K nosecone has a sharp tip and a shorter length and the GRAS nosecone has a longer and blunt leading edge, \citet{soderman2002} has shown that the installation of both types of nosecone configurations will generate essentially negligible installation effect at the frequency range of interest in this work (150 \textasciitilde 20kHz). 
More detailed discussion on in-flow fluctuating pressure measurements using miniature microphones is available in \citet{li2021}.
In the current work, the microphones were mounted on a 3D-printed, airfoil-shaped strut which allowed the microphones to be held at the jet centerline, upper and lower jet liplines, respectively while generating a minimal disturbance. 

To reveal the relationship between pressure and velocity in the flow, pressure fluctuations were synchronously recorded with velocity measured from PIV, but at different sampling rates of 80004 Hz and 12 Hz, respectively. 
A 135 mJ dual-cavity Litron Nano Nd:YAG laser with a wavelength $\lambda = 532$ nm was used as the light source to illuminate the seeding particles.
The flow was seeded by an ATI Laskin nozzle aerosol generator which produced particles with diameters around 0.25 \textmu m, and the ambient air was seeded with a Rosco 1700 fog machine with particle diameters between 0.25 to 0.5 \textmu m.
Processing the raw images in the current experiments was accomplished using DaVis 8.3 from LaVision. 
The raw images were divided into interrogation windows, and the velocity vectors were calculated through a correlation-based multi-pass routine with decreasing interrogation window sizes. 
The interrogating window size for the final pass was 32 pixel $\times$ 32 pixel with $75\%$ overlap which leads to a vector resolution of around 500 \textmu m in all experimental sets.
The uncertainty of PIV measurements was implemented in DaVis 8.3 using the method in \cite{wieneke2015}.
This algorithm leads to an average uncertainty in the jet shear layer at around 4.8 m/s for cross-stream stereoscopic measurements and 2.5 m/s for streamwise planar measurements. 

\begin{figure}[!h]
    \centering
    \includegraphics[width=.5\linewidth]{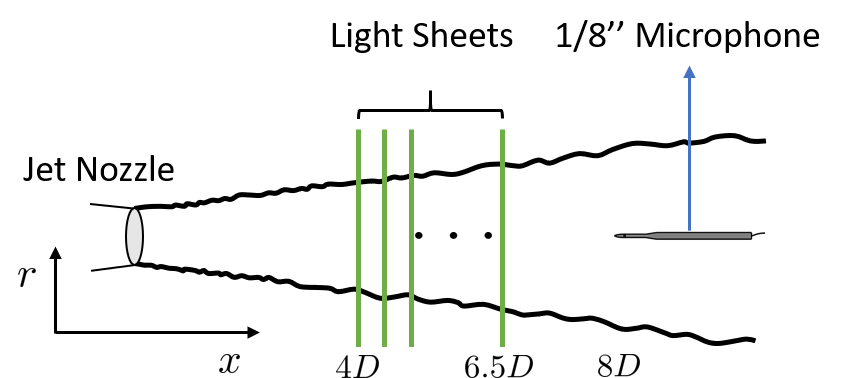}
    \caption{Pressure measurements synchronized with stereo-PIV on cross-stream planes.}
    \label{fig:sync1_setup}
    \includegraphics[width=.5\linewidth]{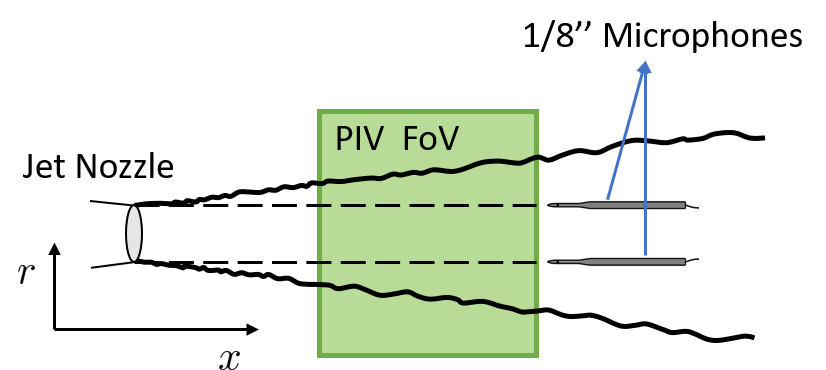}
    \caption{Pressure measurements synchronized with planar PIV on a streamwise plane.}
    \label{fig:sync2_setup}
\end{figure}

Two experimental campaigns were performed in this work. 
The first experimental set synchronizes stereo-PIV with single-point pressure measurements on the jet centerline.
The motivation of this experimental campaign originates from the work of \citet{tinney2008}, in which a volumetric, time-resolved estimation of the turbulent velocity in a Mach 0.85 jet was obtained via stochastic estimation.
In Tinney's work time-resolved pressure inputs were acquired in the upstream hydrodynamic periphery, which are associated with the emergence of the convecting coherent structures. 
Given the ability to directly measure the in-flow fluctuating pressure generated by large-scale structures, the current experimental set aims to perform real-time estimation of velocity informed from turbulent pressure measured downstream of the estimation field. 
However, due to the limited number of microphones that can be placed in the jet, it is not possible to resolve all dominant azimuthal modes in the jet.
On the other hand, the dominance of the axisymmetric mode has been justified in \cite{cavalieri2012,cavalieri2013,jordan2013,camussi2021} with respect to the sound radiation efficiency at shallow emission angles.
Following the argument in \cite{batchelor1962,cavalieri2013}, the kinematic boundary conditions will guarantee finite azimuthal mode-0 pressure on the jet centerline and zero pressure for all higher order modes. 
Therefore, as an initial attempt, the advantage of axisymmetry in round jets was taken by placing the B\&K microphone on the jet centerline to capture the temporal dynamics of the axisymmetric events in pressure. 
The spatial feature of the axisymmetric velocity can be obtained from stereo-PIV measurements on multiple cross-stream planes, and the energetic structures can be extracted in complementary with the application of azimuthal-Fourier POD \cite{tinney2008,li2020} to the mode-0 velocity components.
A schematic of the experimental setup is shown in \Cref{fig:sync1_setup}, where the B$\&$K microphone was placed on the jet centerline at $x/D =8$ and PIV measurements were taken on multiple cross-stream planes between $x/D = 4$ and $6.5$ with a step increment of $\Delta x/D = 0.25$. 
For each streamwise location, 1600 PIV snapshots were recorded synchronously with time-lagged pressure.

The second experimental set aimed to characterize space-time dynamics of two-dimensional velocity fields on a streamwise plane ($x-r$) across the jet centerline.
Planar PIV was utilized to measure velocity vectors within $3\leq x/D\leq6.5$, $\lvert r/D \rvert \leq 1.2$.
Both B\&k and GRAS microphones were employed in this measurement campaign and were placed on the upper and lower jet liplines ($r/D=\pm 0.5$), respectively. 
As displayed in \Cref{fig:sync2_setup}, the nosecone tips were aligned at $x/D=6.6$ which is just downstream of the velocity field-of-view. 
To avoid direct interaction with laser light sheet, an out-of-plane displacement of 5 mm was implemented to the microphones during the positioning process, and a total of 8000 PIV images were acquired in this measurement campaign.

\section{\label{sec:analysis}Analysis Techniques}
\begin{figure}[!b]
    \centering
    \includegraphics[width=0.8\linewidth]{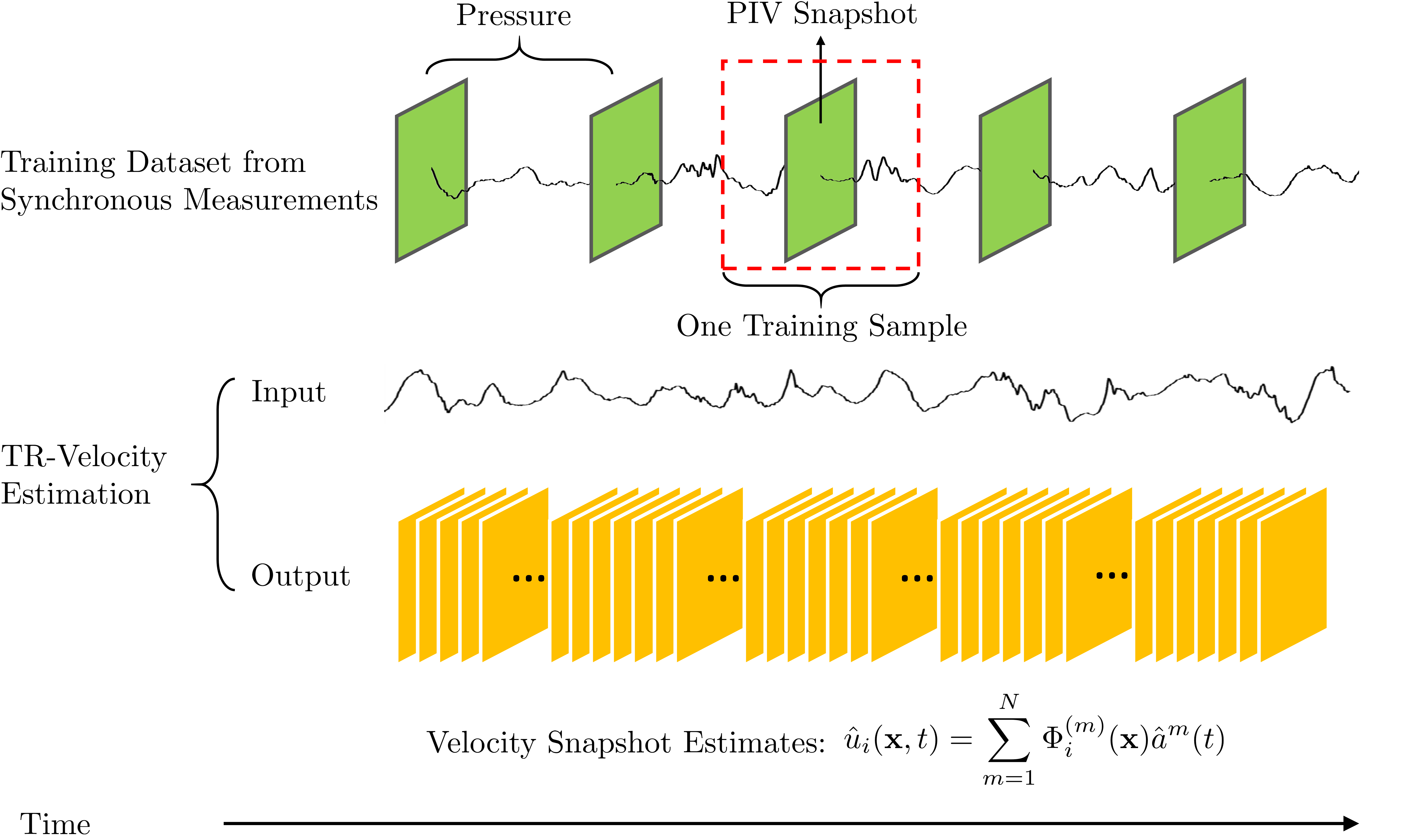}  
    \caption{Visualization of PIV recording versus fluctuating pressure measurements, with the goal of estimating time-resolved velocity fields give pressure inputs. Green frames: PIV snapshots; Yellow frames: velocity estimates via appropriate input-output models.}
    \label{fig:SE_schematic}
\end{figure}

One important goal of this study is to estimate time-resolved velocity fields associated with pressure fluctuations in the flow informed by non-TR PIV measurements in combination with TR-pressure signals. 
A graphical illustration of this procedure is displayed in \Cref{fig:SE_schematic}.
As a data-driven approach, $N_t$ mutually independent velocity snapshots $u_i(\mathbf{x},t_n)$ and the corresponding time-lagged pressure signals $p^k(t_n-\tau\sim t_n+\tau)$ were first collected to construct the experimental dataset. 
Here $k$ represents the $k$th microphone ($k=1,2,...I$) and $\tau$ is the maximum time lag. In both experimental campaigns a fixed time lag $\tau U_{\infty}/D = 38$ was chosen to guarantee that the dominant wave-packet structures are preserved in the cross-correlation functions, and $U_{\infty}$ represents jet exit velocity. 
For each velocity snapshot $u_i$ at any time $t$, the snapshot POD will provide the following reduced-order, energy-ranked representation following \cite{sirovich1987}:
\begin{equation}
{u}_i(\mathbf{x},t) \approx \sum_{j = 1}^N \phi_i^{(j)}(\mathbf{x})a^{(j)}(t)
\label{eqn:SE1}
\end{equation}
where $\phi_i^{(j)}(\mathbf{x})$ is the eigenfunction of the $j$th POD mode, and $a^{(j)}(t)$ represents the corresponding POD expansion coefficient. 
$N$ is the truncation of the first $N$th POD modes. 
Since the eigenfunctions are invariant in time, the original problem is equivalent to the TR-estimation of the POD expansion coefficients $\hat{a}(t)$. 
This problem can be addressed by the proposition of appropriate input-output models between time-lagged pressure and POD expansion coefficients. 
Once the models are well established, they can be deployed to estimate time-varying POD expansion coefficients from pressure inputs, and the results will serve to reconstruct TR-velocity field in combination with spatial POD eigenfunctions.

In this work two types of models were utilized to estimate the evolution of POD expansion coefficients in time. 
The first model is the spectral linear stochastic estimation (SLSE), which is termed "SLSE" in short. 
In addition, two neural network architectures were also introduced in this work to model the input-output relationship. 
Details of both techniques will be presented in the following.

\subsection{Velocity Estimation via SLSE-POD}
The SLSE-POD is a linear approach which aims to model the relationship between $\hat{a}$ and $p^k(t)$ from the following convolution equation:
\begin{equation}
    \hat{a}^{(j)}(t) = \sum_{k}\int_{-\tau}^{\tau} h^{(j)}_k(\tau')p^k(t-\tau')d\tau'.
    \label{eqn:SE2}
\end{equation}
Here $h^{(j)}_k$ is the weighting function which requires to be determined from the experimental dataset. \citet{zhang2020} has shown that this linear model is equivalent to the following form after taking the Fourier transform on both sides of \Cref{eqn:SE2}:
\begin{equation}
A(f) = H(f)P(f).
\label{eqn:SE3}
\end{equation}
Here $A$, $H$, $P$ are Fourier transforms of $\hat{a}$, [$h_1$, ..., $h_k$]$^T$, [$p^1$, ..., $p^k$]$^T$, respectively. 
$A\in\mathbb{C}(N\times 1)$, $H\in\mathbb{C}(N\times I)$ and $P\in\mathbb{C}(I\times 1)$. Multiplying the complex conjugate $P^*$ on both sides of \Cref{eqn:SE3} and taking the ensemble average $E(\cdot)$ over independent realizations, one will get:
\begin{equation}
G_{pa}(f) = H(f)G_{pp}(f).
\label{eqn:SE4}
\end{equation}
$G_{pp}\in\mathbb{C}(I\times I)$ is a Hermitian matrix in which diagonal terms are the auto-spectral density functions of pressure and off-diagonal terms are the cross-spectral density functions among different sensors. $G_{pa}\in\mathbb{C}(N\times I)$ is the cross spectral density function between pressure and POD coefficients. 
A direct solution of $G_{pa}$ requires time-resolved velocity measurements which is not available in the current work. 
As suggested by \citet{tinney2006}, an indirect approach was adopted in this study from the definition of cross-correlation between $a$ and $p$:
\begin{equation}
R_{ap}(\tau) = \langle a(t_n)p(t_n+\tau) \rangle.
\label{eqn:SE5}
\end{equation}
\Cref{eqn:SE5} indicates that $R_{ap}$ can be computed by orderly shifting the input signals with respect to each velocity snapshot and ensemble-averaging from all realizations. Since for stationary flows $R_{pa}(\tau) = R_{ap}(-\tau)$, the cross spectral density function $G_{pa}$ can be calculated by taking Fourier transform of $R_{ap}(-\tau)$. Consequently, the optimized weighting function $H(f)$ can be determined from the following expression:
\begin{equation}
H(f) = {G_{pa}}{G_{pp}}^{-1}.
\label{eqn:SE6}
\end{equation}

\Cref{eqn:SE6} is termed "spectral linear stochastic estimation" (SLSE). The core difference between the time-lagged LSE and the SLSE lies in the fact that the former method minimizes the mean-squared error in a macro sense while the latter performs a series of least-square regressions at all frequencies of interest. A thorough discussion of both techniques can be found in \citet{tinney2006}. Once the linear model $H(f)$ is optimized from the training dataset, a real-time prediction of the velocity field can be performed using pressure signals of length $T$ from the following equations:
\begin{equation}
  P = \text{FFT}(p),
\end{equation}
\begin{equation}
  \hat{A}(f) = H(f)P(f),
\end{equation}
\begin{equation}
  \hat{a}(t) = \text{IFFT}(\hat{A}).
  \label{eqn:SE7}
\end{equation}
The output of \Cref{eqn:SE7}, which is time-resolved estimation of POD expansion coefficients, will be used in \Cref{eqn:SE1} to get spatial-temporal estimation of the velocity field.

\subsubsection*{Wavelet-Based Filtering (WF) of Cross-Correlation Functions}
\begin{figure}[!h]
\centering
\begin{subfigure}{0.49\textwidth}
\includegraphics[width=.99\linewidth]{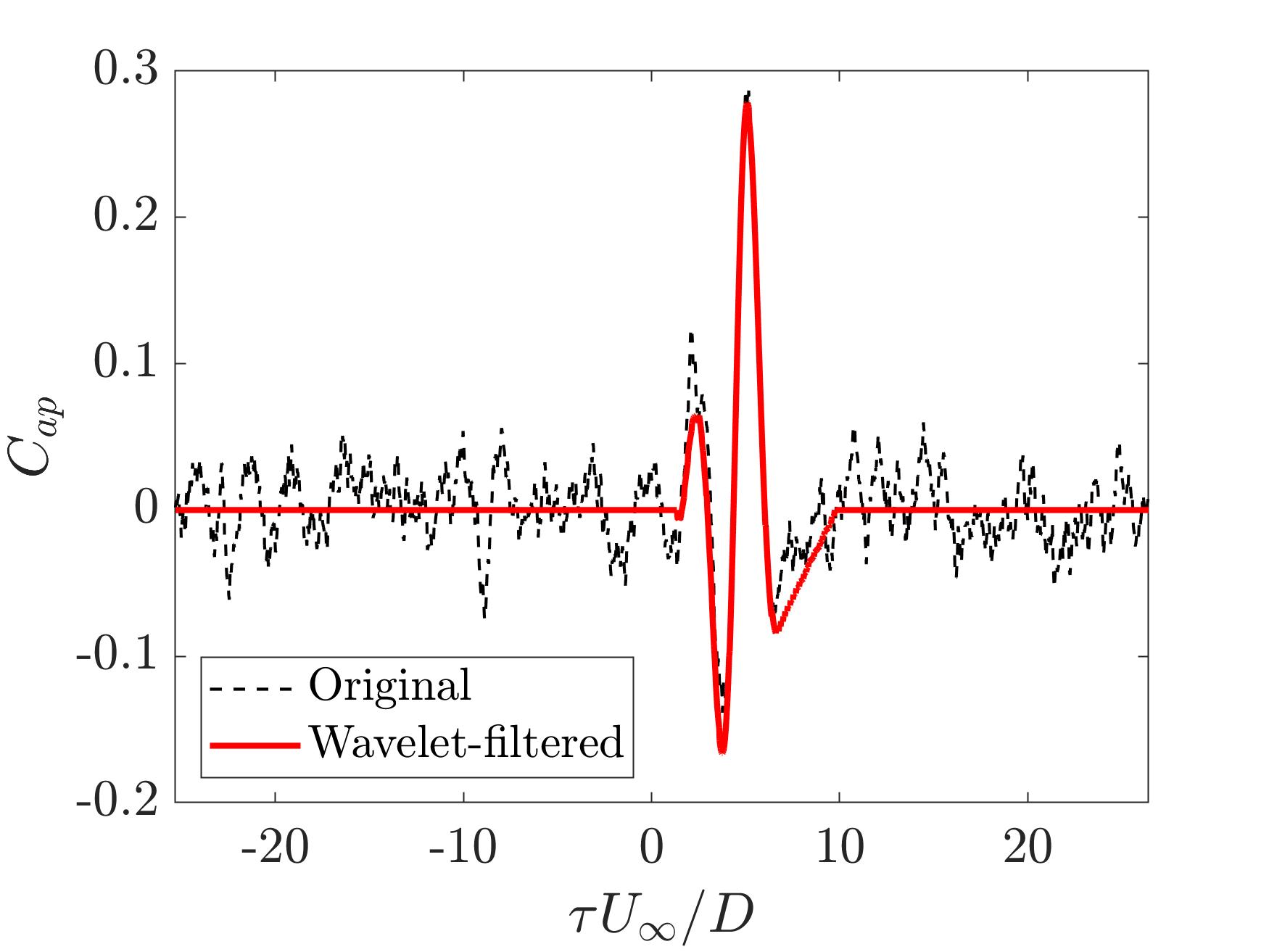}
\caption{}
\label{fig:cap}
\end{subfigure}
\begin{subfigure}{0.49\textwidth}
\includegraphics[width=.99\linewidth]{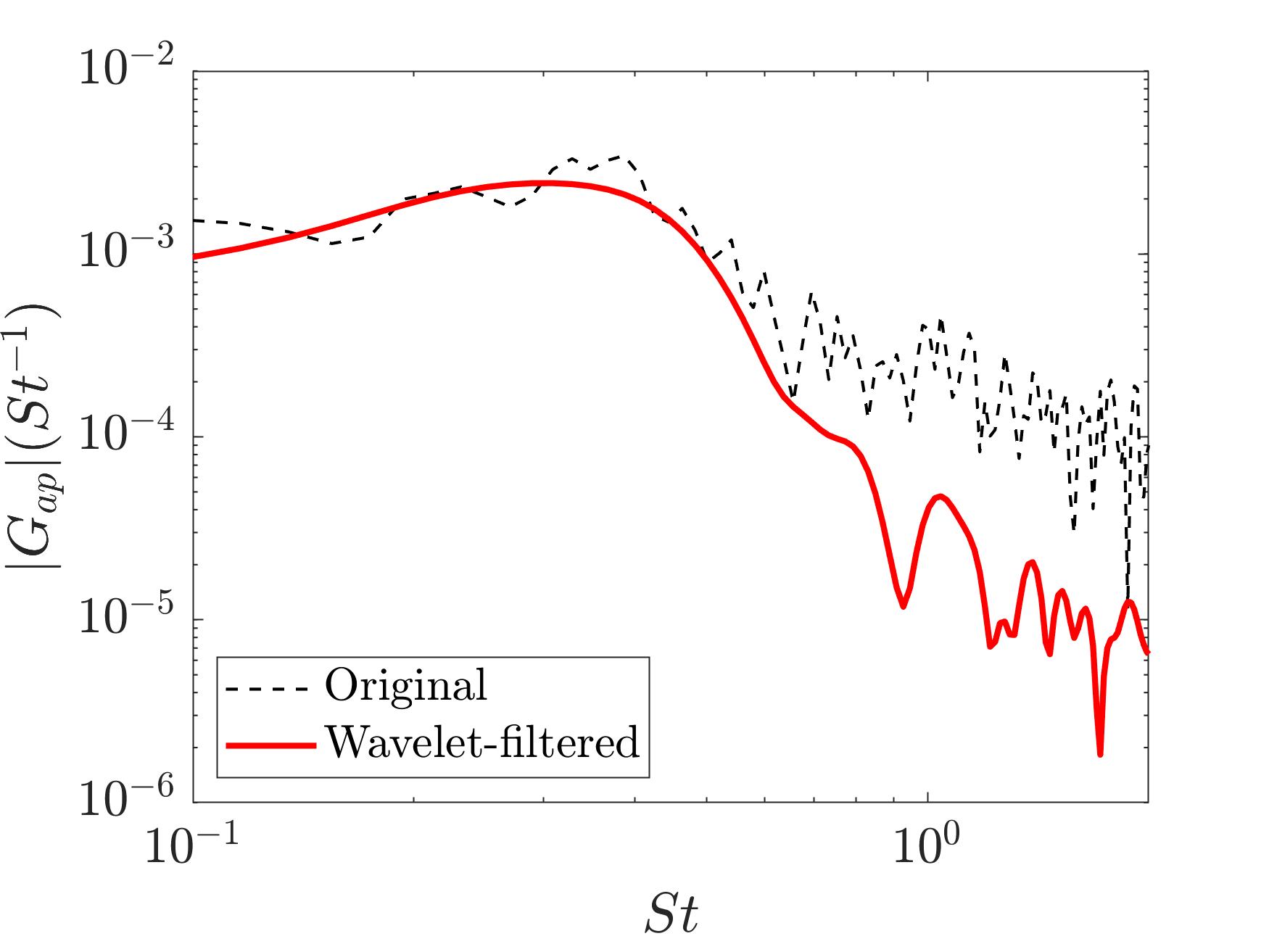}
\caption{}
\label{fig:gap}
\end{subfigure}
\caption{(a) A comparison of original and wavelet-filtered cross-correlation coefficient between the centerline pressure fluctuations at $x/D = 8$ and the first POD mode of the axisymmetric streamwise velocity at $x/D = 4$; (b) same functions in the Fourier domain.}
\label{fig:wavelet-filter}
\end{figure}

\Cref{eqn:SE5} provides an alternative approach to calculate cross-correlation between pressure and POD expansion coefficients from non-TR velocity data. 
However, in practice extraneous noise will appear in the calculation of $R_{ap}$, and will adversely influence the outcome from SE. 
To attenuate the noise level, a wavelet-based filter was implemented in this work to zero out the portion of the cross-correlation function which possesses a weak or none wave-packet shape.
The continuous wavelet transform (CWT) of a time sequence $x$ is defined as a convolution between the original signal and the wavelet function $\psi$ translated to different scales $s$:
\begin{equation}
w(s,t)=\int_{-\infty}^{\infty} x(\tau)\psi^*(\frac{t-\tau}{s})d\tau.
\label{eqn:CWT1}
\end{equation}
Here the asterisk ($^*$) indicates the complex conjugate. Details of the continuous wavelet transform can be found in \cite{mallat1999} and \cite{torrence1998}.
The validity of the wavelet-based filter originated from the fact that if the turbulent velocity (i.e. the dominant POD eigenfunctions) and pressure fluctuations both possess wave-like behaviors driven by the same coherent events in the flow, the cross-correlation function between the two quantities should also be dominated by some wave-packet shape. 
Hence, a wavelet-based filter can be implemented to preserve the major wave-like events in the cross-correlation function and effectively reject the noise. 
The implementation of the wavelet-based filter starts from the transformation of $R_{ap}(\tau)$ into the time-scale domain, which will yield complex-valued wavelet coefficients $w(\tau,s)$. 
Next, a real-valued threshold $T$ is imposed to obtain the filtered wavelet coefficients $w'$ such that:
\begin{equation}
w'(\tau,s) = \left\{
\begin{array}{ll}
  w(\tau,s) \text{, if } \lvert w(\tau,s) \rvert >T;\\
  0 \text{, otherwise.}
\end{array}
\right.
\end{equation}
In this work, the complex-valued Morlet wavelet function with the angular frequency $\omega_0=6$ was selected to better highlight the wave-packet events in the correlation results.
The threshold was empirically selected as $T = 0.3 \text{ max}(|w|)$, and the filtered cross-correlation function $R'_{ap}$ was obtained from the inverse wavelet transform of $w'$ \cite{mallat1999}.
An exemplary comparison between raw and filtered cross-correlation coefficients (see \Cref{eqn:xcorrcoeff} for definition) is displayed in \Cref{fig:cap}, from which the filtered function is seen to better highlight the portion of the signal with prominent wave-packet shape and high correlation level. 
The comparison is also performed in the Fourier space in \Cref{fig:gap}, where the cross-spectrum is found to be significantly smoothed between $0.1\leq St\leq 1$ after filtering.
Although the spectral shape at higher frequencies doesn't exhibit significant improvement, the noise amplitude is still significantly brought down. 
Since current discussions primarily focus on the coherent structures in the jet shear layer which governs the hydrodynamic hump in the low frequency range, the filtered results remains satisfying and will be employed in the following discussions.

\subsection{Velocity Estimation via Neural Networks}

\begin{figure}[!h]
  \centering
  \includegraphics[width=0.5\linewidth]{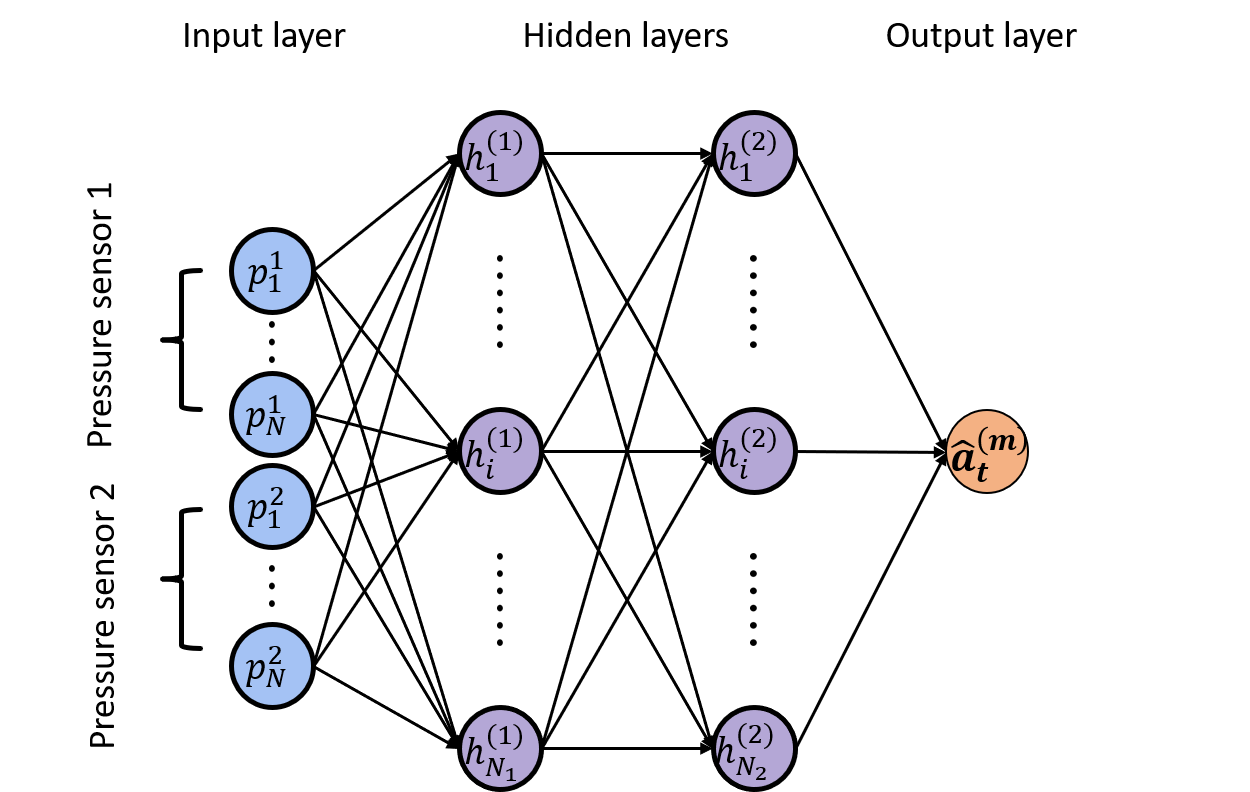}  
  \caption{POD coefficient estimation based on the MLP.}
  \label{fig:mlp}
\end{figure}

For the experimental dataset with synchronized streamwise PIV and two-point pressure measurements, two neural network architectures were proposed in this study to establish the connection between pressure and POD expansion coefficients as an alternative approach. 
The first NN architecture is a many-to-one model which originates from the multi-layer perceptron (MLP) and the schematic is presented in \Cref{fig:mlp}. For each POD mode $m$, the input layer collects time-lagged pressure data from two in-flow microphones $(p_1^1\sim p_N^1,p_1^2\sim p_N^2)$.
The subscripts $1\sim N$ denote the time-lagged discrete data points and the superscripts $1,2$ are the indices of the in-flow microphones. 
Two fully connected linear hidden layers are included in the architecture, each contains 1024 hidden units ($N_1=N_2=1024$). 
In the hidden layers, linear operators are applied to upscale the input data to a high dimensional space. 
The output layer compresses the feature outputs from the last hidden layer into a $1\times 1$ scalar with a nonlinear tanh activation function, which represents the prediction of the POD expansion coefficient ($\hat{a}^{(m)}_t$).

\begin{figure}[!h]
  \centering
  \begin{subfigure}{0.49\textwidth}
  \includegraphics[width=0.9\linewidth]{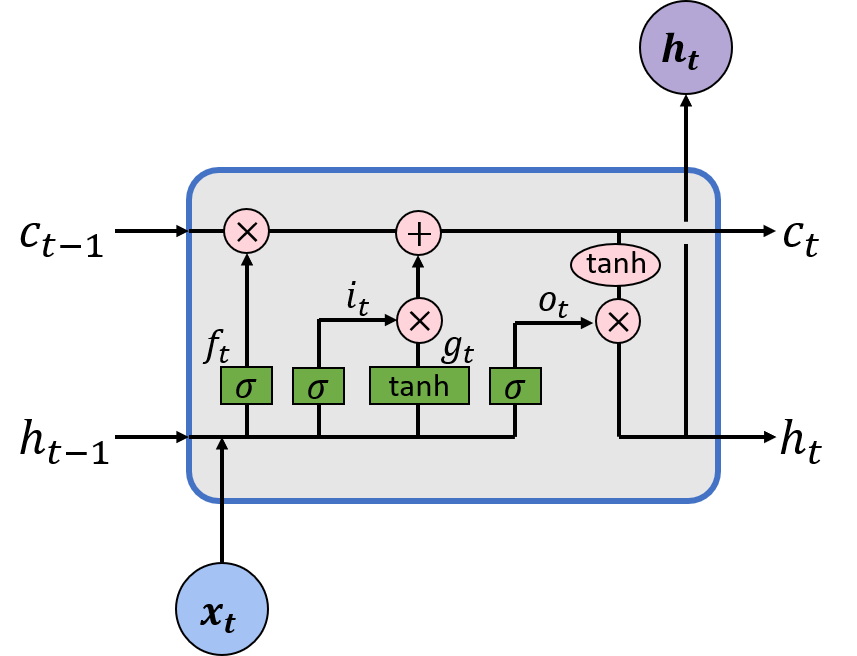} 
  \caption{}
  \label{fig:lstm_1}
  \end{subfigure}
  \begin{subfigure}{0.49\textwidth}
  \includegraphics[width=0.9\linewidth]{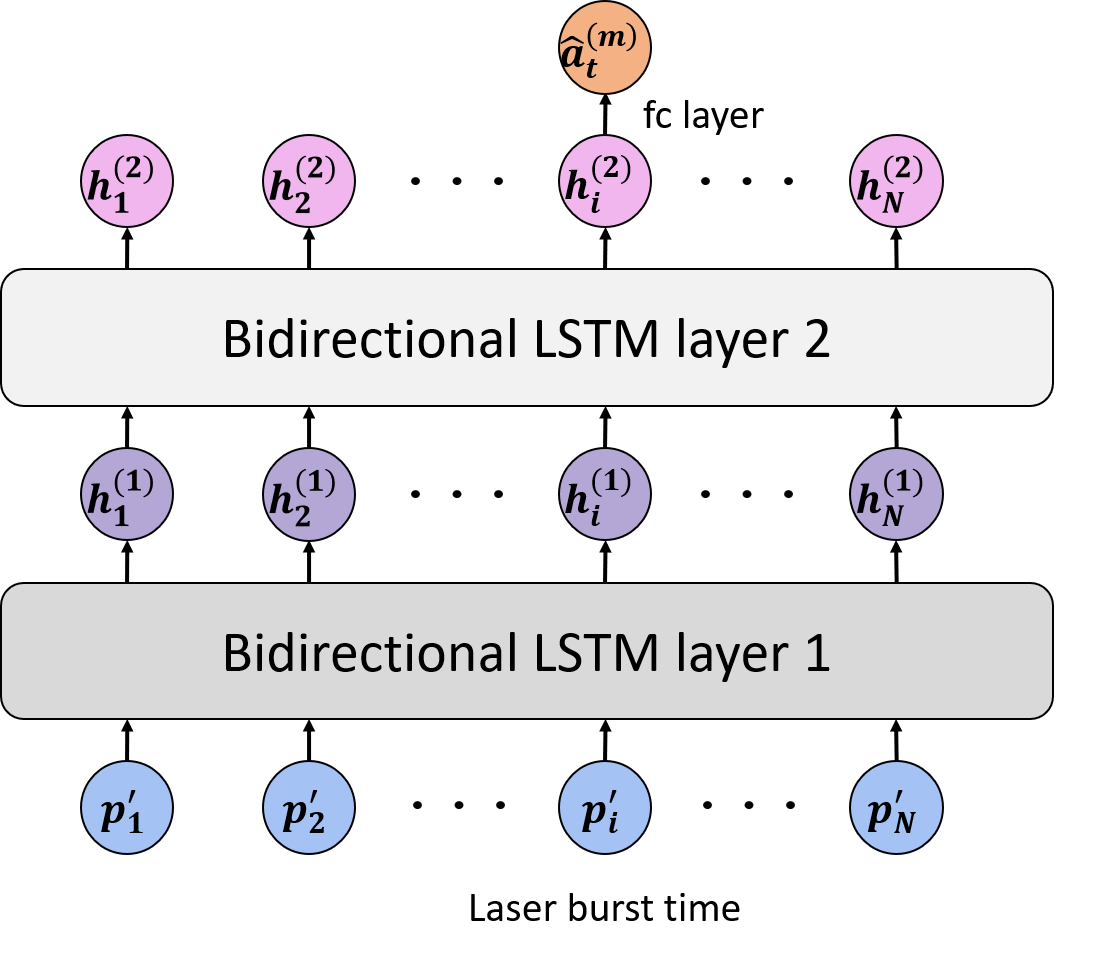} 
  \caption{}
  \label{fig:lstm_2}
  \end{subfigure}
  \caption{(a) An LSTM cell at time $t$; (b) The architecture to estimate POD coefficient based on bidirectional LSTM.}
  \label{fig:lstm}
\end{figure}

The second NN architecture is constructed based on the bidirectional long-short-term-memory (LSTM) \cite{hochreiter1997}. 
An LSTM cell is composed of three gates: an input gate $i_t$, an output gate $o_t$, and a forget gate $f_t$. 
These gates allow LSTM cells to retain and process information over long periods of time, hence minimizing the effect of a vanishing gradient \cite{aggarwal2018}.
Following the configuration in \Cref{fig:lstm_1}, an LSTM cell at time $t$ computes the following functions: 
\begin{equation}
i_t = \sigma(W_ix_t+U_ih_{t-1}+b_i),
\label{eqn:lstm1}
\end{equation}
\begin{equation}
f_t = \sigma(W_fx_t+U_fh_{t-1}+b_f),
\label{eqn:lstm2}
\end{equation}
\begin{equation}
g_t = \text{tanh}(W_gx_t+U_gh_{t-1}+b_g),
\label{eqn:lstm3}
\end{equation}
\begin{equation}
o_t = \sigma(W_ox_t+U_oh_{t-1}+b_o),
\label{eqn:lstm4}
\end{equation}
\begin{equation}
c_t = f_t\odot c_{t-1}+ i_t\odot g_t,
\label{eqn:lstm5}
\end{equation}
\begin{equation}
h_t = o_t\odot \text{tanh}(c_t).
\label{eqn:lstm6}
\end{equation}
In the equations above, $x_t$ is the input at time $t$, $h_t$ and $c_t$ are the hidden state and the cell state at time $t$.
$g_t$ is the new memory cell. 
$i_t$, $f_t$, $o_t$ are the input, forget, and output gates, respectively. 
$\sigma$ represents the sigmoid function, and $\odot$ denotes the element-wise product. 
$W$, $U$ and $b$ are unknown parameters which need to be learned from the training dataset. 
In an LSTM layer the hidden state $h_t$ only receive information from its past states yet the future states are not included. 
The bidirectional LSTM (\cite{schuster1997}) overcomes this issue by concatenating the outputs from two LSTM layers of opposite directions.
With this form, outputs from a bidirectional LSTM layer will encompass information from both past and future states simultaneously. 

Taking the advantage of LSTM on handling time sequence inputs, the second NN includes two bidirectional LSTM layers and a feed-forward output layer, and the schematic is presented in \Cref{fig:lstm_2}. 
Pressure signals measured from two microphones at any time $t_j$ are concatenated into a $2\times 1$ input vector $p'_j= [p_j^1,p_j^2]^T$. 
The laser burst time $t_i$ is indicated by the subscript $i$ for all variables. 
In this network, sequentially arranged inputs $p'_j\sim p'_N$ are fed into the first bidirectional LSTM layer which generates a series of hidden state vectors $h_1^{(1)}\sim h_N^{(1)}$. 
Then the second bidirectional LSTM layer takes in these vectors and computes the new hidden state vectors $h^{(2)}$s in a similar approach. 
The output sizes of both hidden layers were set to 64. 
To better highlight the information around the PIV burst moment which is highly related to the expected output, $h_i^{(2)}$ is connected to the fully-connected output layer to obtain POD coefficient estimates of the $m$th mode $\hat{a}^{(m)}_t$, in which a linear mapping function as well as a tanh activation function is included in the output layer.

Both NN architectures were trained based on the mini-batch gradient descent optimization \cite{goodfellow2016}. 
The experimental dataset contains the modal expansion coefficients of the leading POD modes from $N_s=8000$ mutually independent velocity snapshots as well as the time-lagged pressure signals. 
These samples were randomly shuffled, 80$\%$ were used to train the NN models and the rest for validation. 
For each mode, pressure and POD coefficients were re-scaled into $[-1,1]$. 
To evaluate the model performance, estimation results from the training data $\hat{a}^{(m)}$ were compared to the known POD coefficients from PIV snapshots during the iterative training process.
The mean squared error (MSE) loss was utilized as the criterion of the above-mentioned regression problem: 
\begin{equation}
  L^m = \frac{1}{N_s}\sum_{i=1}^{N_s}(a_i^{(m)}-\hat{a}_i^{(m)})^2.
\label{MSE_loss}
\end{equation}
In contrast to the "flow" direction of the input data, gradients of the loss function were back-propagated from the output layer to the neutral network to optimize the unknown weights and biases in the architectures. The Adaptive Moment Estimation (ADAM, \cite{kingma2014}) was chosen as the optimizer, and the initial learning rate was set to 1e-3. 
Each model was trained with 200 epochs for every POD mode and the size of every mini-batch was set to 64. 
Models with the best performance on the validation set were recorded during iterations and were used for the velocity estimation from pressure inputs. 
The training process in this study was carried out on a cloud-based platform with four Normalized Graphics Processor Units (NGUs).

\section{\label{sec:results}Results and Discussions}
In the following section the results of the modal analysis of the experimentally measured velocity fields and pressure informed velocity estimation will be presented.
\subsection{Estimation of Axisymmetric Velocity from Cross-Stream PIV and In-Flow Pressure Measurements}
As introduced in the previous section, single-point pressure measurements were taken synchronously with stereo-PIV on a series of cross-stream planes to investigate the space-time dynamics of axisymmetric velocity components in the jet. 
To extract the axisymmetric velocity component from the PIV snapshots, velocity fields measured under the Cartesian coordinate were first mapped onto a polar grid such that the axisymmetric velocity $(u^{(0)}_x,u^{(0)}_r)$ can be extracted accordingly. 
We note that although small portion of the azimuthal velocity component $u^{(0)}_\theta$ still exists after the grid transformation, theoretical work from \citet{batchelor1962} has shown that the mode-0 azimuthal velocity should be zero in round jets. 
Hence the existing azimuthal component are believed to be the residue introduced from the measurement uncertainty and is excepted from the following analysis. 

After obtaining 1600 samples of $(u^{(0)}_x,u^{(0)}_r)$ at each streamwise location, the azimuthal-Fourier POD (\cite{tinney2008,li2020}) was performed to extract the most energetic spatial structures in the radial direction.
The only difference between the azimuthal-Fourier POD and the snapshot POD originates from an additional weighting factor $r$ in the eigenvalue decomposition problem, and the corresponding strategy to tackle this problem has been thoroughly explained in \cite{citriniti2000jfm}.

\begin{figure}[!b]
\centering
    \begin{subfigure}{0.49\textwidth}
    \includegraphics[width=0.99\textwidth]{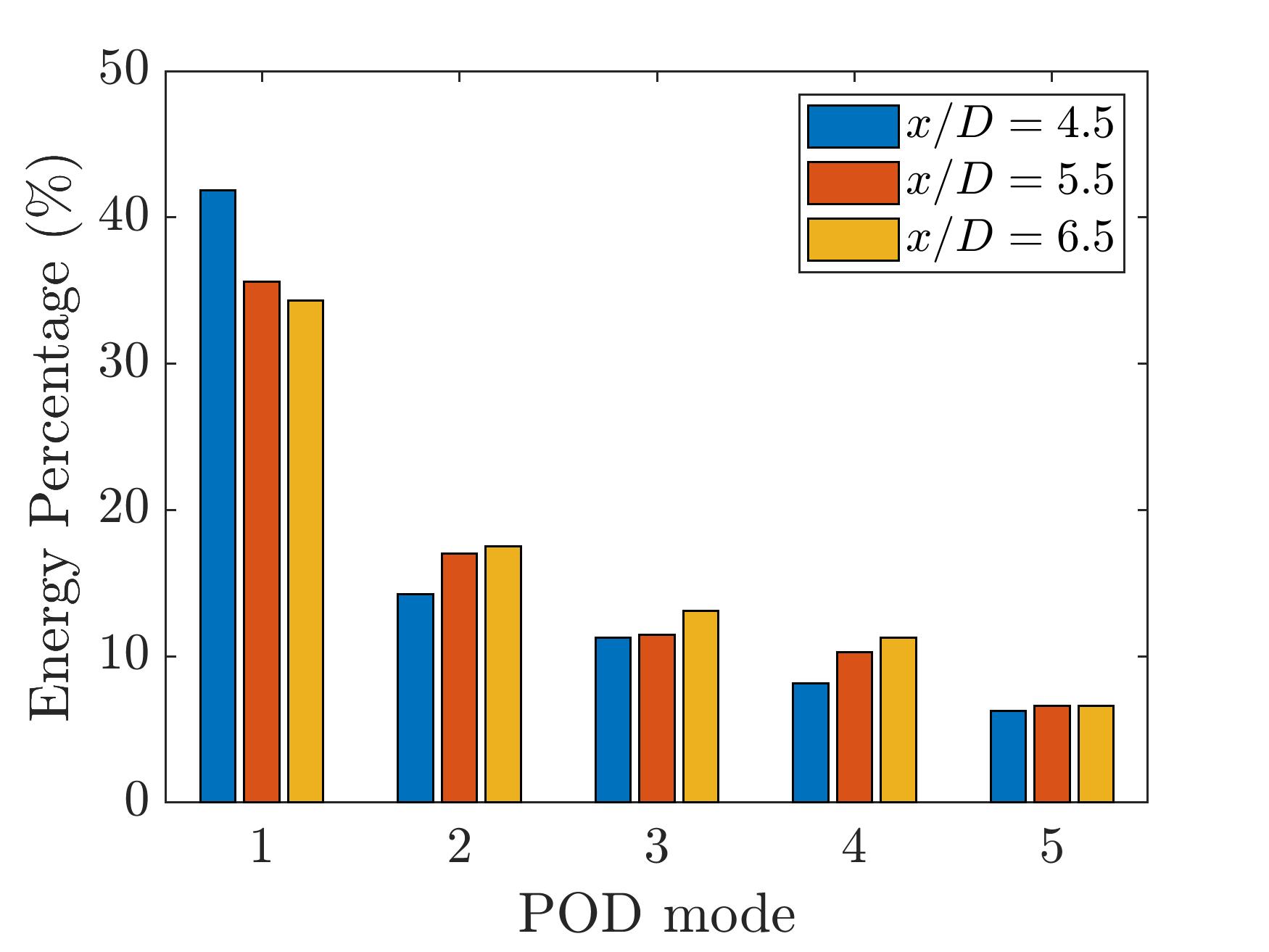}
    \caption{Streamwise POD}
    \end{subfigure}
    \begin{subfigure}{0.49\textwidth} 
    \includegraphics[width=0.99\textwidth]{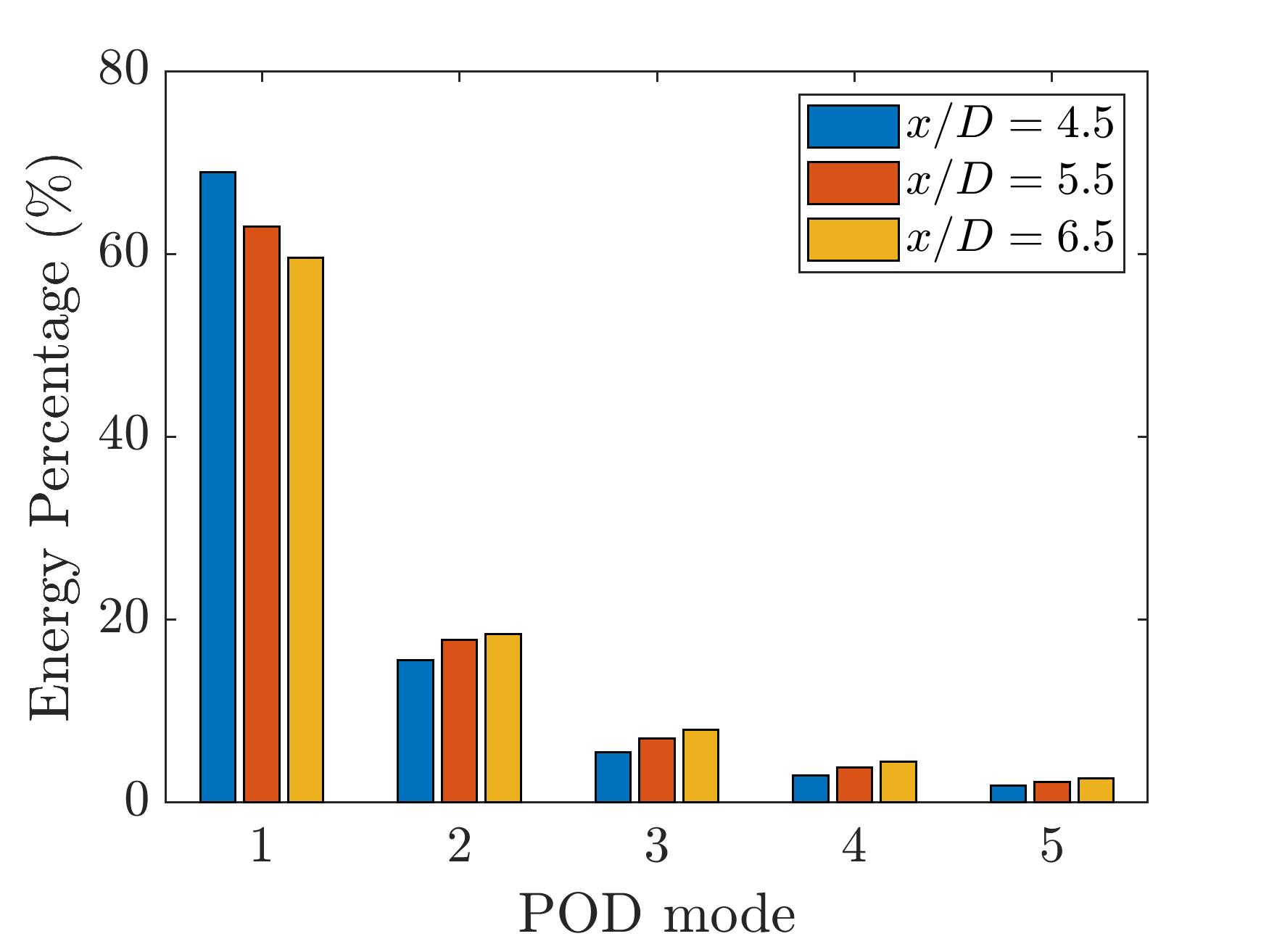}
    \caption{Radial POD}
    \end{subfigure}
\caption{Energy distribution of the first 5 POD modes at $x/D = 4.5$, $5.5$, and $6.5$.}
\label{fig:energy_3d}
\end{figure}

As suggested by \citet{tinney2008}, a $90^\circ$ phase difference exists between the streamwise velocity and its radial counterpart. 
To better preserve the phase difference between both velocity components and maintain the completeness of the turbulent structures in each direction, the scalar-based azimuthal-Fourier POD was performed on each velocity component individually.
\Cref{fig:energy_3d} presents the energy distribution of the first 5 POD modes at three representative axial locations from the streamwise and the radial POD, respectively. 
Strong mode-1 dominance for both velocity components can be observed at all three streamwise locations, and the superposition of the first 2 POD modes takes up around $60\%$ of the energy in the streamwise direction and $80\%$ in the radial direction. 
These results imply that at each cross-stream station, the dominant spatial features of the axisymmetric turbulent velocity can be effectively captured from the first few modes of the reduced-order representation, and this property will be utilized to greatly simplify the computational cost of the stochastic estimation.

\begin{figure}[!b]
\centering
    \begin{subfigure}{0.49\textwidth}
    \includegraphics[width=0.99\textwidth]{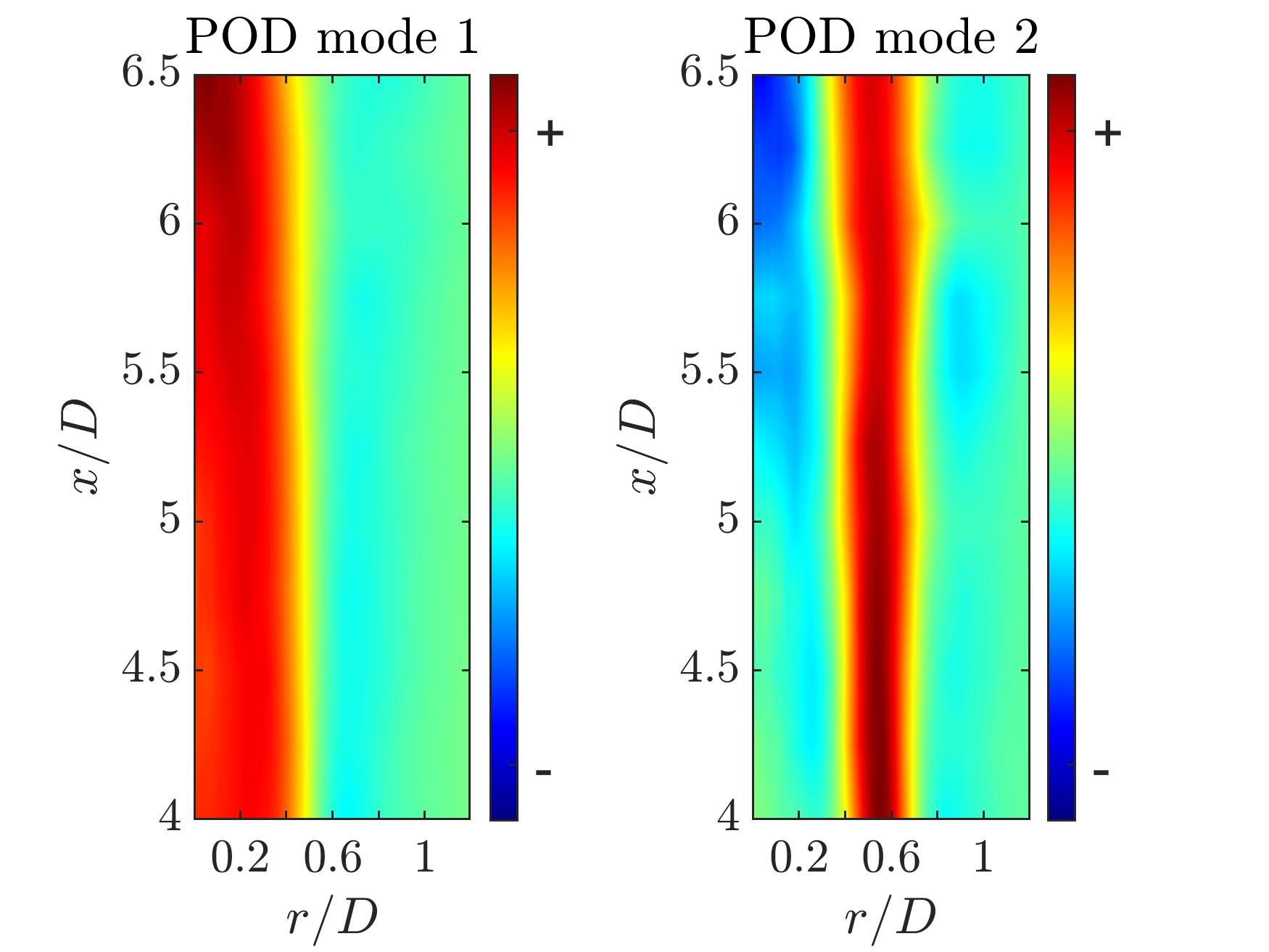}
    \caption{$\phi_x^{(0)}$ from streamwise POD}
    \end{subfigure}
    \begin{subfigure}{0.49\textwidth}
    \includegraphics[width=0.99\textwidth]{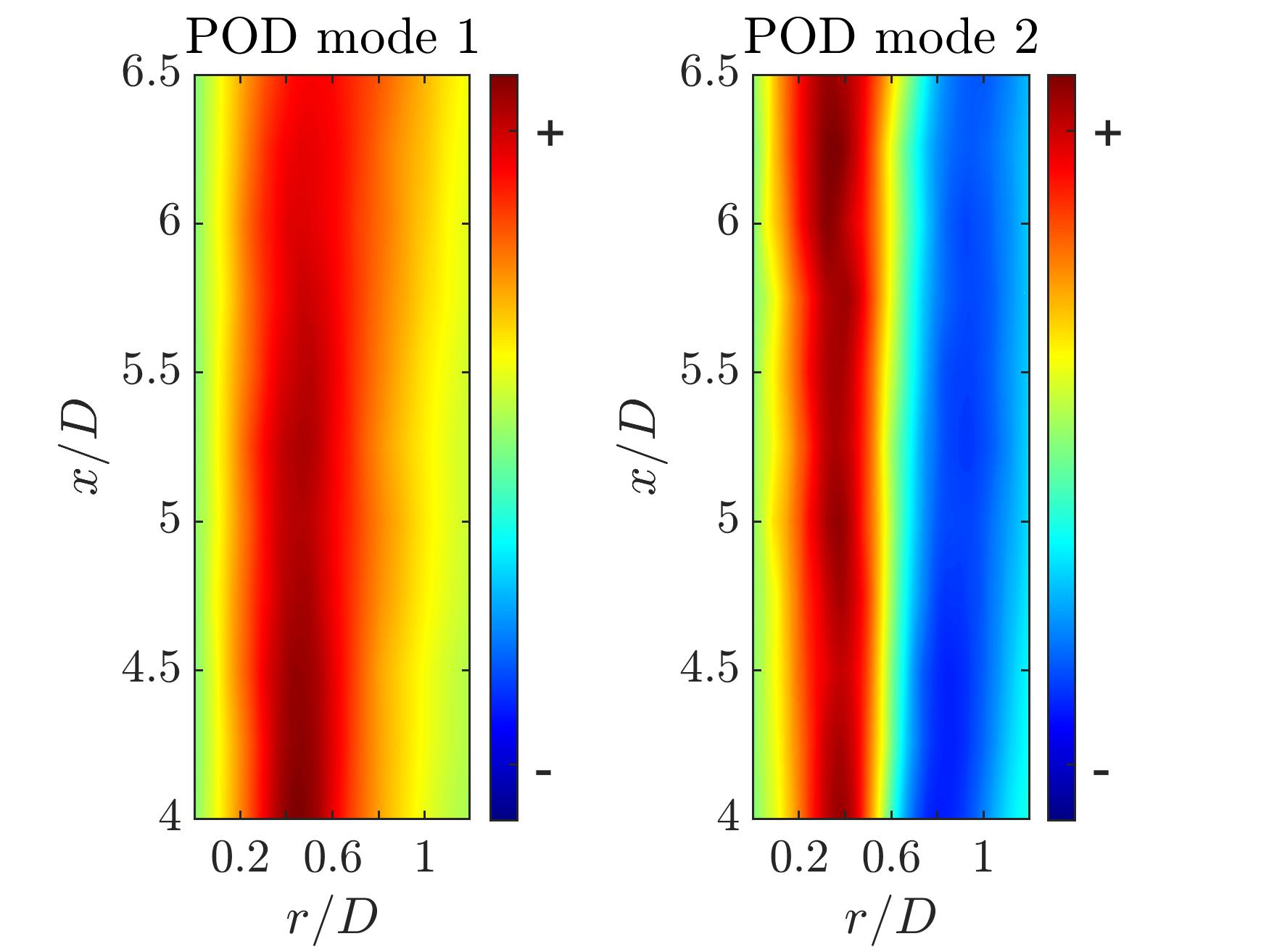}
    \caption{$\phi_r^{(0)}$ from radial POD}
    \end{subfigure}
\caption{Streamwise evolution of the first two POD eigenfunctions.}
\label{fig:eign_3d}
\end{figure}

\Cref{fig:eign_3d} shows the shapes of the first two eigenfunctions obtained from the azimuthal-Fourier POD. The streamwise POD eigenfunctions are seen to evolve along the jet axis while the radial modes will slightly tilt toward the ambient side at downstream locations. 
The most energetic structures of the first streamwise POD mode are mostly concentrated within $r/D \leqslant 0.5$, and the first radial mode highlights energetic events between $r/D = 0.2 \sim 0.8$. 
Furthermore, the second streamwise mode represents the convection of some radially compact turbulent structures around the jet lipline, accompanied by two opposite-signed side-lobes at relatively low amplitudes. Meanwhile the second radial eigenfunction depicts the appearance of injection-ejection events across the jet lipline.

\begin{figure}[!h]
\centering
\begin{subfigure}{0.45\textwidth}
    \includegraphics[width=0.99\textwidth]{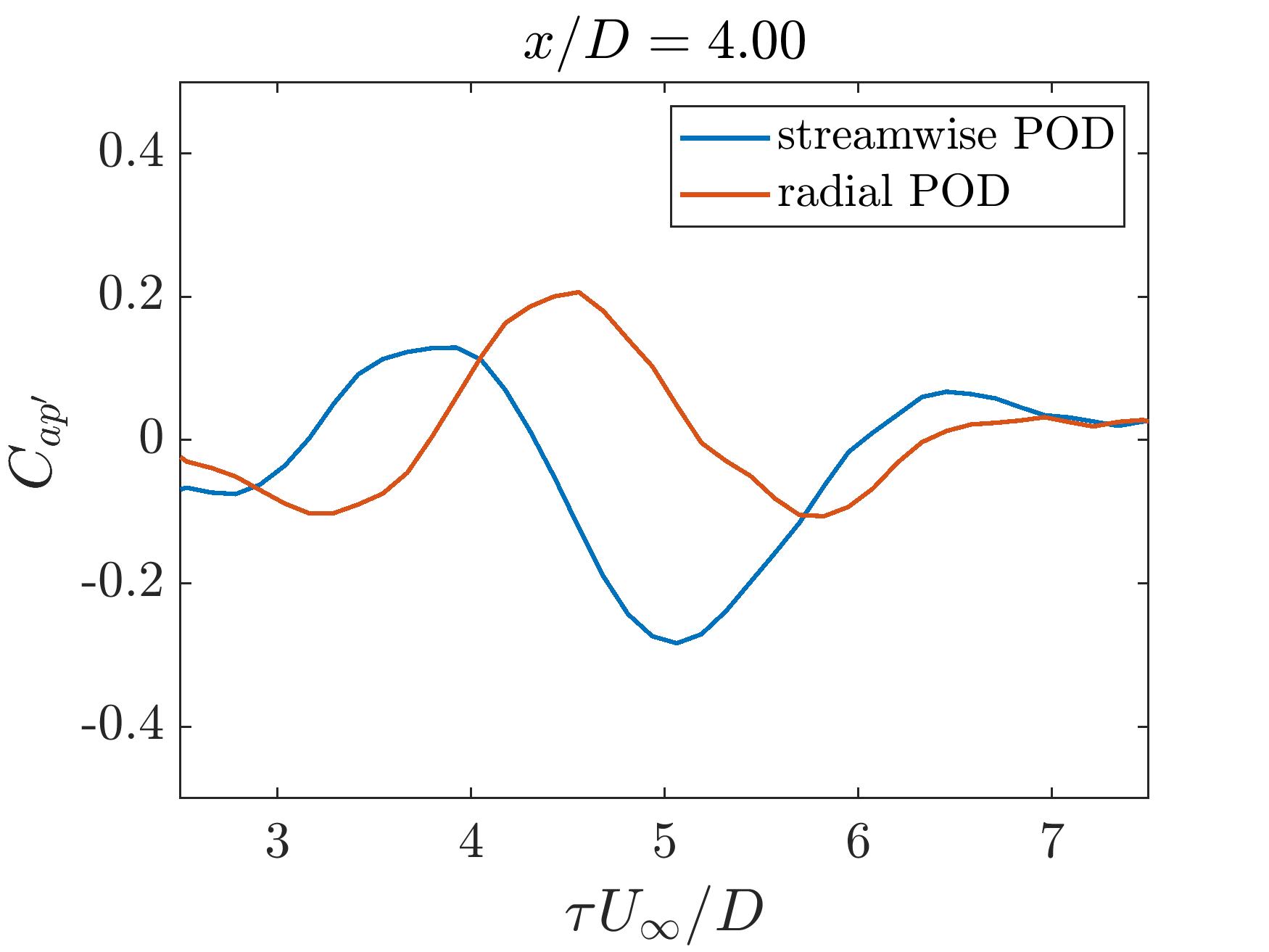}
    \caption{}
\end{subfigure}
\begin{subfigure}{0.45\textwidth}
    \includegraphics[width=0.99\textwidth]{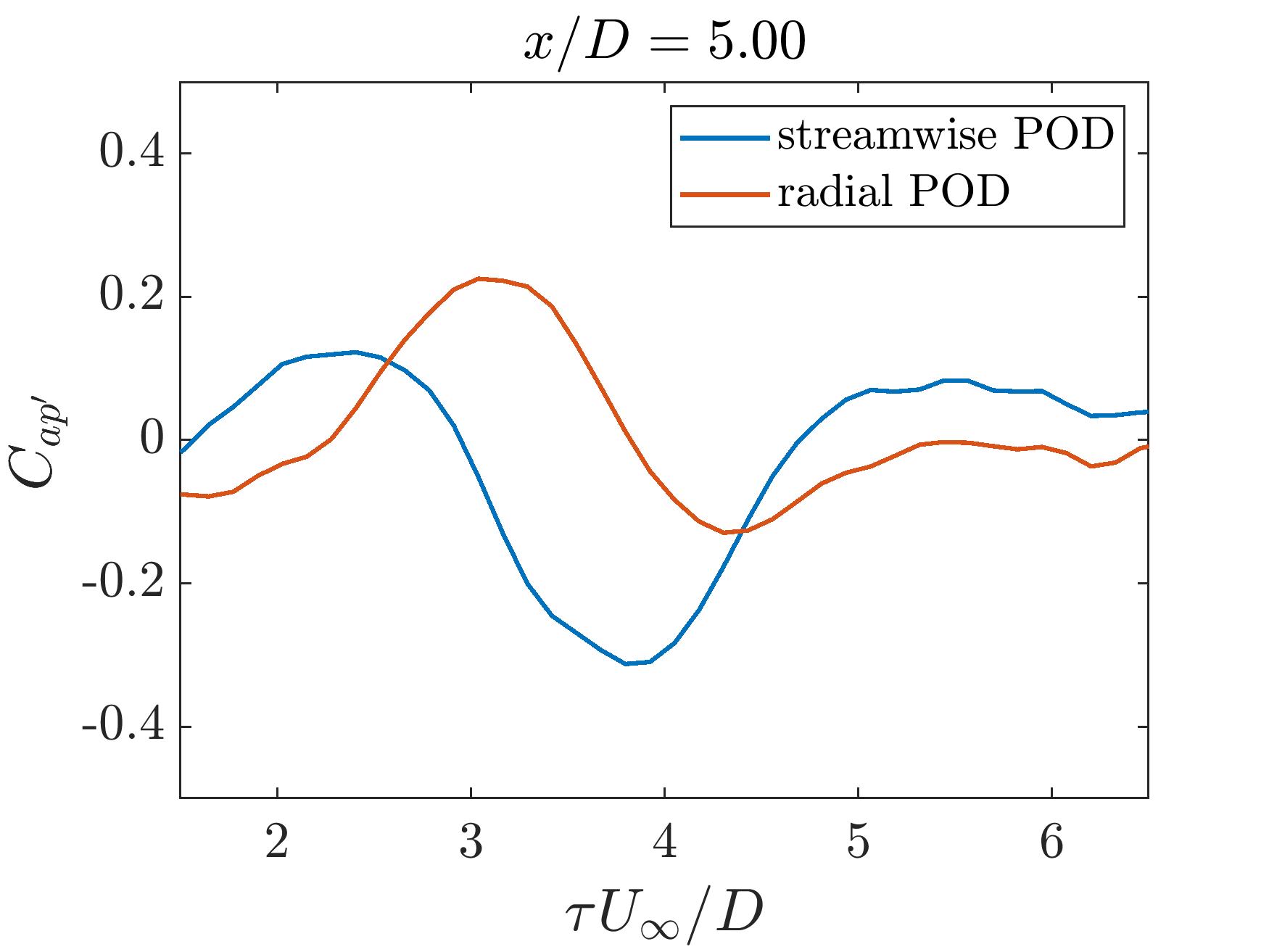}
    \caption{}
\end{subfigure}\\
\begin{subfigure}{0.45\textwidth}
    \includegraphics[width=0.99\textwidth]{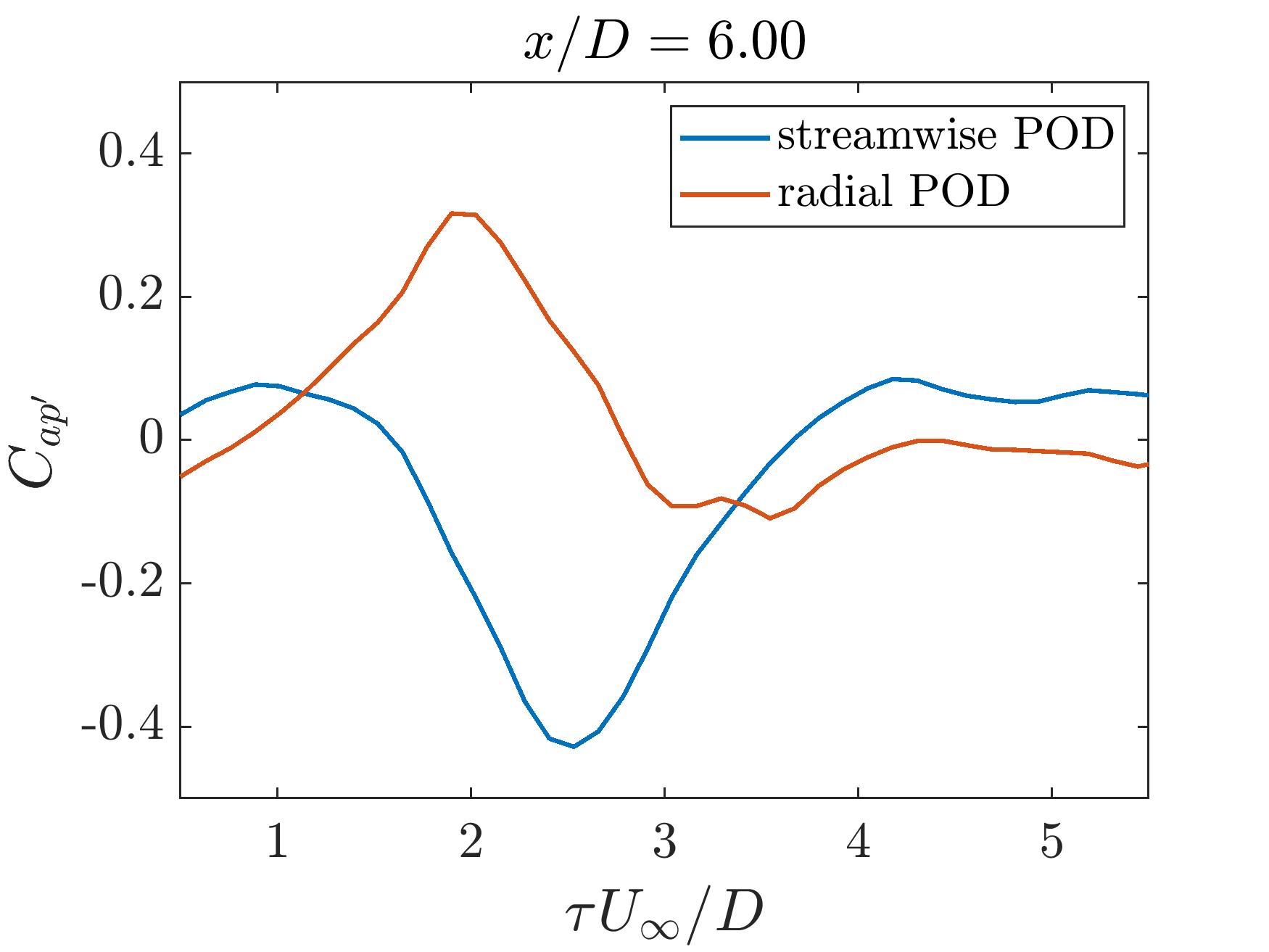}
    \caption{}
\end{subfigure}
\caption{Cross-correlation coefficients between the first POD expansion coefficients and pressure ($C_{ap}$) at $x/D = 4,5,6$. Pressure lowpassed at 2000 Hz ($St \approx 1$).}
\label{fig:xcorr_3d}
\end{figure}

Since SLSE-POD is a correlation-based technique, the time-lagged cross-correlation coefficients between the first POD modal expansion coefficients and the in-flow pressure fluctuations ($C_{ap}$) are first examined, and the results at $x/D = 4,5,6$ are presented in \Cref{fig:xcorr_3d}. Here $C_{ap}$ is defined as:
\begin{equation}
C_{ap}(\tau) = \frac{\langle a({t_n})p(t_n+\tau)\rangle_n}{{rms}(a){rms}(p)}.
\label{eqn:xcorrcoeff}
\end{equation}   
In an attempt to focus on the coherent, large-scale motions inside the flow, pressure in \Cref{fig:xcorr_3d} was lowpass filtered at 2000 Hz which corresponds to a Strouhal number of $St \approx 1$. 
In all instances, wave-like patterns can be clearly observed from cross-correlation curves, and the correlation level gradually increases as the velocity measurement plane moves closer to the pressure probe. 
For both the first streamwise POD mode and the first radial POD mode, the peak values of $|C_{ap}|$ always exhibit similar amplitudes at each axial location, and an apparent phase difference of $\sim 90^\circ$ between two curves can be clearly identified. 
This is consistent to the observation in \cite{lau1972} where the axial velocity was found to be in antiphase with in-flow pressure fluctuations whereas the radial velocity was $90^{\circ}$ out-of phase to pressure.

\begin{figure}[!t]
\centering
\begin{subfigure}{0.45\textwidth}
\includegraphics[width=0.99\textwidth]{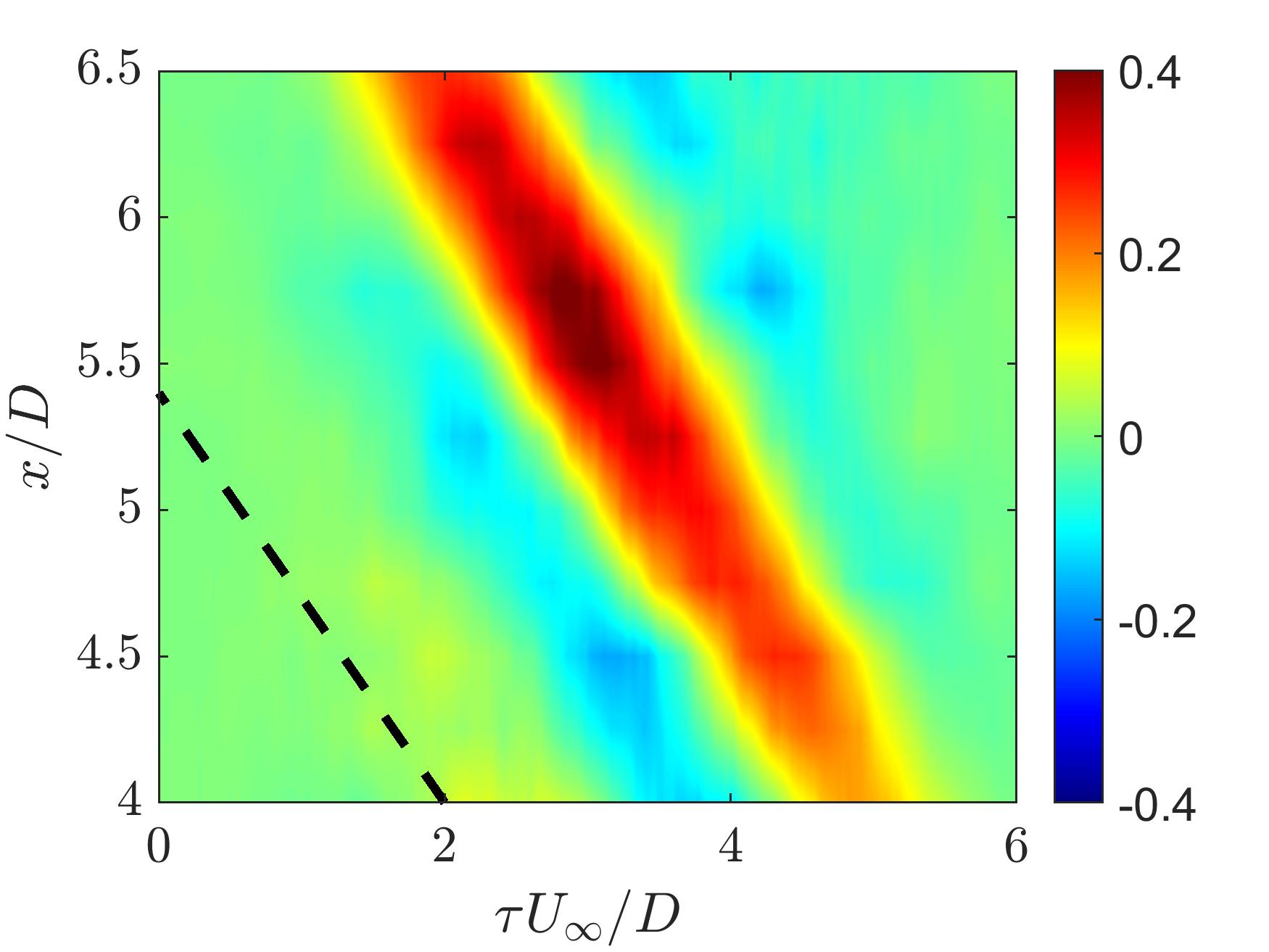}
\caption{Streamwise POD}
\end{subfigure}
\begin{subfigure}{0.45\textwidth}
\includegraphics[width=0.99\textwidth]{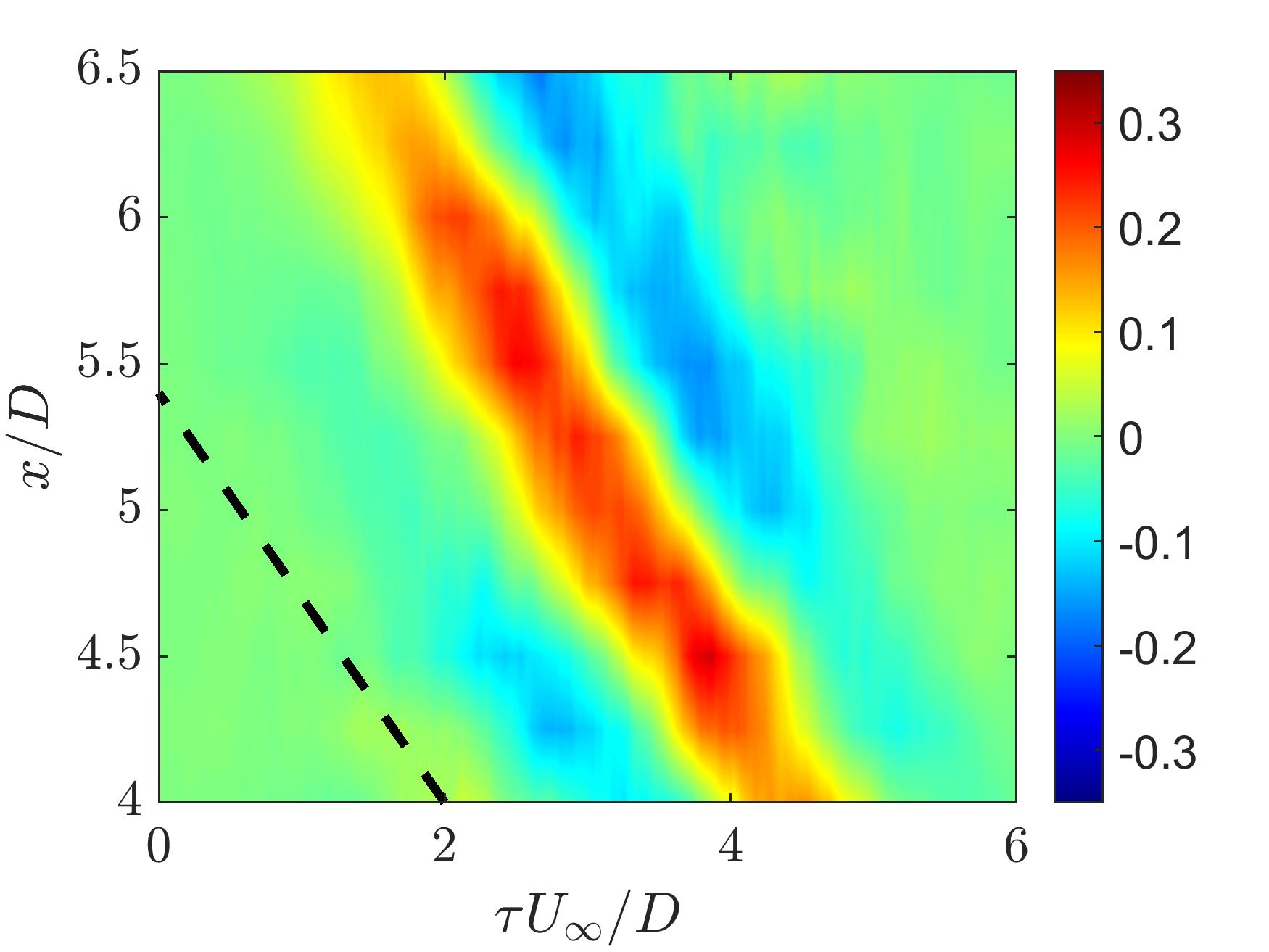}
\caption{Radial POD}
\end{subfigure}          
\caption{Evolution of cross-correlation coefficient $C_{ap}$ between the first POD mode and pressure fluctuations. Pressure lowpassed at 2000 Hz ($St\approx 1$) and dashed lines represent an empirical convective speed of $U_c = 0.7 U_{\infty}$.}
\label{fig:xcorr_3d_global}
\end{figure}

The space–time correlations $C_{ap}$ calculated from all 11 cross-stream measurement planes are presented in \Cref{fig:xcorr_3d_global}.
Since the correlation with higher order POD modes will decay rapidly, only results from the first POD mode is shown in the figure. 
Similar to what has been observed in \Cref{fig:xcorr_3d}, wave-packet structures can be found at all streamwise locations with a nearly uniform convective speed.
To quantify this convective behavior, the propagative pattern in \Cref{fig:xcorr_3d} is compared to an empirical convective speed of $U_c = 0.7U_{\infty}$ \citep{breakey2014} which is represented by dashed lines. 
Both velocity components yield good agreements with the empirical slope, which indicates that $U_c = 0.7U_{\infty}$ is a good representation of the convective speed of the axisymmetric wave-packets.

\begin{figure}[!t]
\centering
\begin{subfigure}{0.45\textwidth}
\includegraphics[width=0.99\textwidth]{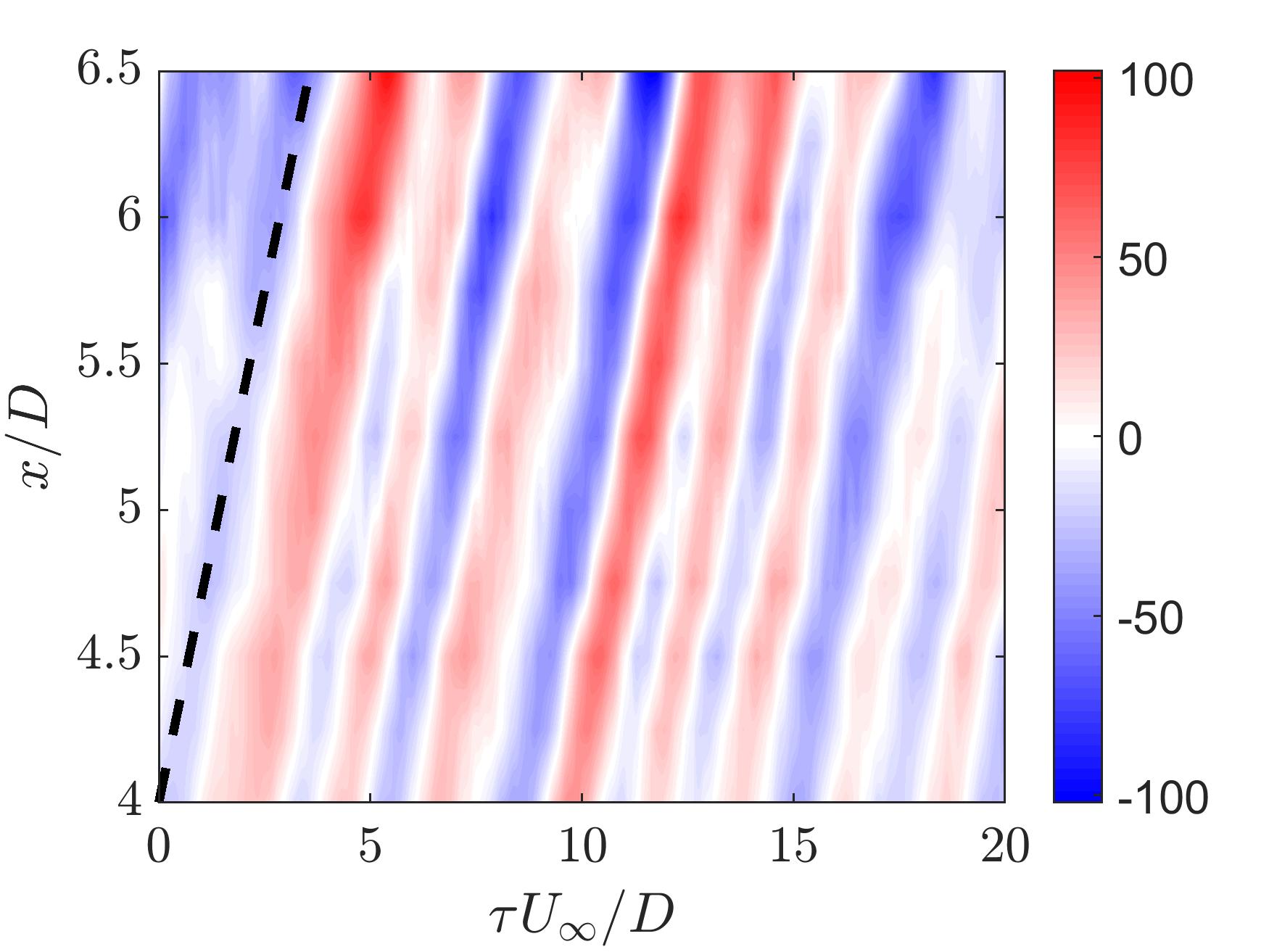}
\caption{Streamwise POD}
\end{subfigure}
\begin{subfigure}{0.45\textwidth}
\includegraphics[width=0.99\textwidth]{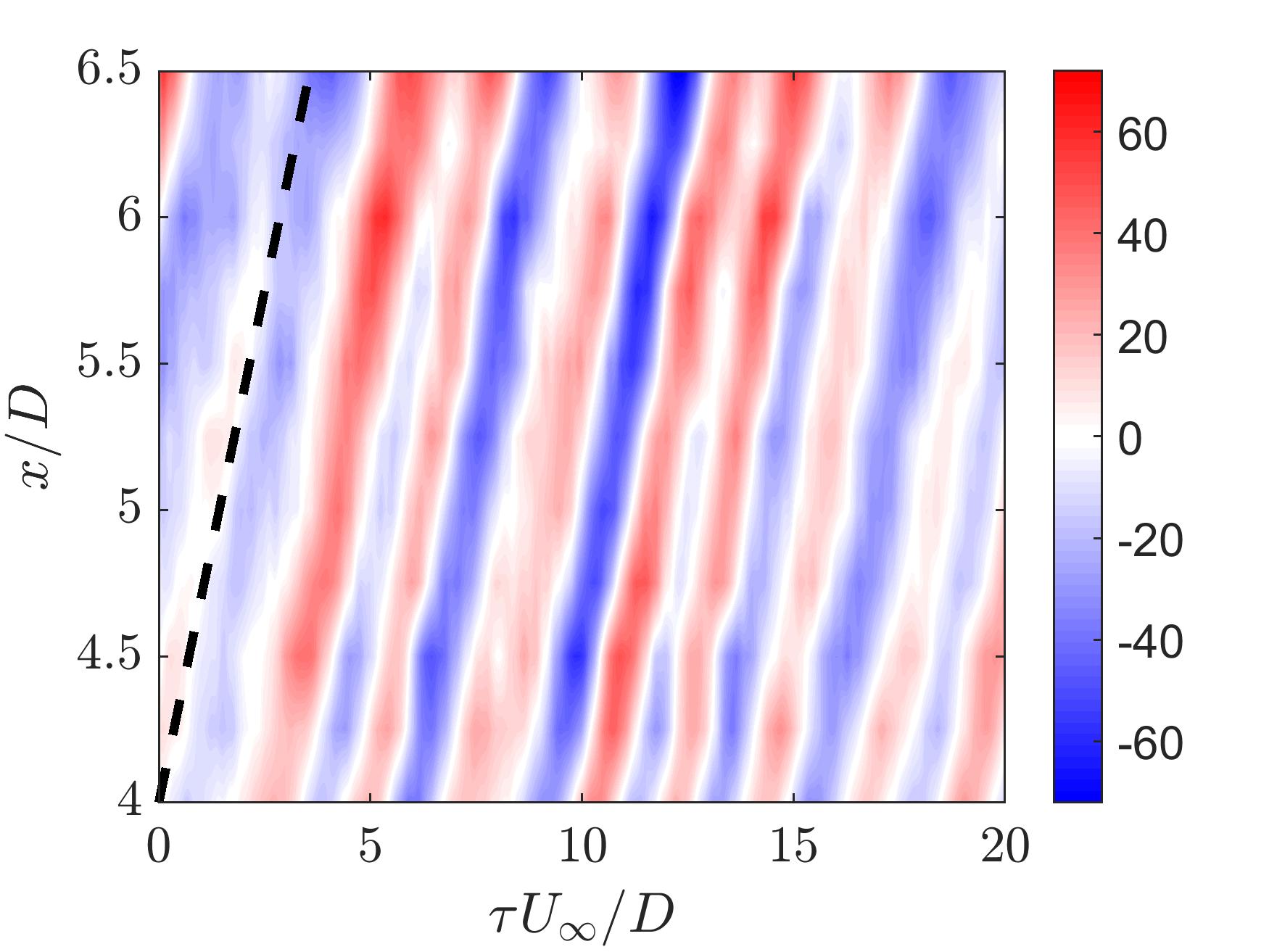}
\caption{Radial POD}
\end{subfigure}          
\caption{Estimation of the space–time dynamics of the mode-1 POD expansion coefficients via SLSE-POD. Dashed lines represent $U_c=0.7U_{\infty}$.}
\label{fig:a_se}
\end{figure}

After obtaining cross-correlation functions between the POD expansion coefficients and in-flow pressure fluctuations at all cross-stream planes, 
the SLSE-POD was employed to obtain time-resolved estimation of the POD expansion coefficients with the use of the the wavelet-based filter discussed above.
\Cref{fig:a_se} presents the time-resolved estimates of the first POD expansion coefficients from this hybrid strategy.
To make the results comparable, the estimation was realized by using the same pressure signal segment as the unconditional input for both streamwise and radial estimates. 
The most dominant feature from the reconstruction is the streamwise propagation of wavy patterns in time, which confirms that the convective nature of the dominant axisymmetric turbulent structures can be precisely captured from this correlation-based estimation technique. 
In addition, the $\sim 90^\circ$ phase difference is consistently observed between the streamwise expansion coefficients and the radial ones. 
The candidate of the convective speed, $U_c= 0.7 U_{\infty}$, is also plotted in the figures by the dashed lines, which are seen to match well with the slope of the travelling waves. 
However, since SLSE-POD grounds on the correlation between pressure and POD expansion coefficients, the estimation can only extract the portion of the input that is linearly correlated with the output. 
Therefore, as the PIV plane moves upstream, the coherency between the downstream pressure and the POD expansion coefficients is gradually weakened, which will directly lead to an attenuation of the output amplitude.
Hence, the spatial envelope of the axially evolving wave-packet shape can not be fairly captured from this set of experiments. In order to better describe the real-time amplitudes, more input information from upstream locations might be necessary.

\begin{figure}[!ht]
\centering
\mbox{
    \includegraphics[width=0.32\textwidth]{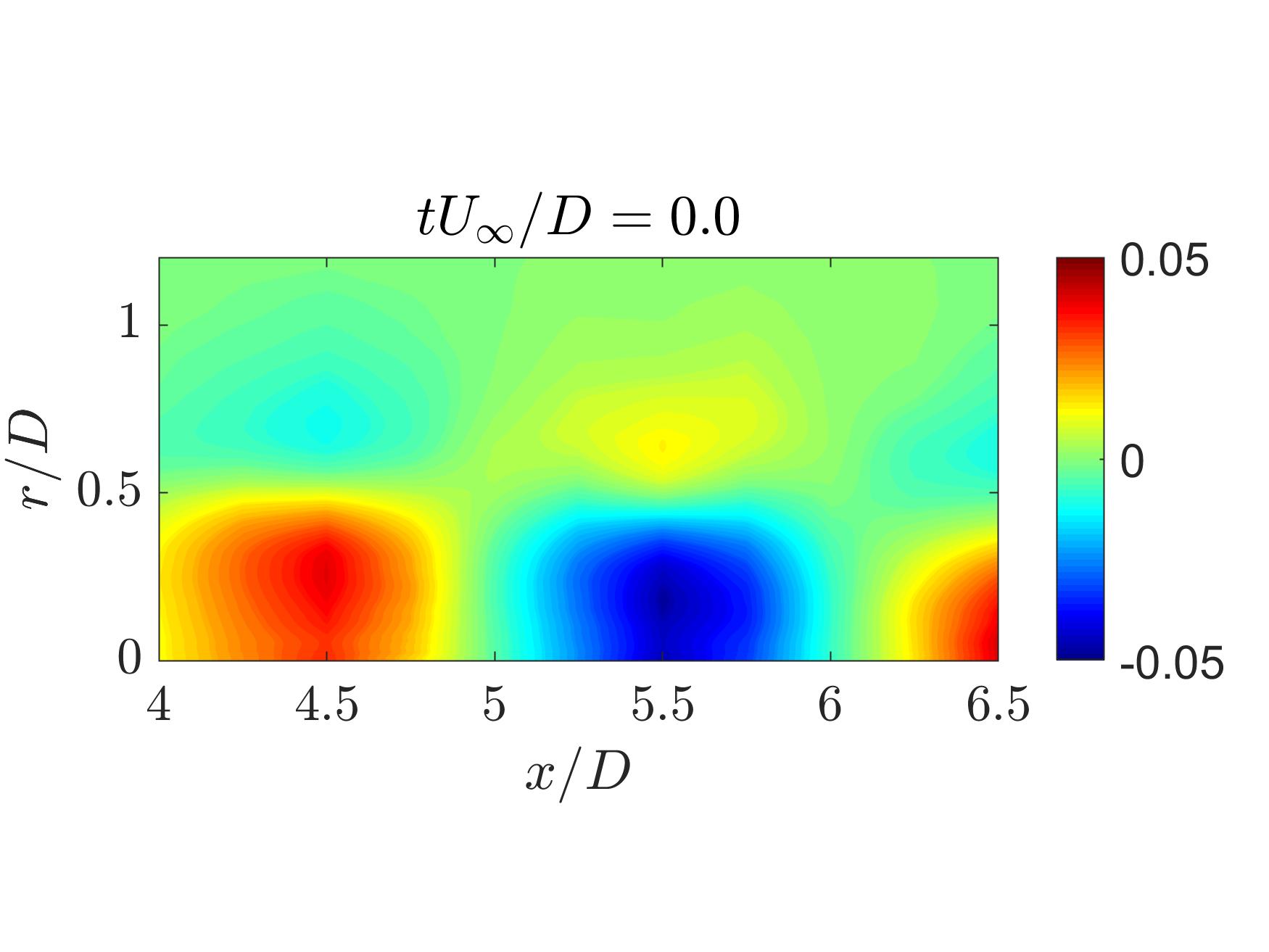}
    \includegraphics[width=0.32\textwidth]{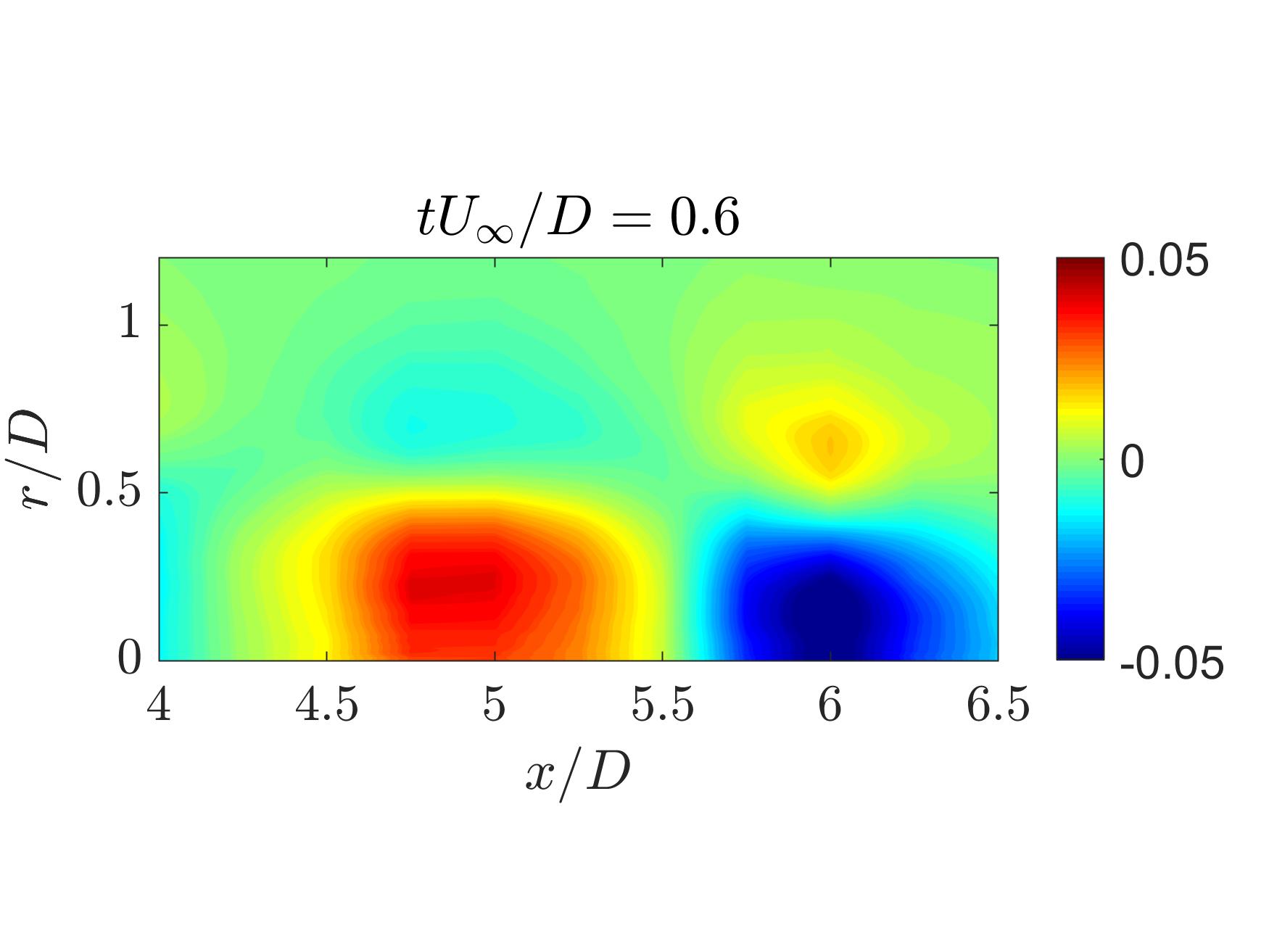}
    \includegraphics[width=0.32\textwidth]{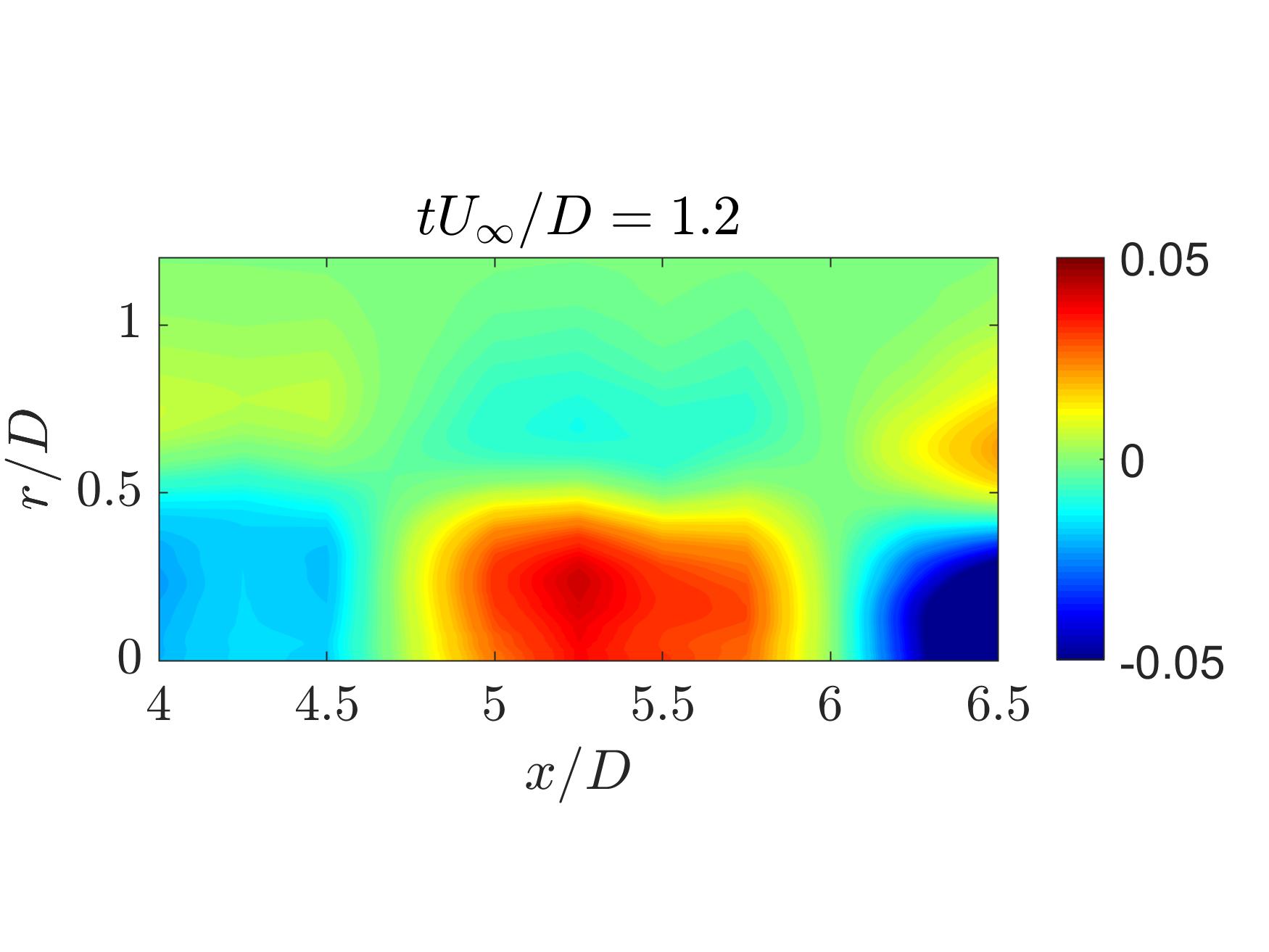}}
\mbox{
    \includegraphics[width=0.32\textwidth]{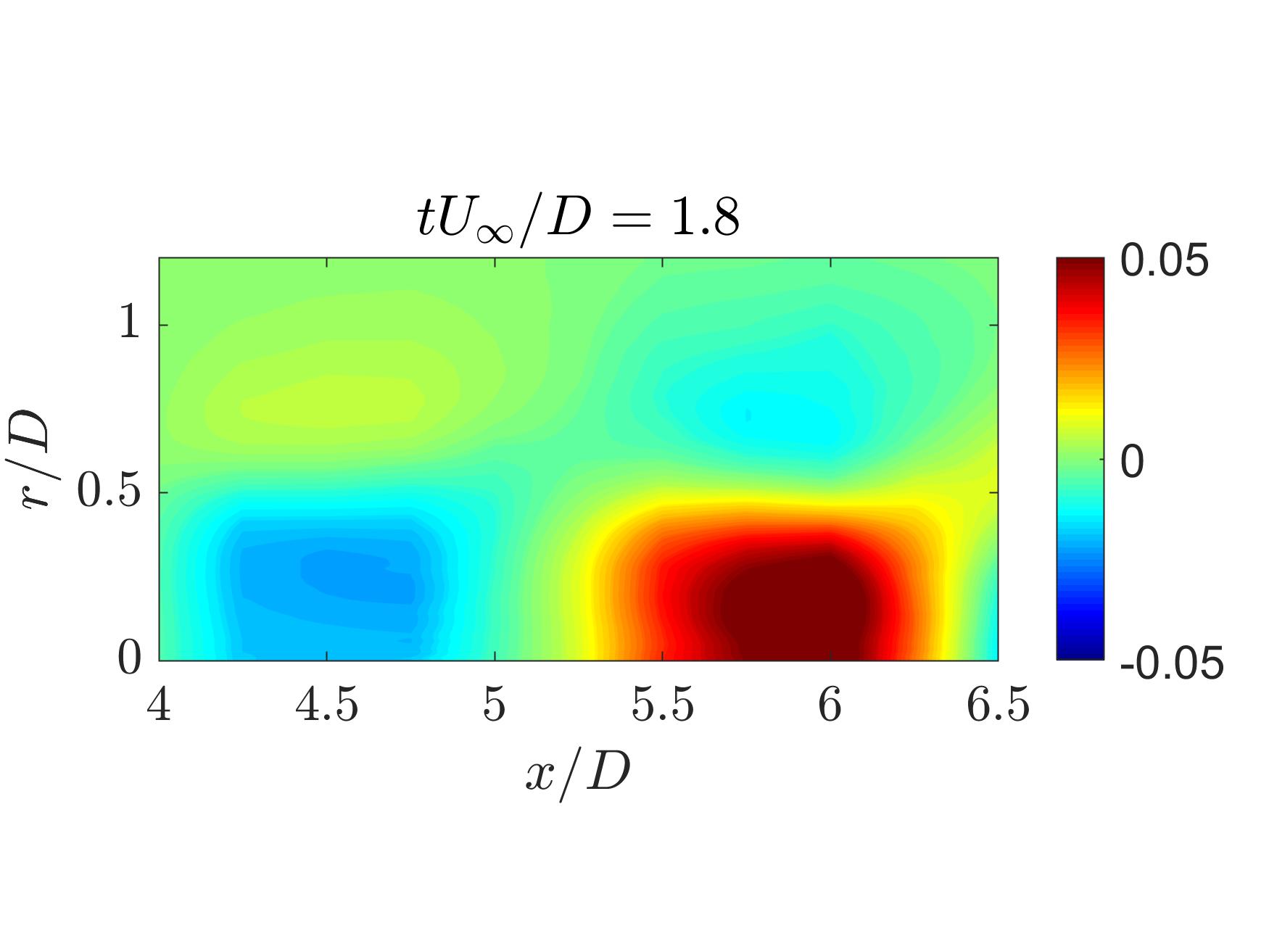}
    \includegraphics[width=0.32\textwidth]{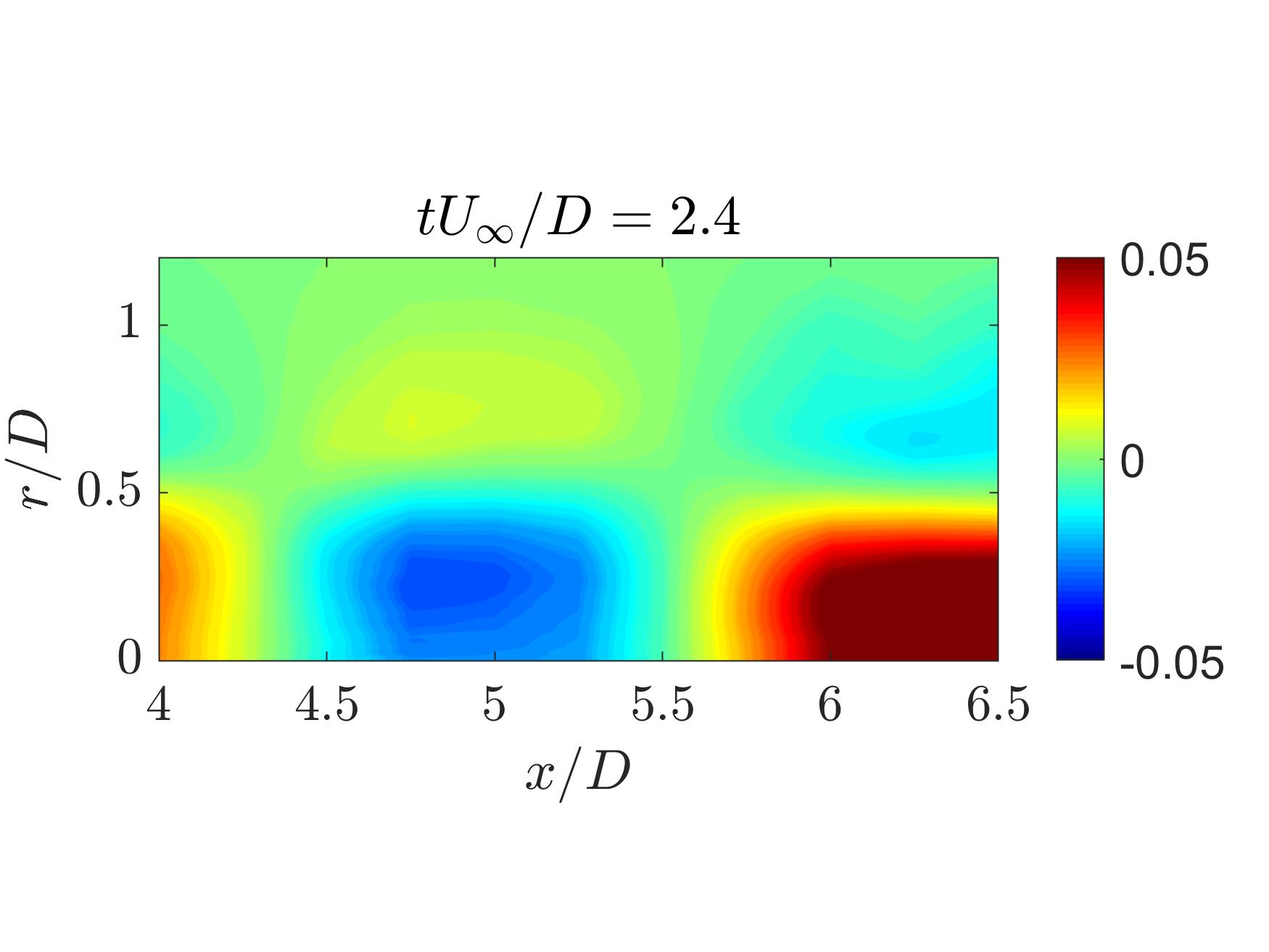}
    \includegraphics[width=0.32\textwidth]{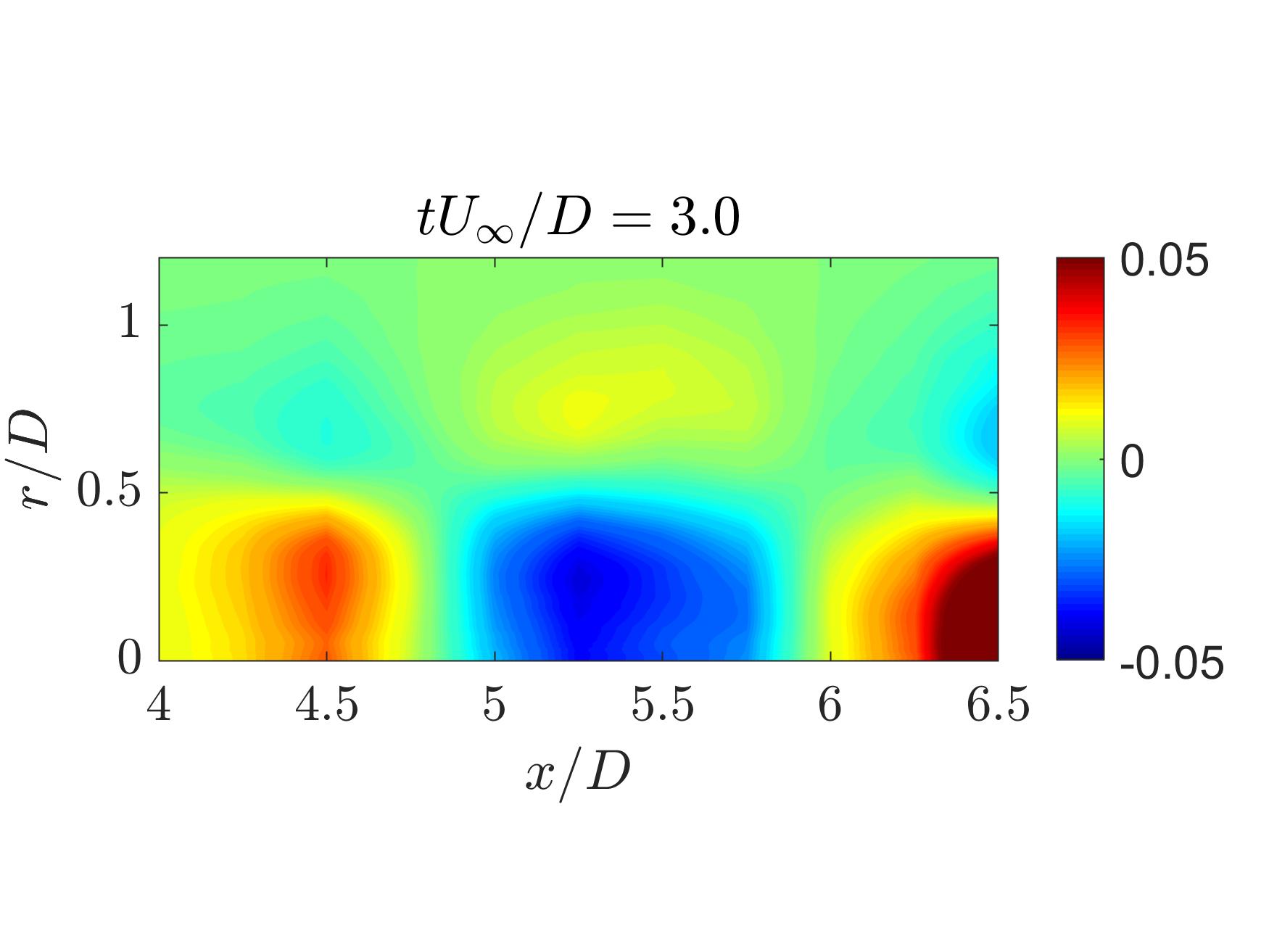}}             

\caption{Five-mode reconstruction of the streamwise turbulent velocity. Velocity is scaled by the jet exit velocity $U_{\infty}$.}
\label{fig:movie_ux}
\end{figure}

\begin{figure}[!ht]
\centering
\mbox{
    \includegraphics[width=0.32\textwidth]{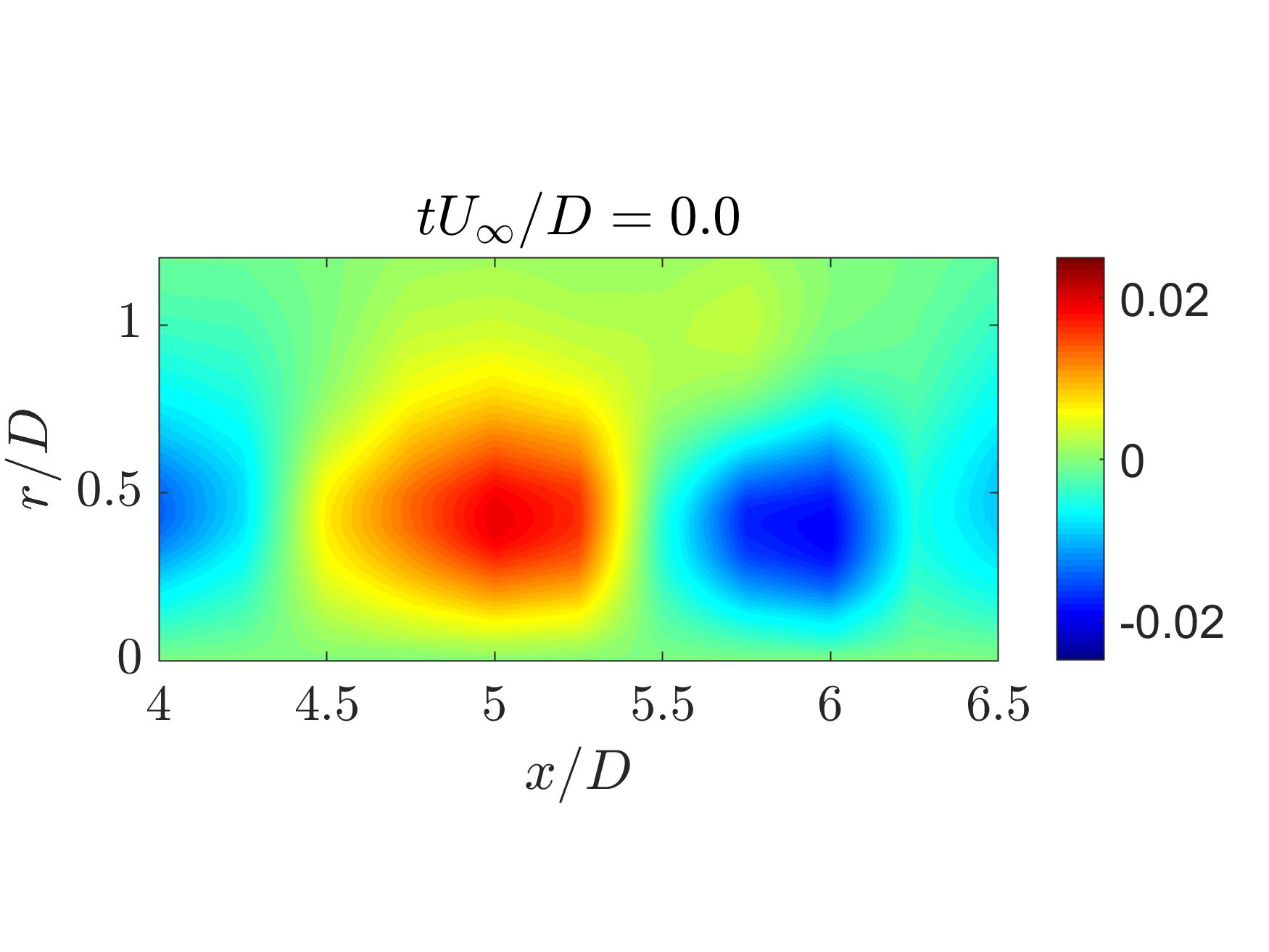}
    \includegraphics[width=0.32\textwidth]{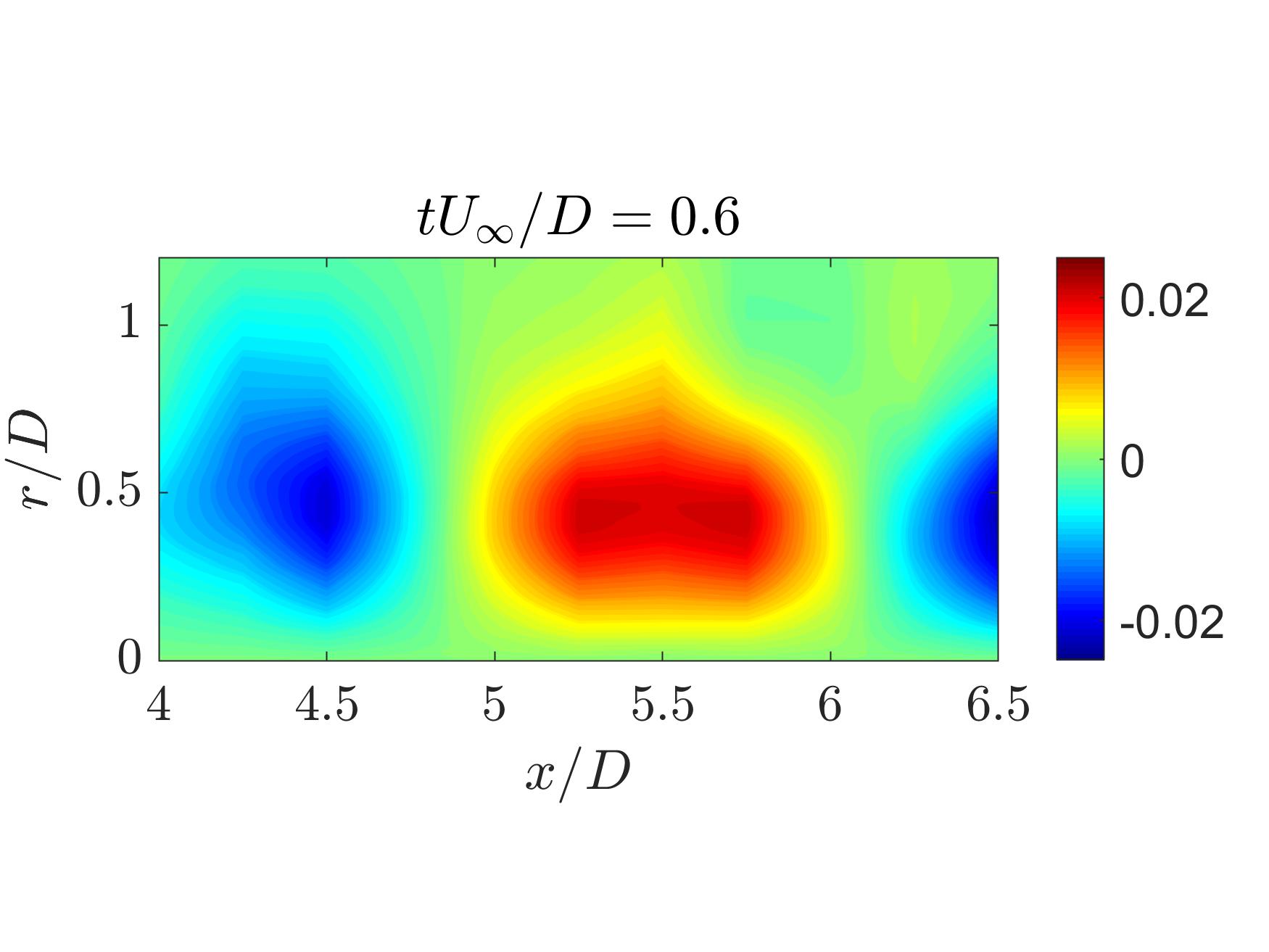}
    \includegraphics[width=0.32\textwidth]{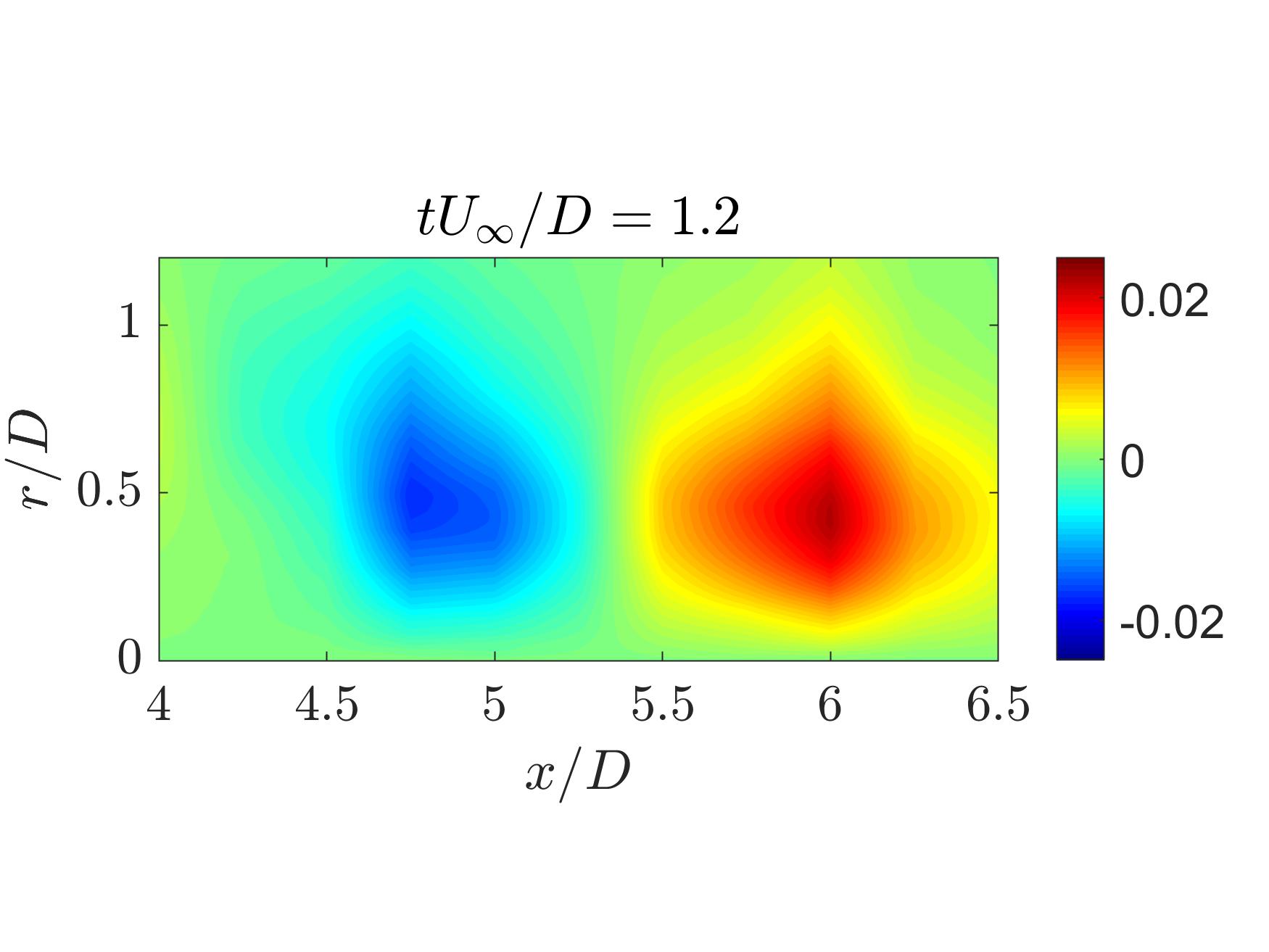}}
\mbox{
    \includegraphics[width=0.32\textwidth]{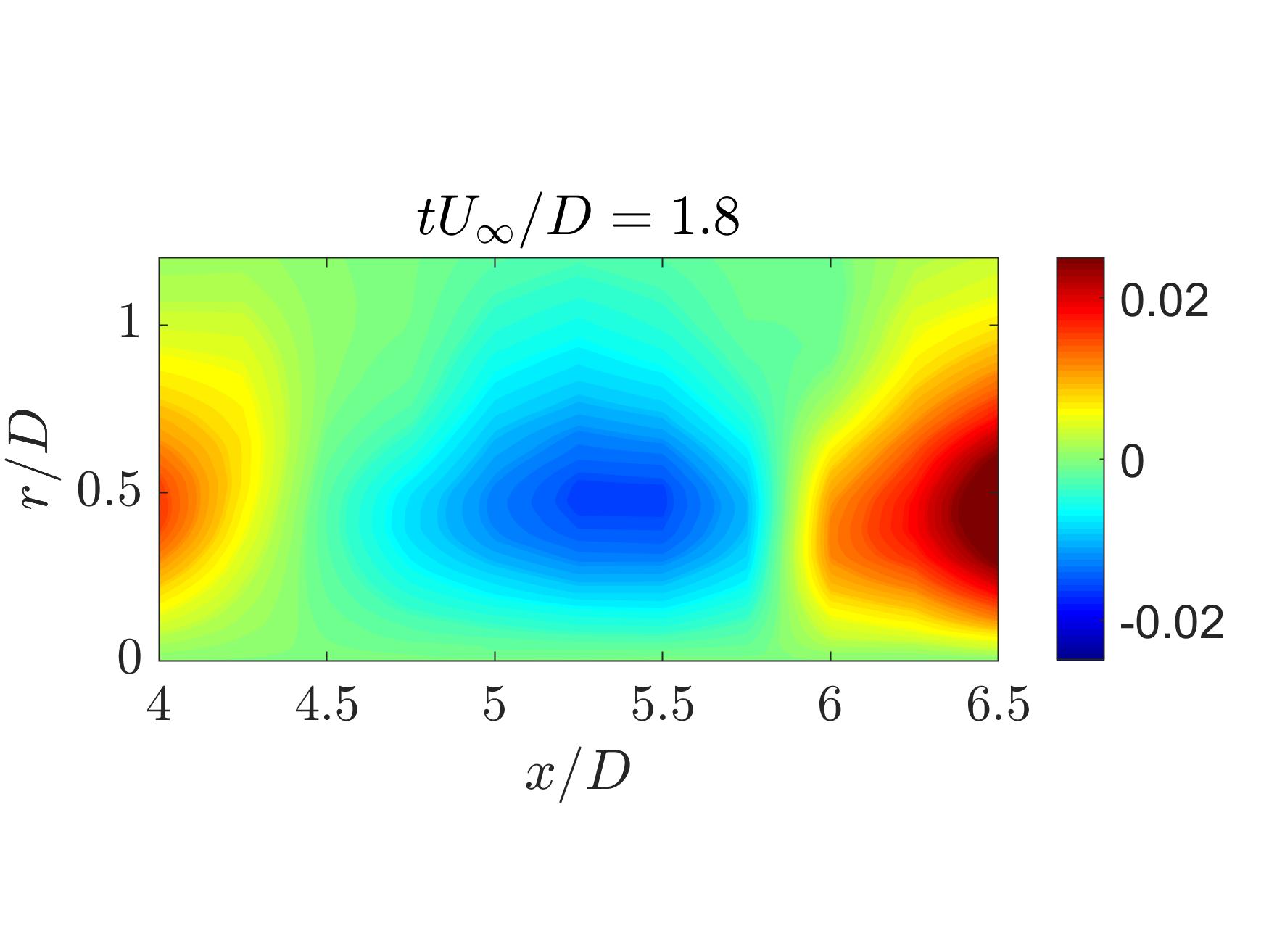}
    \includegraphics[width=0.32\textwidth]{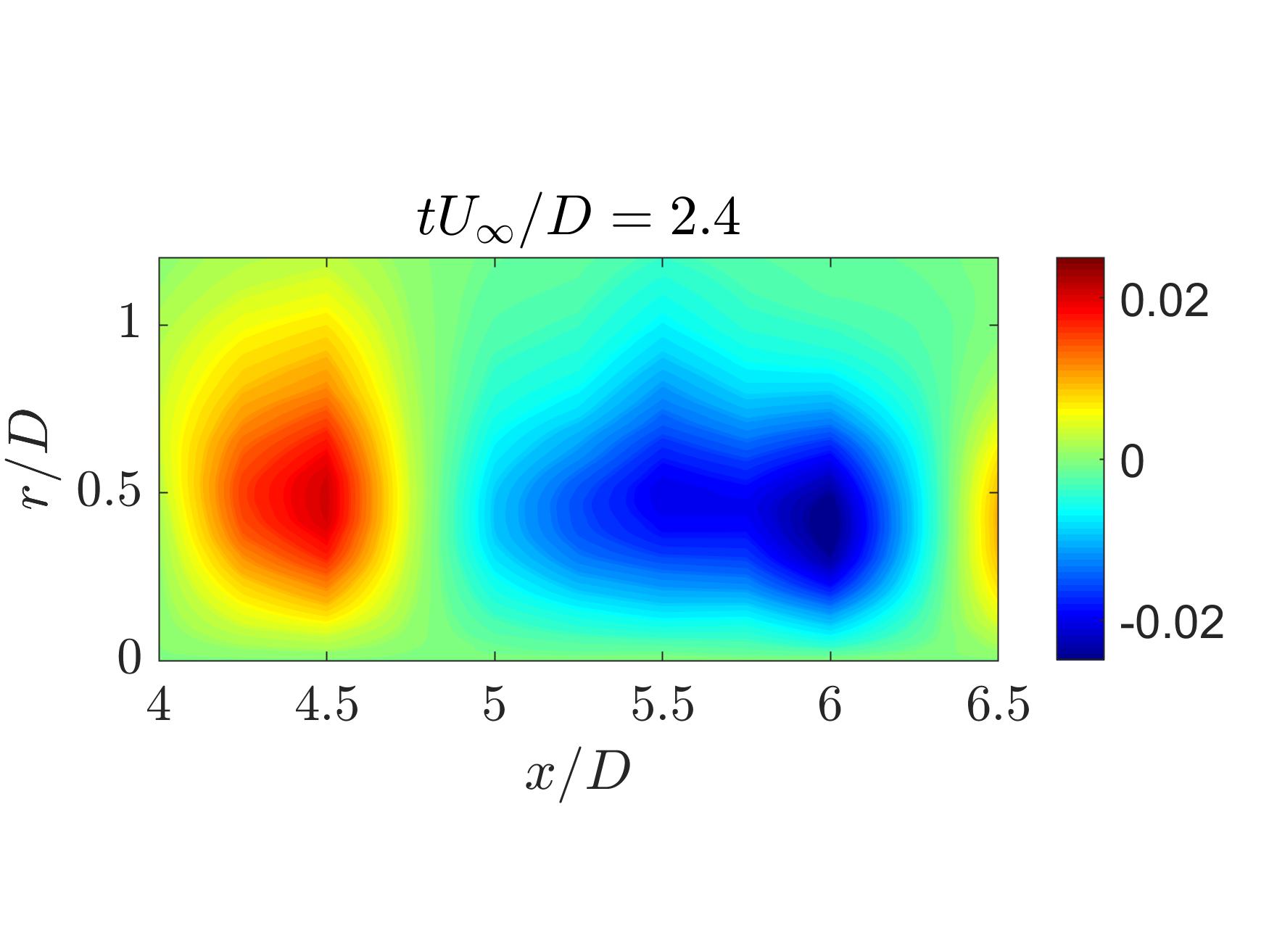}
    \includegraphics[width=0.32\textwidth]{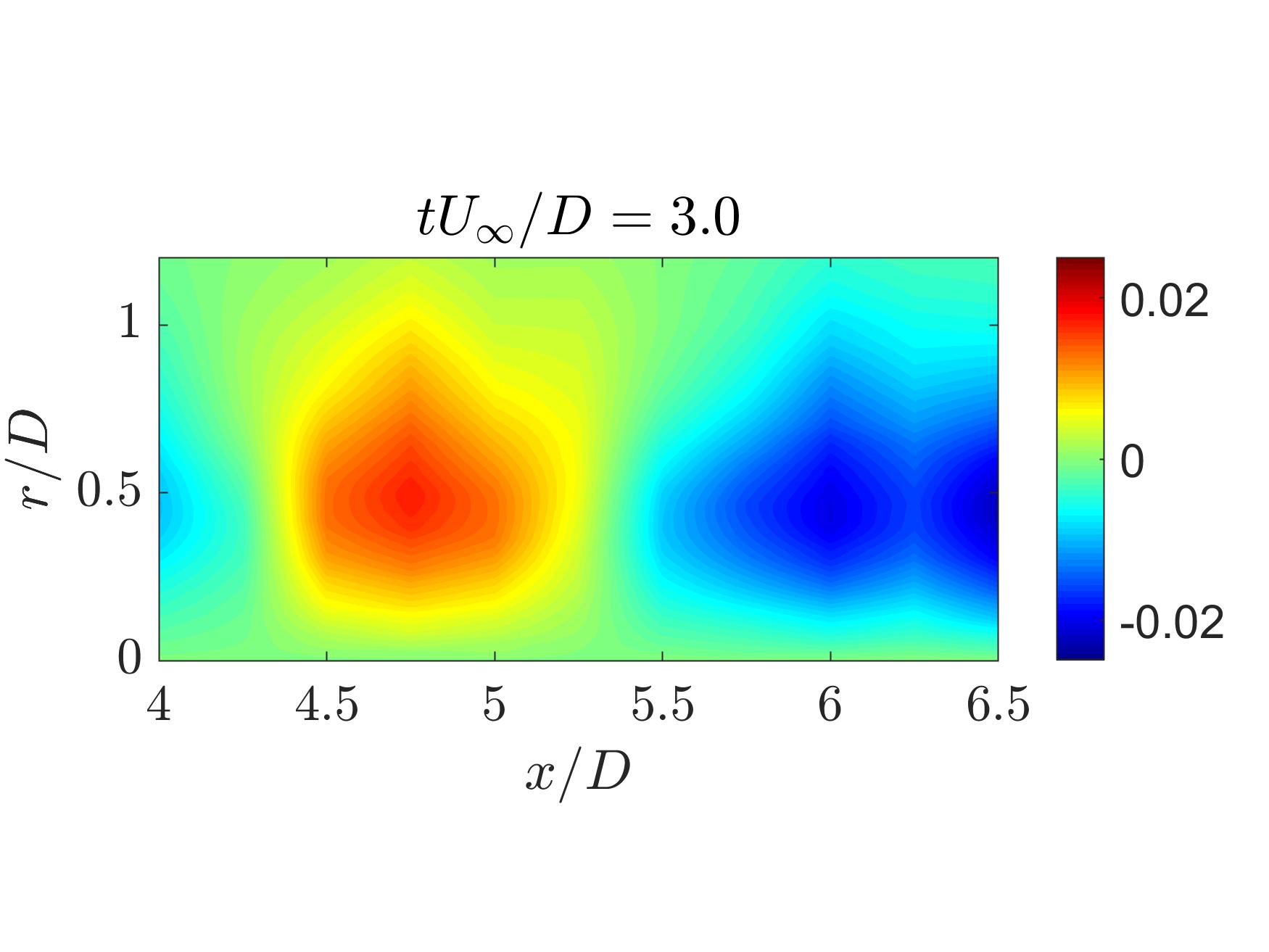}}             

\caption{Five-mode reconstruction of the radial turbulent velocity. Velocity is scaled by the jet exit velocity $U_{\infty}$.}
\label{fig:movie_ur}
\end{figure}

\Cref{fig:movie_ux} and \Cref{fig:movie_ur} display the global reconstruction of time-resolved axisymmetric velocity using the first five POD modes.
The space-time evolution of coherent structures can be clearly observed in the figures across the sequential instances. 
The size of each eddy is on the order of the jet diameter $D$. 
In addition, the streamwise velocity is oppositely signed across the jet lipline as displayed in \Cref{fig:movie_ux}, which is a result of the strong shear events inside the flow. 

\begin{figure}[!h]
\centering
\begin{subfigure}{0.45\textwidth}
    \includegraphics[width=0.99\textwidth]{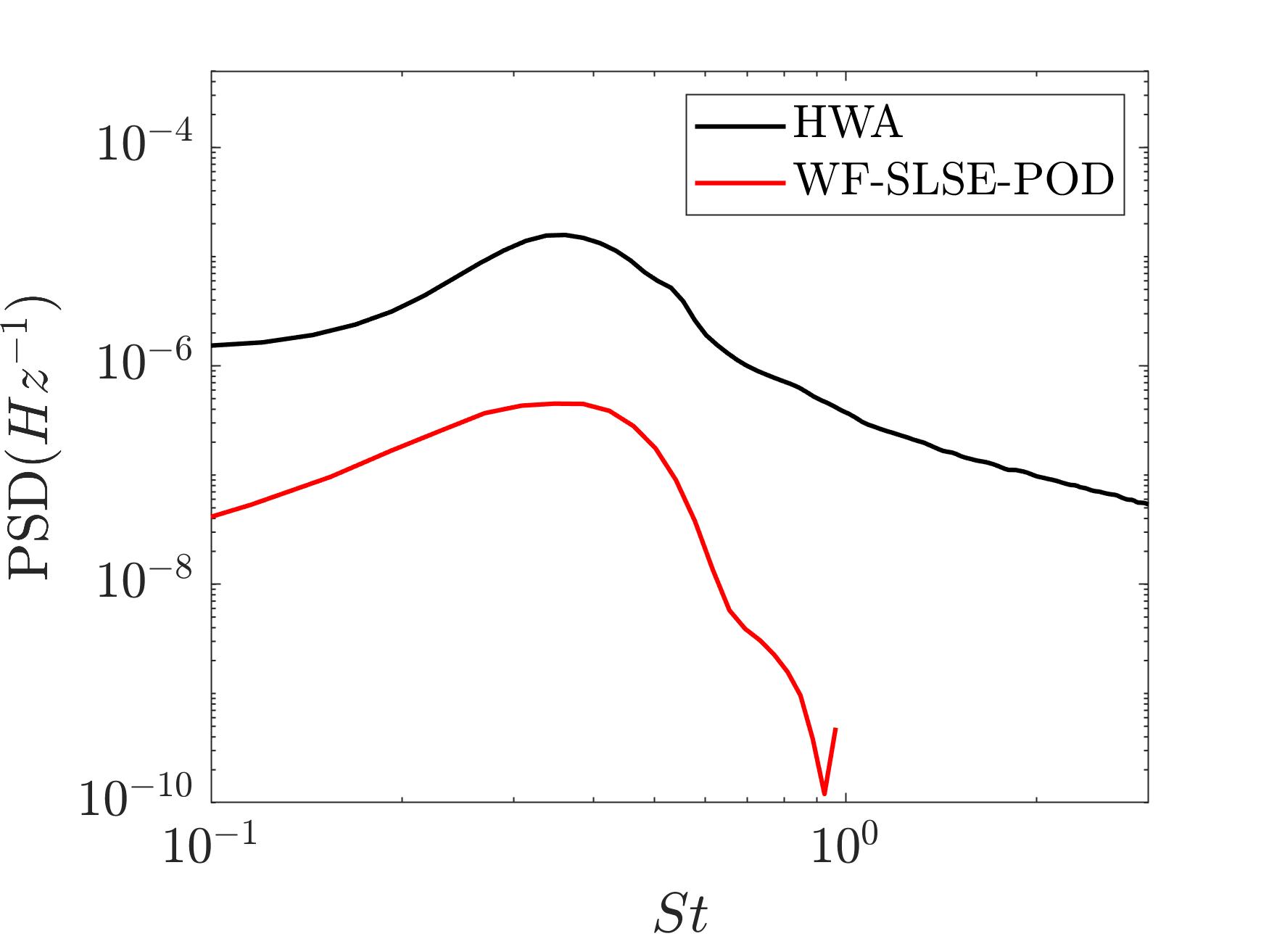}
    \caption{$x/D = 4$}
\end{subfigure}
\begin{subfigure}{0.45\textwidth}
    \includegraphics[width=0.99\textwidth]{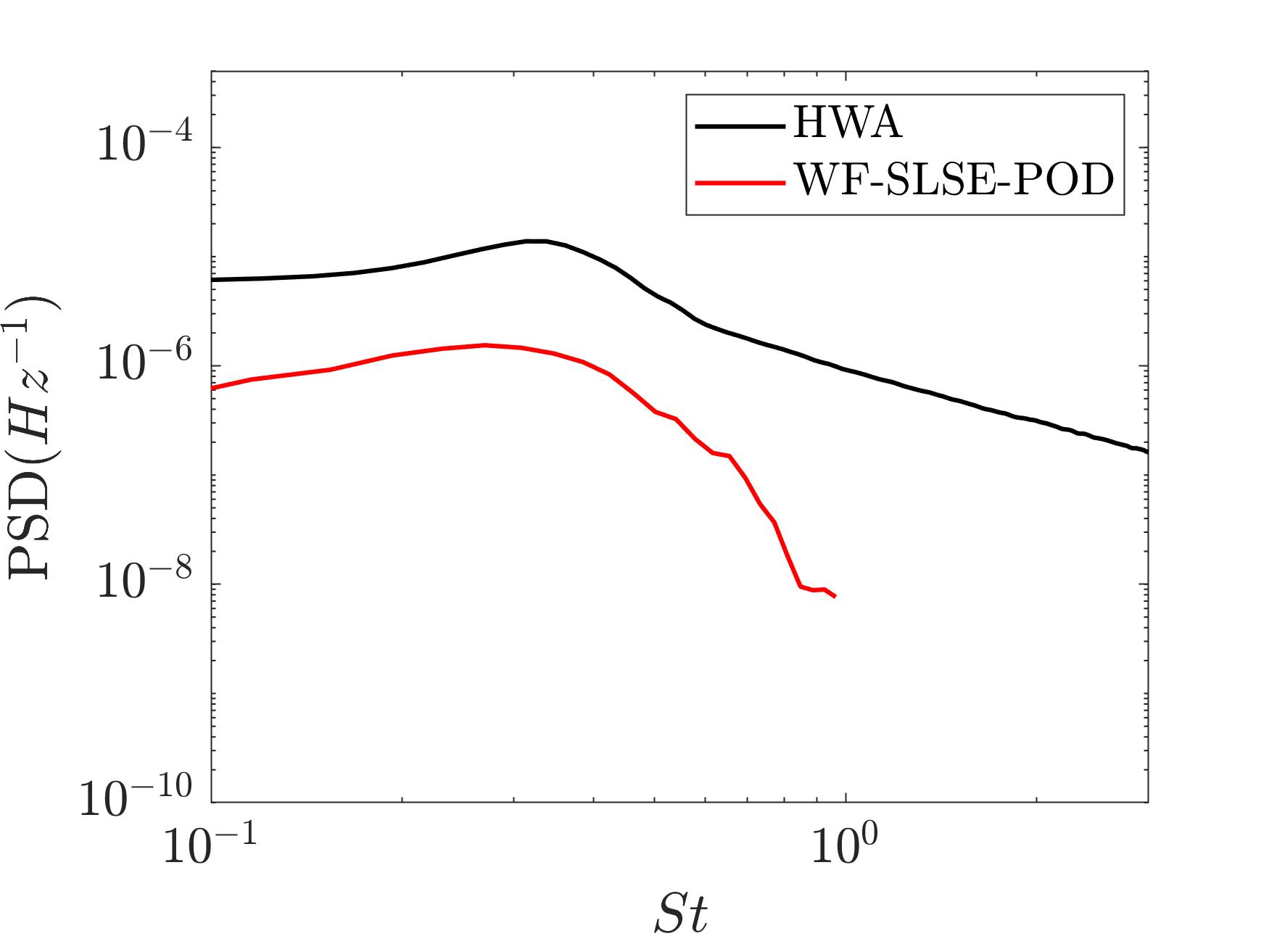}
    \caption{$x/D = 5$}
\end{subfigure}\\
\begin{subfigure}{0.45\textwidth}
    \includegraphics[width=0.99\textwidth]{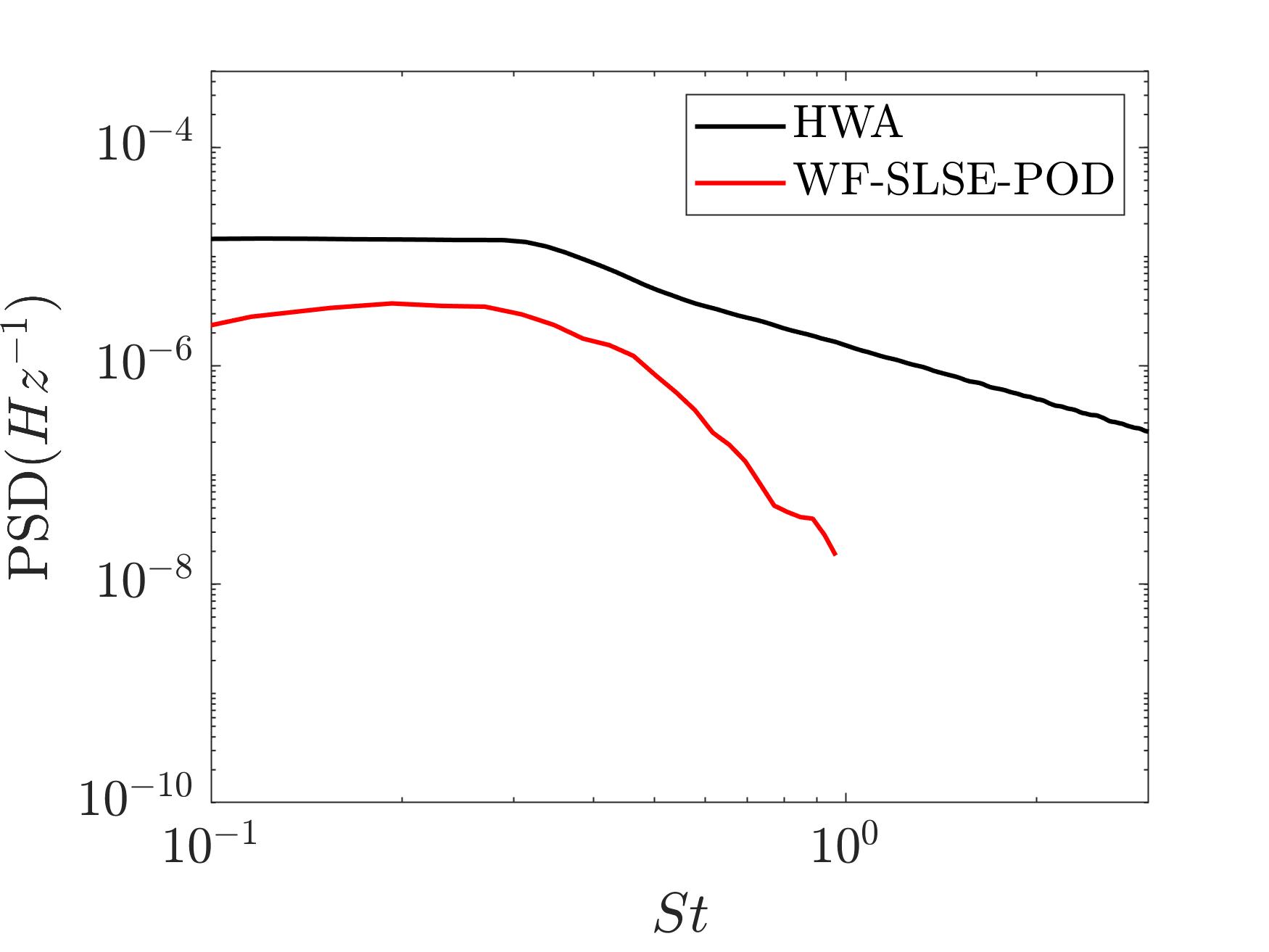}
    \caption{$x/D = 6$}
\end{subfigure}
\caption{A comparison of the measured streamwise velocity spectra versus spectral estimates from wavelet-filtered (WF) SLSE-POD on the jet centerline. Velocity spectra are scaled by the jet exit velocity $U_{\infty}$.}
\label{fig:spectra_se}
\end{figure}

\Cref{fig:spectra_se} presents the comparison between the estimated streamwise velocity spectra as well as the velocity spectra measured from the hotwire anemometry at three different locations on the jet centerline. 
Since the high frequency content can not be fairly retrieved from wavelet-filtered cross-correlations, only the low frequency portion ($St<1$) of the estimated spectra are presented. 
In general, the velocity spectra from estimation can faithfully reflect the trend of the measured spectra in the range of $0.1\leq St\leq 0.5$, which can be described as a broadband hump governed by large-scale structures. The discrepancy of the overall amplitude between the estimation results and the measured ones comes from the fact that only the portion of velocity correlated to pressure inputs will be encompassed in the estimated spectra. The closer the PIV plane is to the pressure probe, the larger the coherency between the pressure and dominant POD modes, and it will result in less amplitude discrepancy between measurement and estimation. At $St>0.5$, the estimated velocity spectra exhibit a steeper decaying rate than the real spectra. Since the leading POD modes only contribute to the large-scaled turbulent structures in the flow, the rolling-off of the spectra, which is dominated by the fine-scale turbulence, can't be well captured from the estimation technique.

\subsection{Estimation of 2D Velocity from Streamwise PIV and In-Flow Pressure Measurements}

\begin{figure}[!b]
    \centering
    \includegraphics[width=0.6\linewidth]{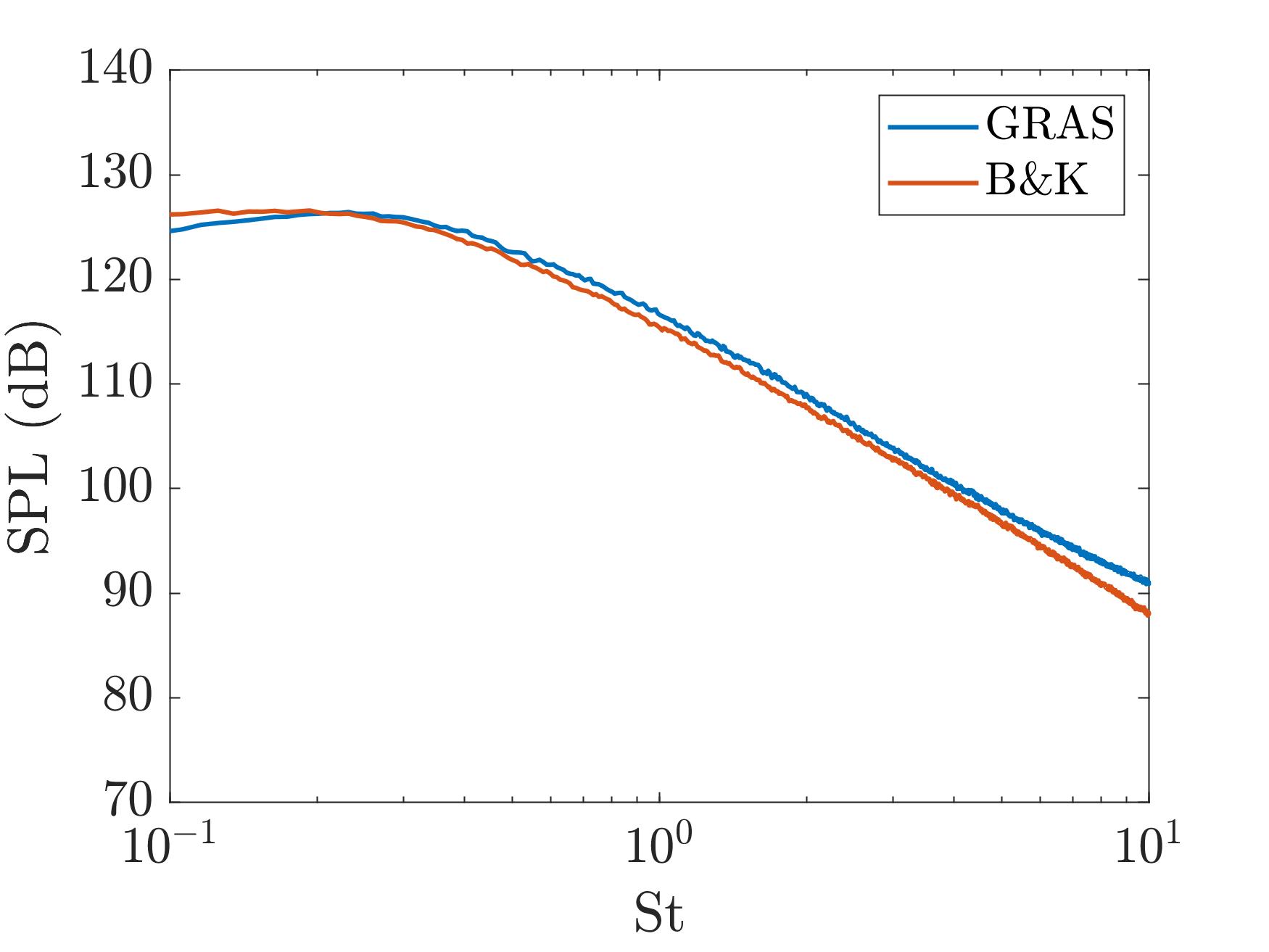}
    \caption{A comparison between pressure spectrum measured from the B\&K microphone at $x/D = 6.6$, $r/D = 0.5$ and the one from the GRAS microphone at $x/D = 6.6$, $r/D = -0.5$.}
    \label{fig:mics_comparison}
\end{figure}

In this subsection experimental results from synchronously measured two-compnent PIV and in-flow pressure signals on the upper and lower jet liplines will be utilized to provide time-resolved, pressure-informed velocity estimation on the streamwise plane. 
As a priority to detailed analysis, pressure spectra from both GRAS and B\&K microphones were first examined and results are displayed in \Cref{fig:mics_comparison}. 
Although different aerodynamic designs are employed to the nosecones of the two microphones, careful examination shows high level of consistency with respect to the spectral shapes from both microphones.
The slight discrepancy between the two pressure spectra, which is less than 2 dB throughout the frequency range shown in the figure, might arise from the positioning uncertainties of microphones ($\pm 0.5$mm in all directions and $1^{\circ}$ in angle), the different frequency response characteristics of microphones, and the different aerodynamic design of nosecones. 
Furthermore, self-noise induced from one in-flow microphone is not noticed to distort the pressure spectrum measured from the other microphone, indicating the mutual interaction between both microphones is negligible compared to the intensive fluctuating pressure level generated by the jet turbulence. 

\begin{figure}[!b]
    \centering
    \includegraphics[width=0.6\linewidth]{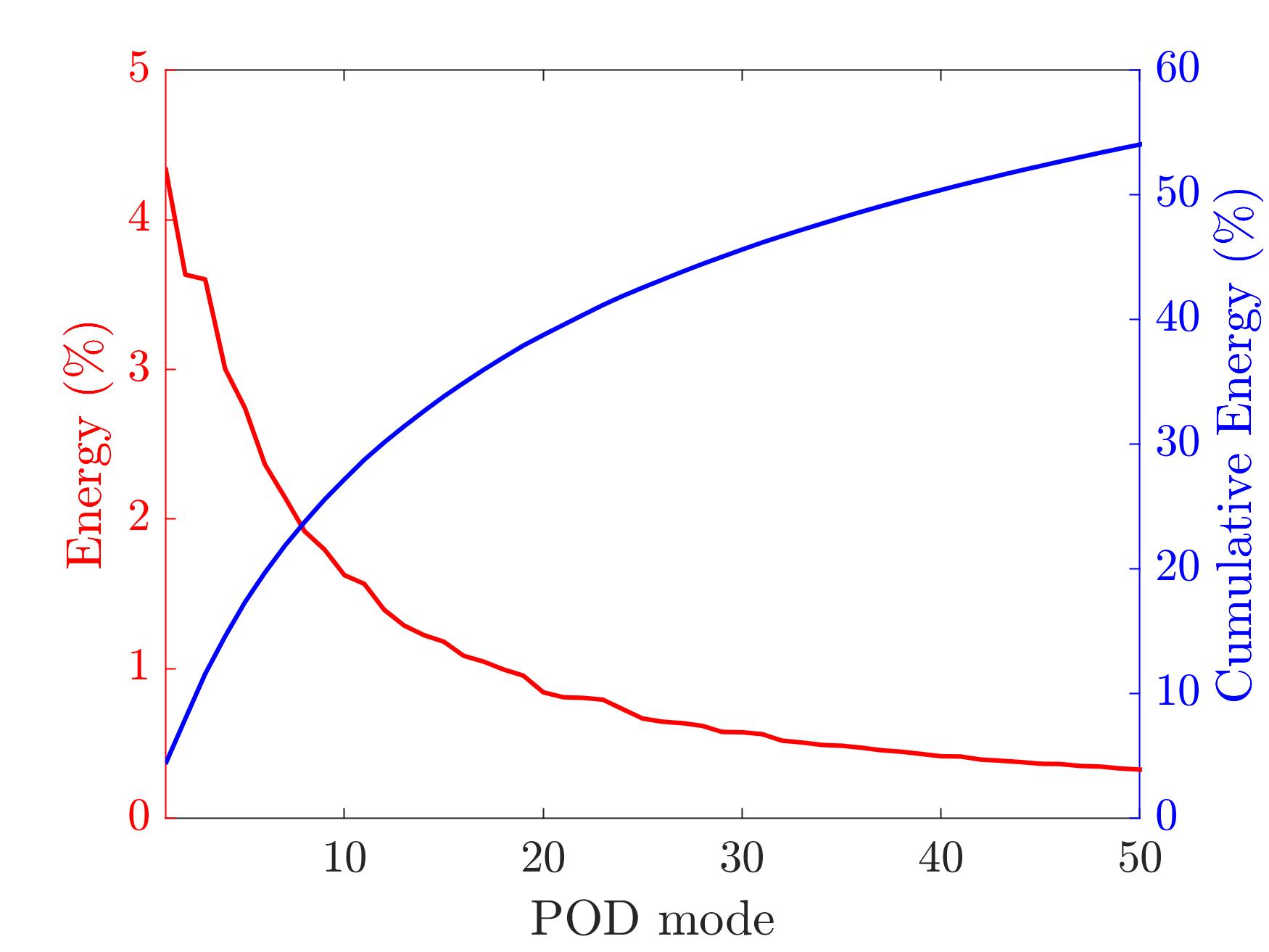}
    \caption{Energy distribution and the cumulative sum of the first 50 POD modes under the current field of view.}
    \label{fig:pod_energy}
\end{figure}

\begin{figure}[!h]
    \centering
    \includegraphics[width=0.99\linewidth, trim={3cm 0.5cm 2.5cm 0.5cm },clip]{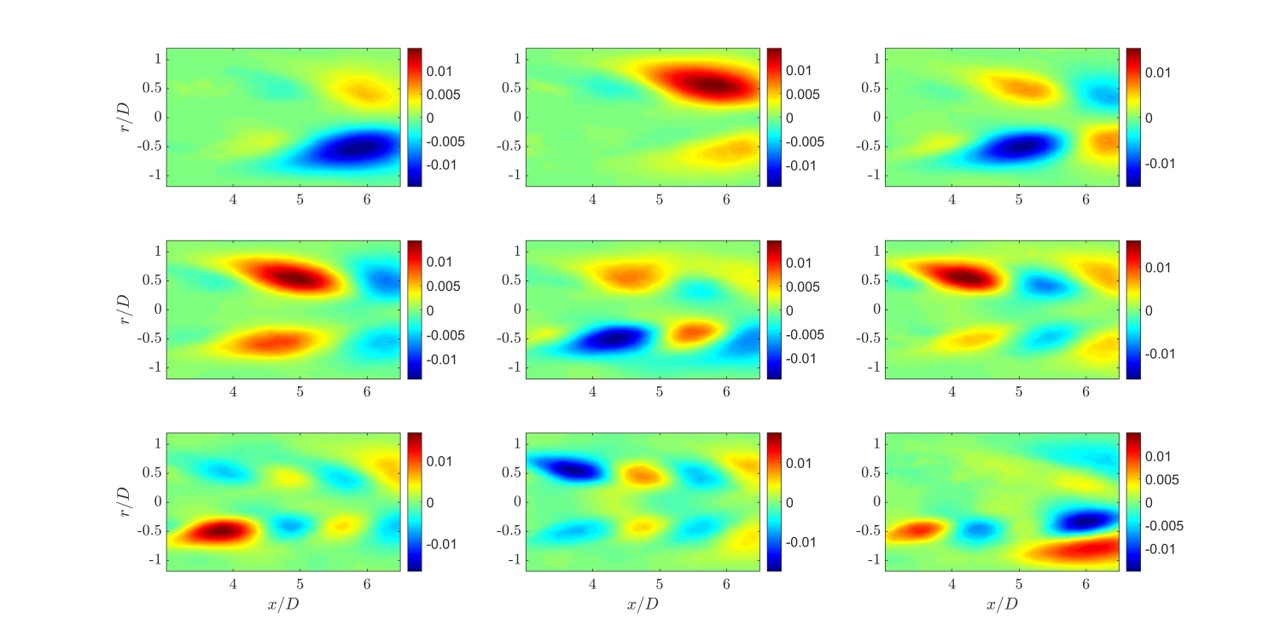}
    \caption{Streamwise eigenfunctions of the leading POD modes.}
    \label{fig:pod_u}
    \includegraphics[width=0.99\linewidth, trim={3cm 0.5cm 2.5cm 0.5cm },clip]{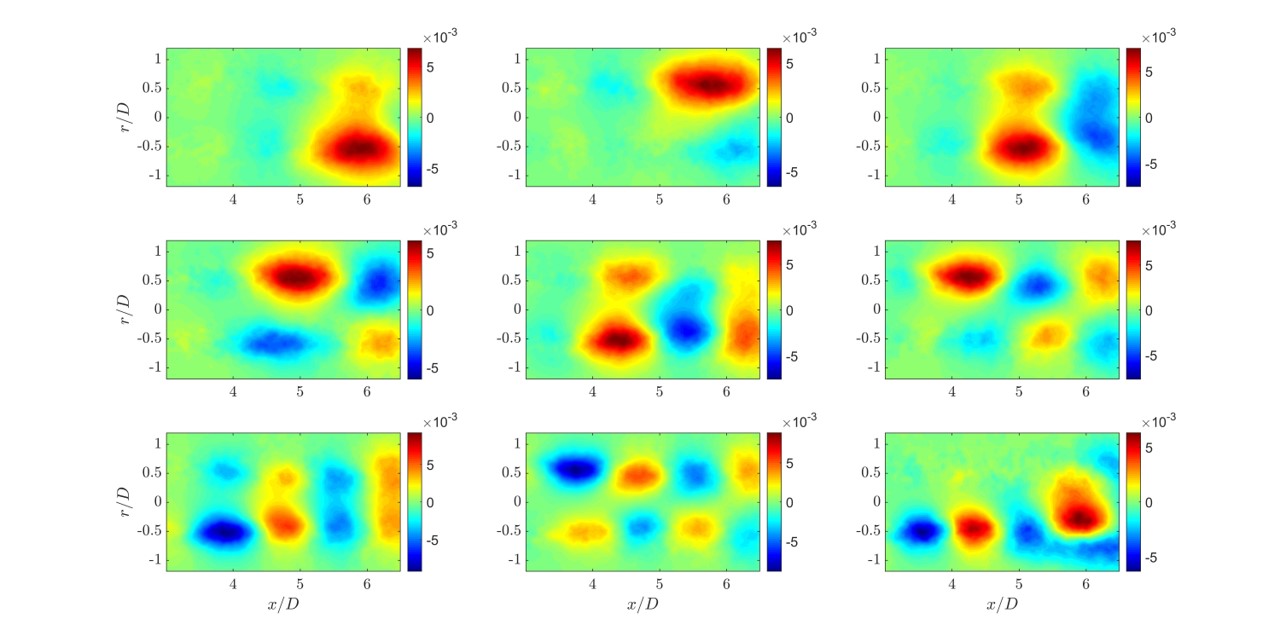}
    \caption{Radial eigenfunctions of the leading POD modes.}
    \label{fig:pod_v}
\end{figure}

To extract dominant spatial features from the PIV measurements, the snapshot POD was performed to decompose the two-dimensional velocity vectors into a series of energy-ranked modes. 
\Cref{fig:pod_energy} presents the energy distribution as well as the cumulative sum of the first 50 POD modes out of the total of 8000 modes. 
The energy is seen to be distributed over a wide range of POD modes in a descending order, and around 55\% of the total energy is contained in the first 50 modes which will be used to reconstruct the time-resolved velocity fields from SLSE-POD and machine learning approaches. 
Shapes of the streamwise and radial components of the leading POD modes are presented in \Cref{fig:pod_u} and \Cref{fig:pod_v}, respectively. 
The positive and negative regions of eigenfunctions in \Cref{fig:pod_u} demonstrates that the most energetic region under the current field of view concentrates inside the jet shear layer. 
The first two POD modes, for example, represents the appearance of large-scale structures spanning between $x/D = 5$ to $6.5$. 
Each of them highlights a structure on one side of the jet axis such that the combination of the two modes is capable of representing axisymmetric and antisymmetric patterns of the structures at the largest scale.
Successively, higher order POD modes are gradually governed by smaller-scaled features which are still centered around the jet liplines. Although none of the leading POD modes are strictly symmetric or antisymmetric about the jet axis, the combination of the dominant POD modes is capable to represent turbulent structures with any symmetrical properties in a similar approach.

\begin{figure}[!b]
    \centering
    \includegraphics[width=0.99\linewidth, trim={3.55cm 0.5cm 2.5cm 0.5cm },clip]{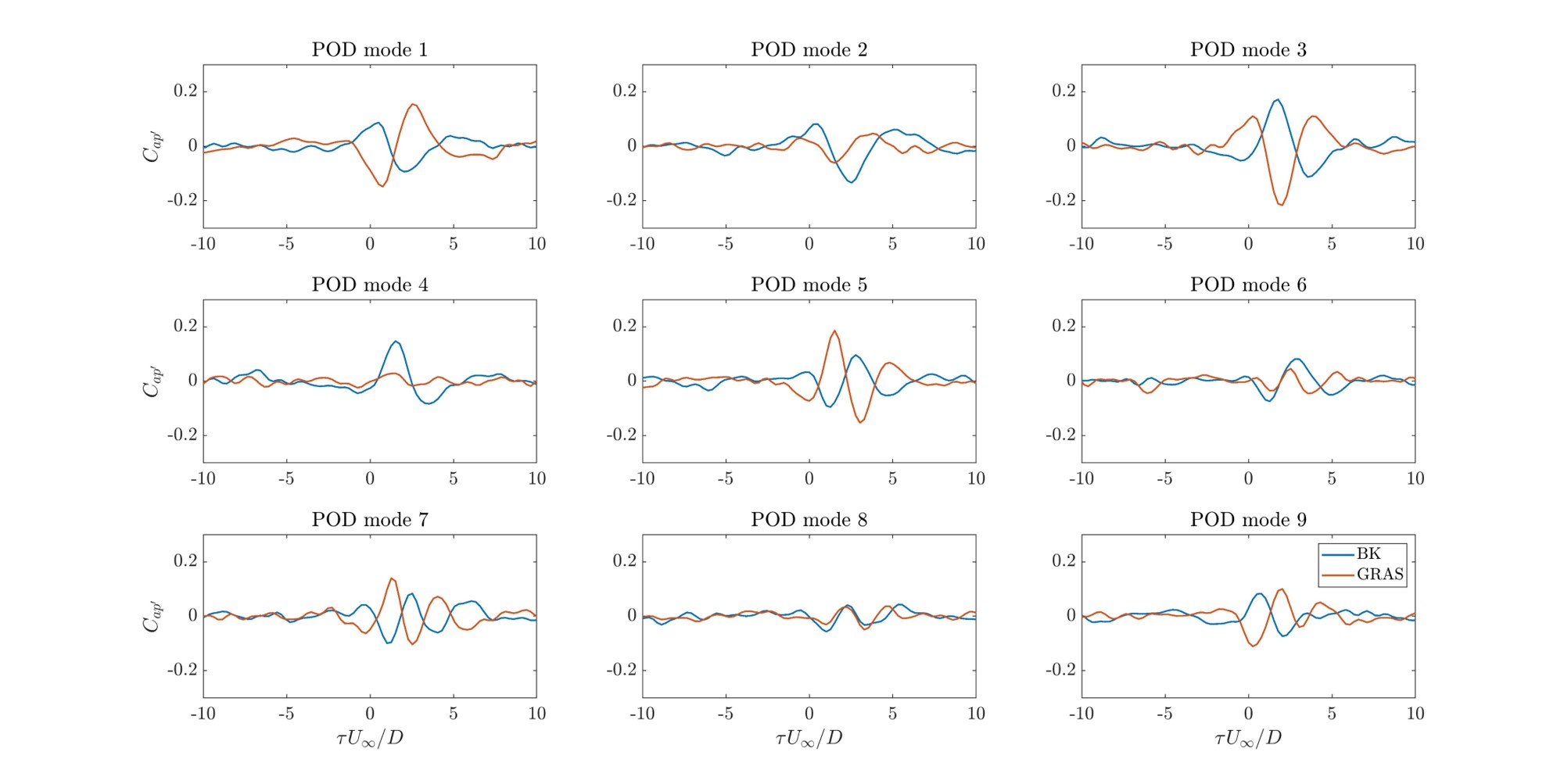}
    \caption{Cross-correlation coefficient $C_{ap}$ between POD expansion coefficients and pressure fluctuations. The B\&K microphone is on the upper lipline and the GRAS microphone is on the lower lipline.}
    \label{fig:xcorr_2d}
\end{figure}

The cross-correlation coefficient, $C_{ap}$, reveals the relationship between the dominant POD modes and the measured pressure fluctuations and are displayed in \Cref{fig:xcorr_2d}. 
Since the pressure measurements were performed downstream of the PIV window, the maximum magnitudes of cross correlation always appear at some positive time lag. 
In general, the amplitude of the correlation coefficient is relatively low (less than 0.22), and will gradually decrease for higher order POD modes as the dominant spatial structures become smaller-scaled vortices that are less influential to the pressure on the jet liplines. 
Since symmetric properties can not be established from individual POD modes, correlations with lipline pressure measurements are not fully symmetric/antisymmetric as well. 
One representative example is POD mode 4, in which a clear wave-like pattern with a peak value around 0.2 can be observed in the correlation with pressure on the upper lipline, however the correlation with pressure on the lower lipline is at a low level overwhelmed by the background noise.

\begin{figure}[!h]
\centering
\begin{subfigure}{0.49\textwidth}
    \includegraphics[width=0.99\textwidth]{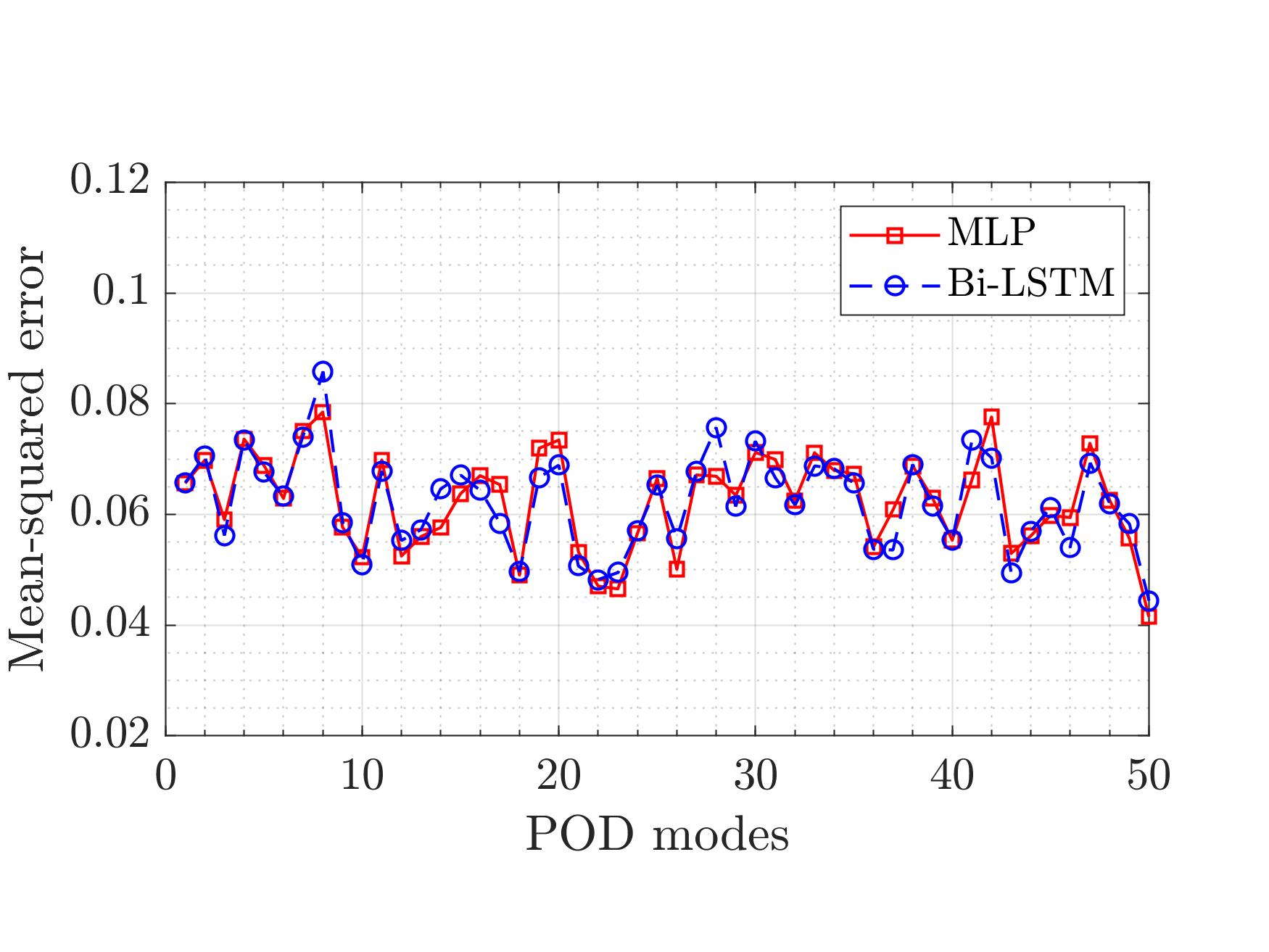}
    \caption{Training loss}
\end{subfigure}
\begin{subfigure}{0.49\textwidth}
    \includegraphics[width=0.99\textwidth]{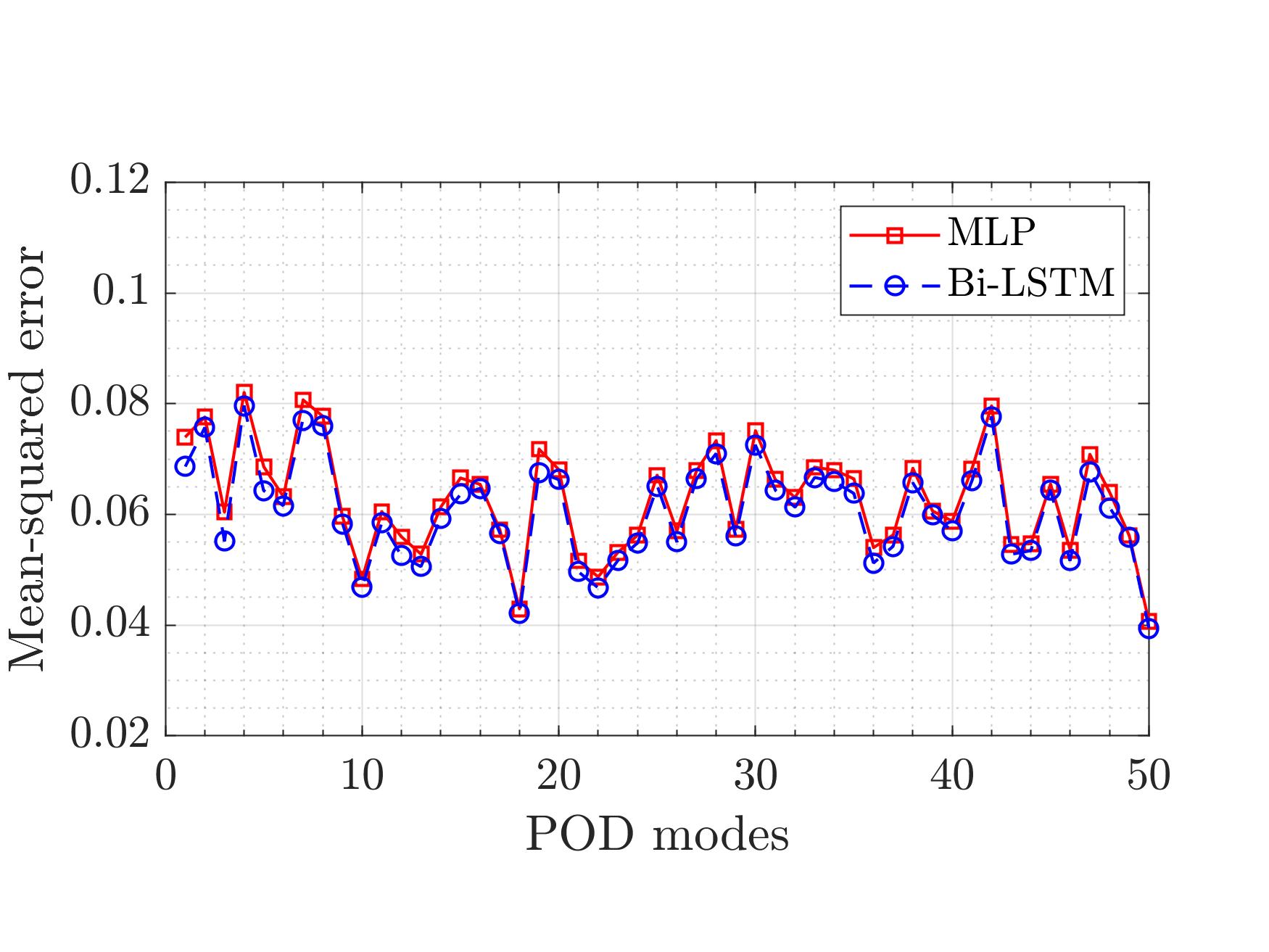}
    \caption{Validation loss}
\end{subfigure}
\caption{Evaluation of training and validation losses from both NN architectures.}
\label{fig:losses}
\end{figure}

Results of velocity estimation from stochastic estimation and machine learning approaches will be discussed in the following.
The learning outcome of the two neural networks are first examined by evaluating the training and validation losses of each POD mode, as presented in \Cref{fig:losses}. 
For all modes, training and validation losses for both the MLP and bidirectional LSTM architectures are always within the range of 0.04 to 0.08. 
However, since the POD expansion coefficients are energy-ranked, errors appearing in the leading POD modes will be more likely to influence the estimation results in velocity than higher order modes. 
The evaluation of validation losses, which is not directly involved in the training process, will constitute the generalization error of the training progress. 
Validation losses for the both NN schemes are almost identical to each other although errors from the bidirectional LSTM scheme are always slightly lower comparatively. 
The pronounced similarity of loss functions between two neural networks indicates that the primary relationship between POD expansion coefficients and time-lagged pressure have been effectively captured from both well-trained networks.

\begin{figure}[!h]
\centering
\begin{subfigure}{0.49\textwidth}
    \includegraphics[width=0.99\textwidth,trim={1cm 3cm 3cm 7.6cm },clip]{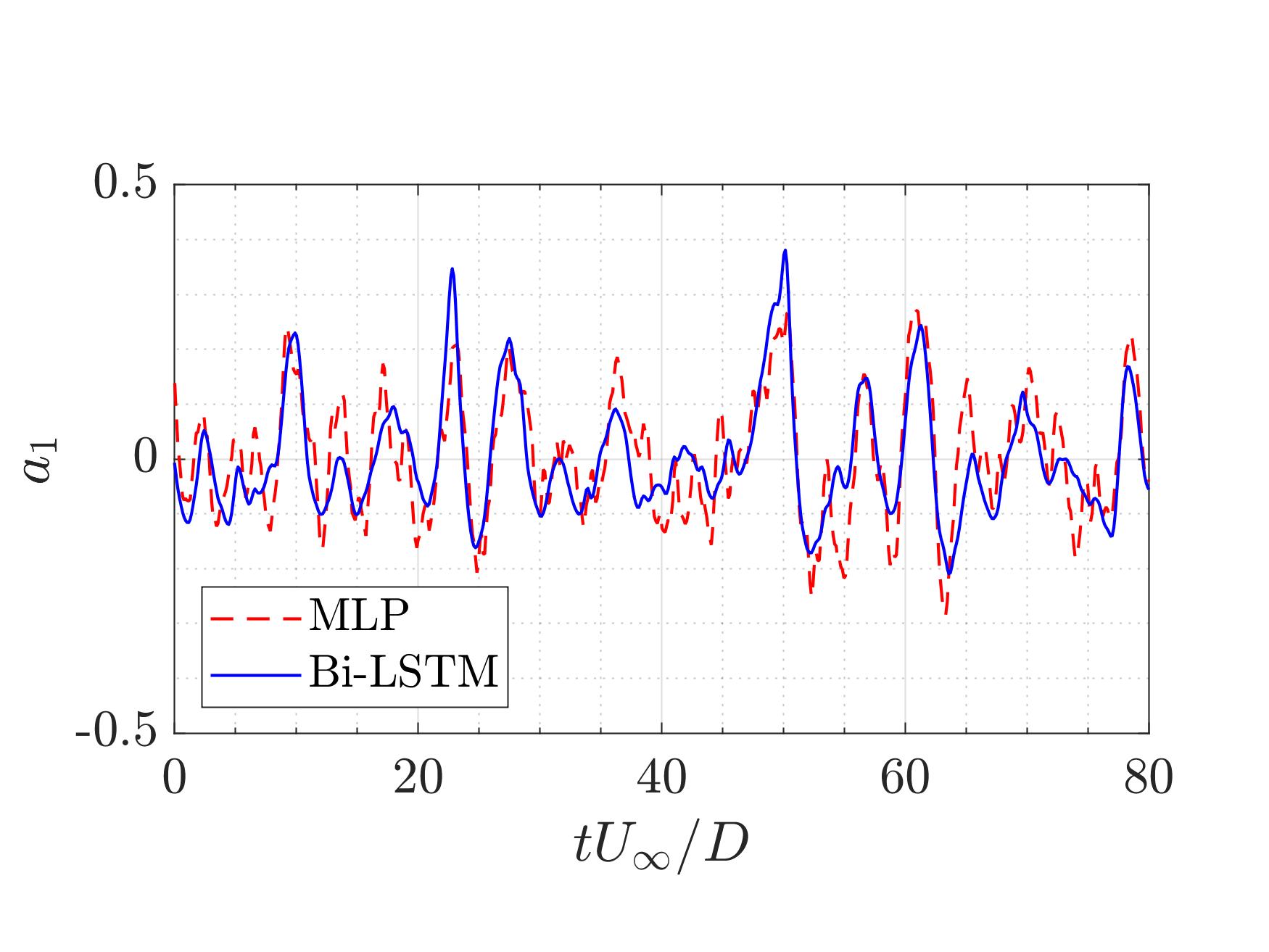}
    \caption{$a_{1}$}
\end{subfigure}
\begin{subfigure}{0.49\textwidth}
    \includegraphics[width=0.99\textwidth,trim={1cm 3cm 3cm 7.6cm },clip]{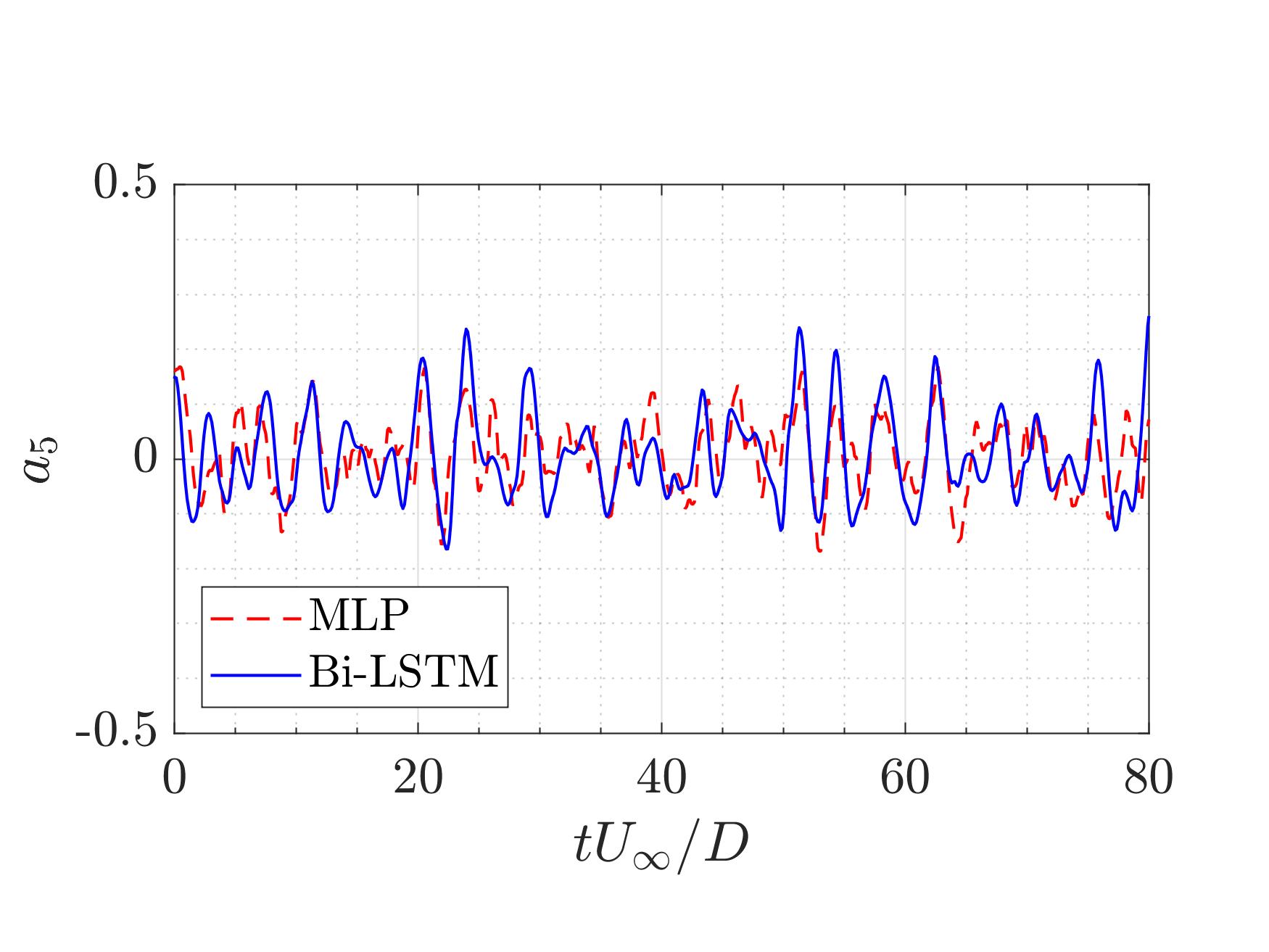}
    \caption{$a_{5}$}
\end{subfigure}\\
\begin{subfigure}{0.49\textwidth}
    \includegraphics[width=0.99\textwidth,trim={1cm 3cm 3cm 7.6cm },clip]{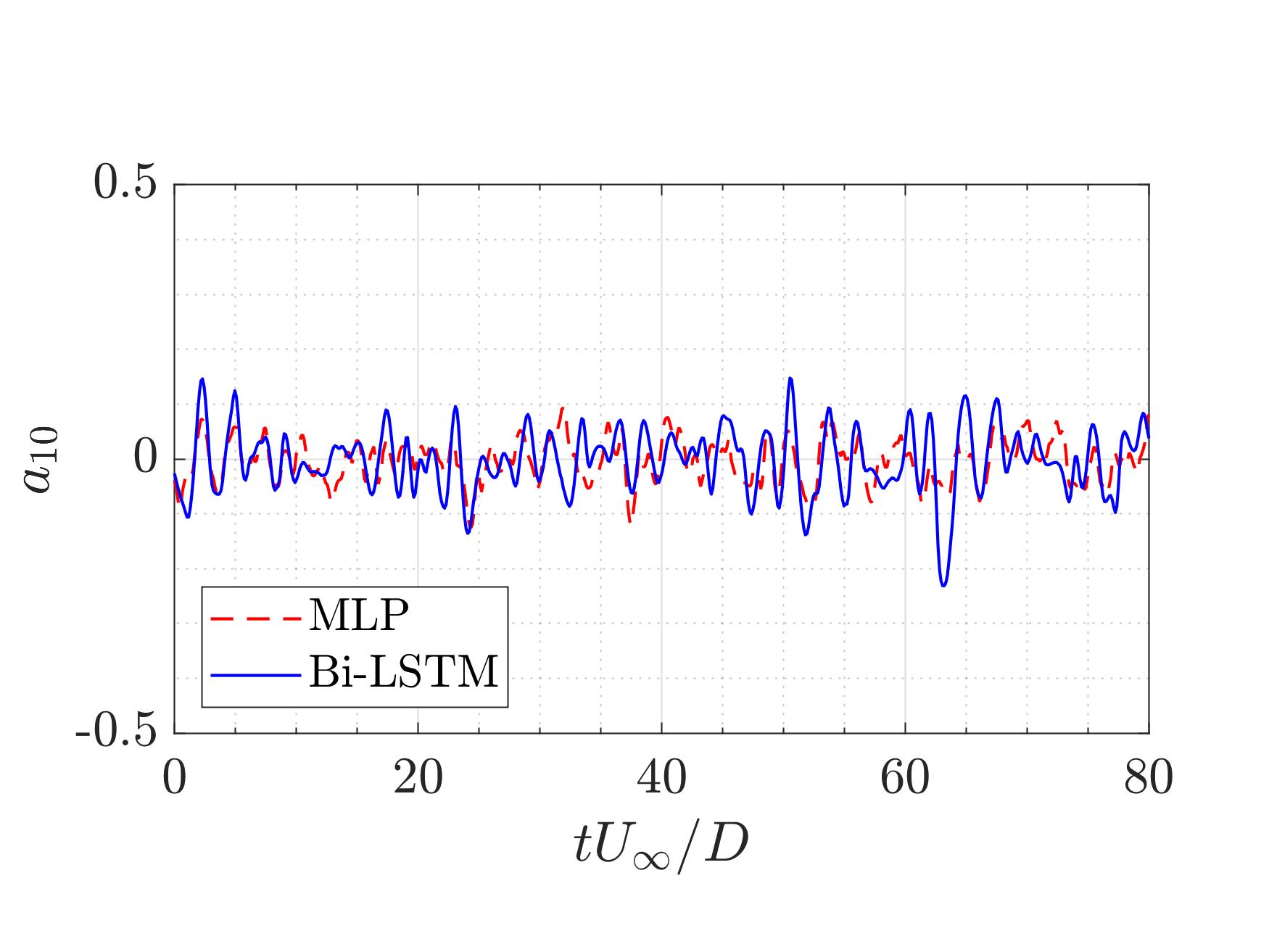}
    \caption{$a_{10}$}
\end{subfigure}
\begin{subfigure}{0.49\textwidth}
    \includegraphics[width=0.99\textwidth,trim={1cm 3cm 3cm 7.6cm },clip]{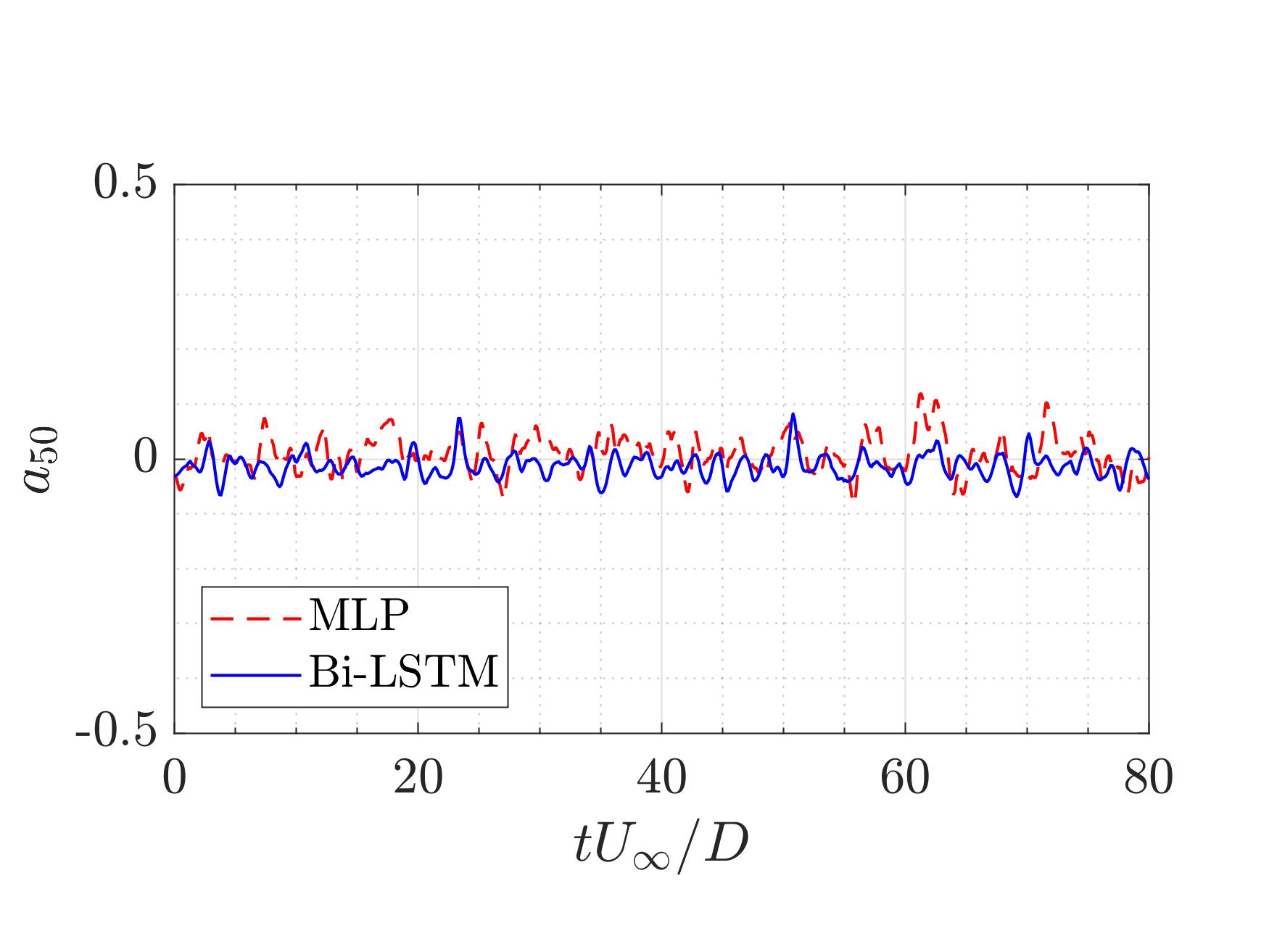}
    \caption{$a_{50}$}
\end{subfigure}
\caption{Time-varying estimation of POD modal expansion coefficients.}
\label{fig:a_est}
\end{figure}

\begin{figure}[!b]
    \centering
    \includegraphics[width=0.6\linewidth]{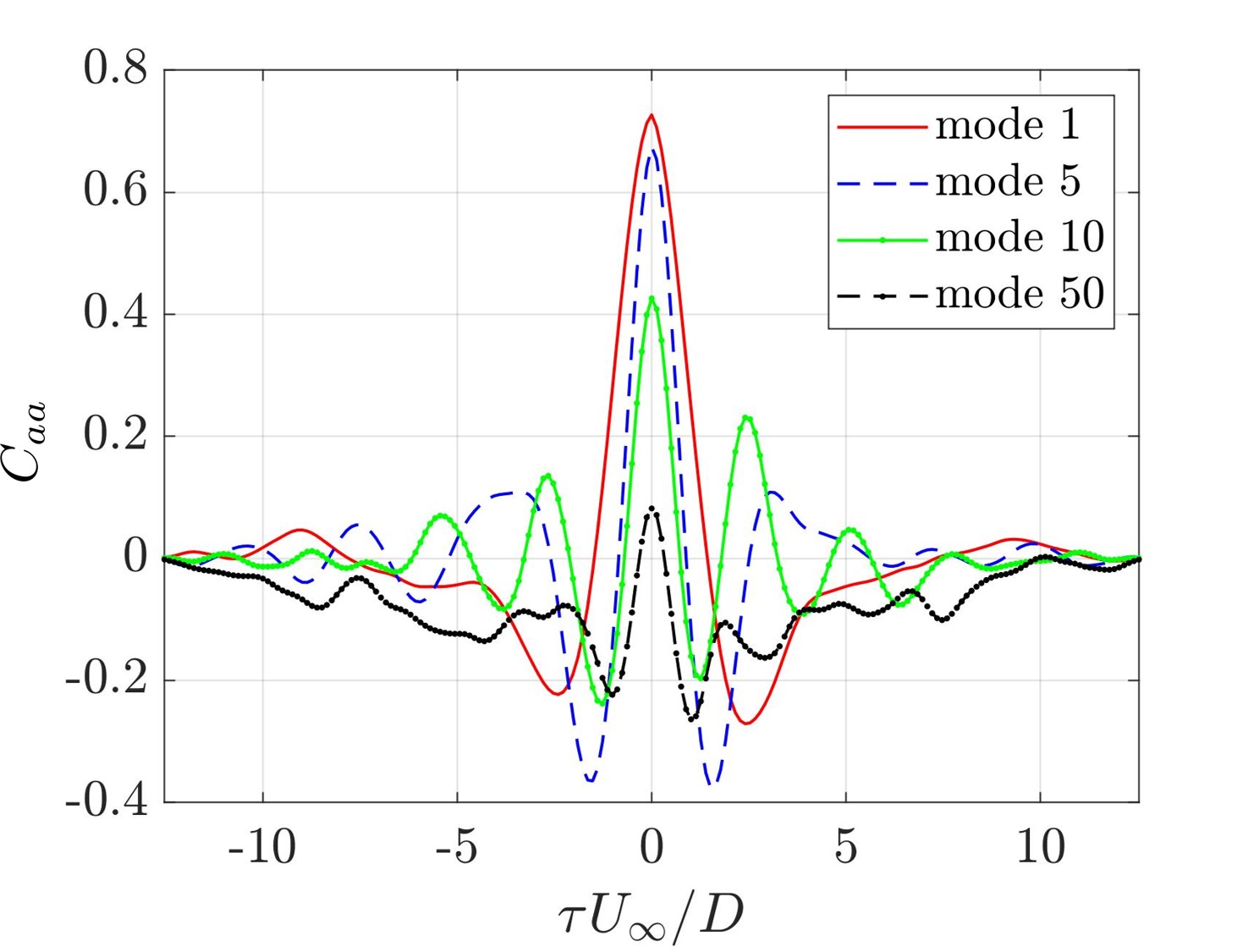}
    \caption{Cross-correlation coefficient between estimated expansion coefficients from two machine learning schemes.}
    \label{fig:xcorr_aa}
\end{figure}

\Cref{fig:a_est} displays pressure-informed estimation of POD expansion coefficients from both NN architectures for modes 1, 5,10, 50, respectively. 
Despite the expansion coefficients for each mode were scaled into $[-1,1]$ before training, \Cref{fig:a_est} manifest that the amplitude of prediction results will gradually reduce with increased POD modes. 
This is consistent to the fact that higher POD modes representing smaller-scaled turbulent structures are less coherent to the pressure fluctuations in the flow. 
For POD mode 1 and mode 5, prediction results from MLP and bidirectional LSTM shows remarkable similarity with respect to large scale fluctuations, in which locations of crests and troughs in time are mostly identical to each other although discrepancies on the amplitudes can be observed. 
Prediction results from both schemes become less similar for higher order modes, and the temporal evolution hardly matches each other at mode 50. 
As a measure of similarity between estimations from the two NN architectures, time-lagged cross-correlation coefficients between outputs from both networks are presented in \Cref{fig:xcorr_aa} using the same pressure input.
High correlation levels can be observed for the POD mode 1 and 5, which suggests strong consistency between estimations of the leading POD modes from both neural network architectures, and the dominant input-output relation between pressure and the most energetic coherent structures can be well captured from both trained networks. 
A decreased correlation level is observed for higher order POD modes, in which the correlation level decays from 0.42 to 0.08 from mode 10 to mode 50.
Since the correlation between pressure and smaller-scaled vortices becomes weaker, the influence of difference NN architectures to the prediction outcomes starts to dominate and greatly reduces the similarity level.
Nevertheless, the wave-packet shapes, as well as the high symmetricity shown in the cross-correlation coefficients, both indicate that both NN schemes are able to represent the dynamics of the leading POD modes (i.e. dominant coherent structures) associated with in-flow pressure to some extent.

\begin{figure}[!h]
\centering
\begin{subfigure}{0.49\textwidth}
    \includegraphics[width=0.9\textwidth]{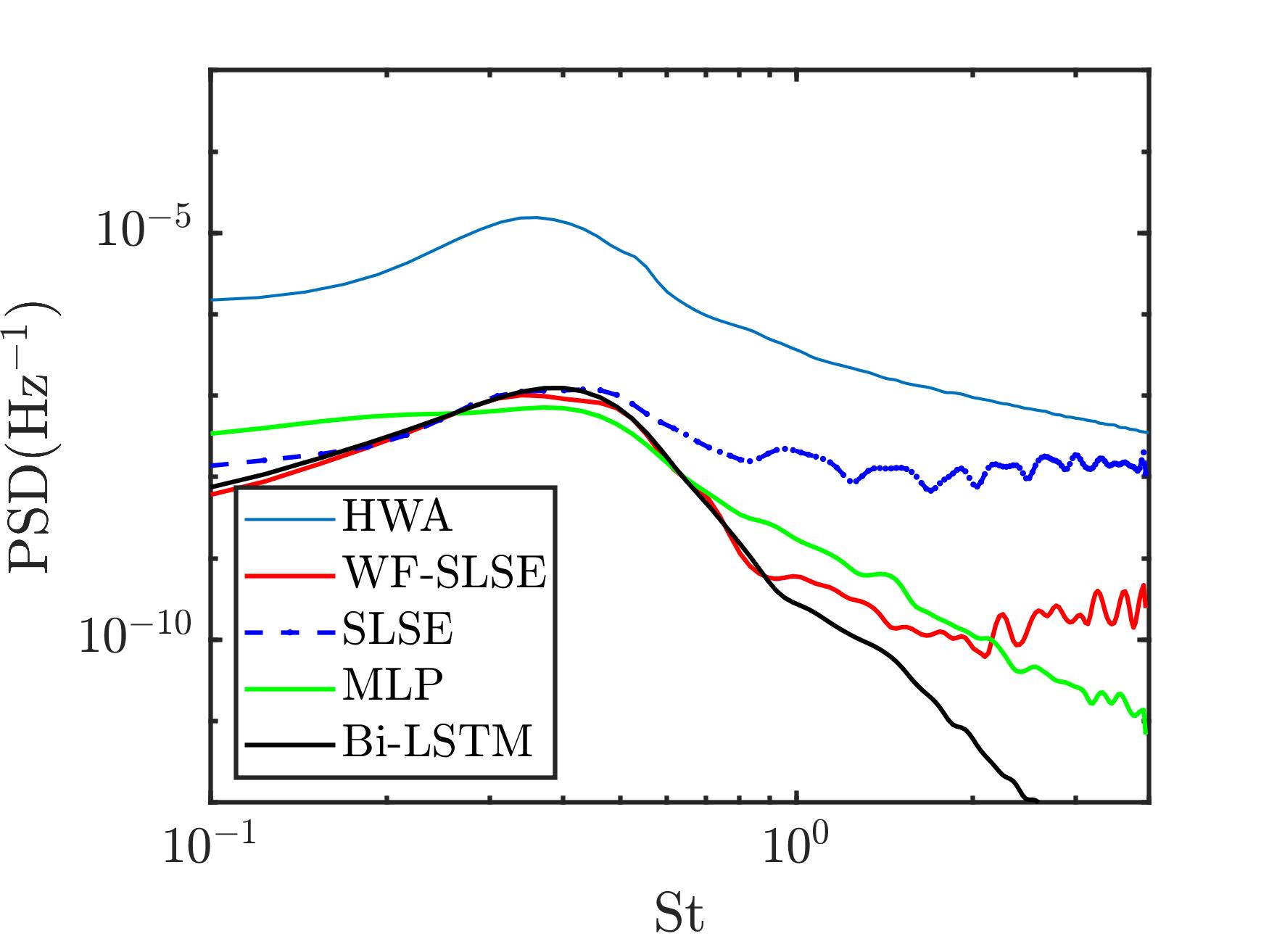}
    \caption{$x/D=4$, $r/D = 0$}
    \label{fig:4C_est}
\end{subfigure}
\begin{subfigure}{0.49\textwidth}
    \includegraphics[width=0.9\textwidth]{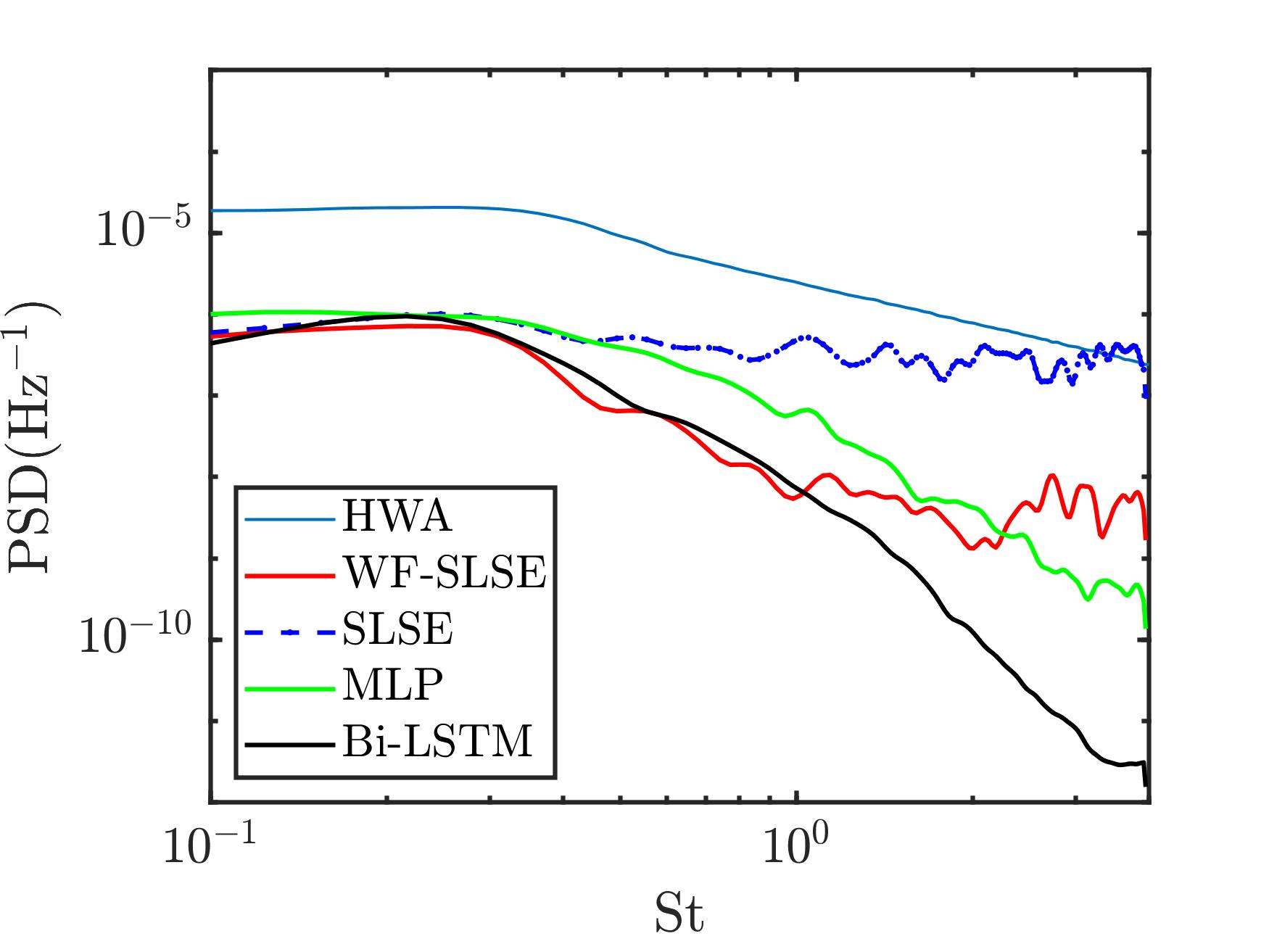}
    \caption{$x/D=4$, $r/D = 0.5$}
    \label{fig:4S_est}
\end{subfigure}\\
\begin{subfigure}{0.49\textwidth}
    \includegraphics[width=0.9\textwidth]{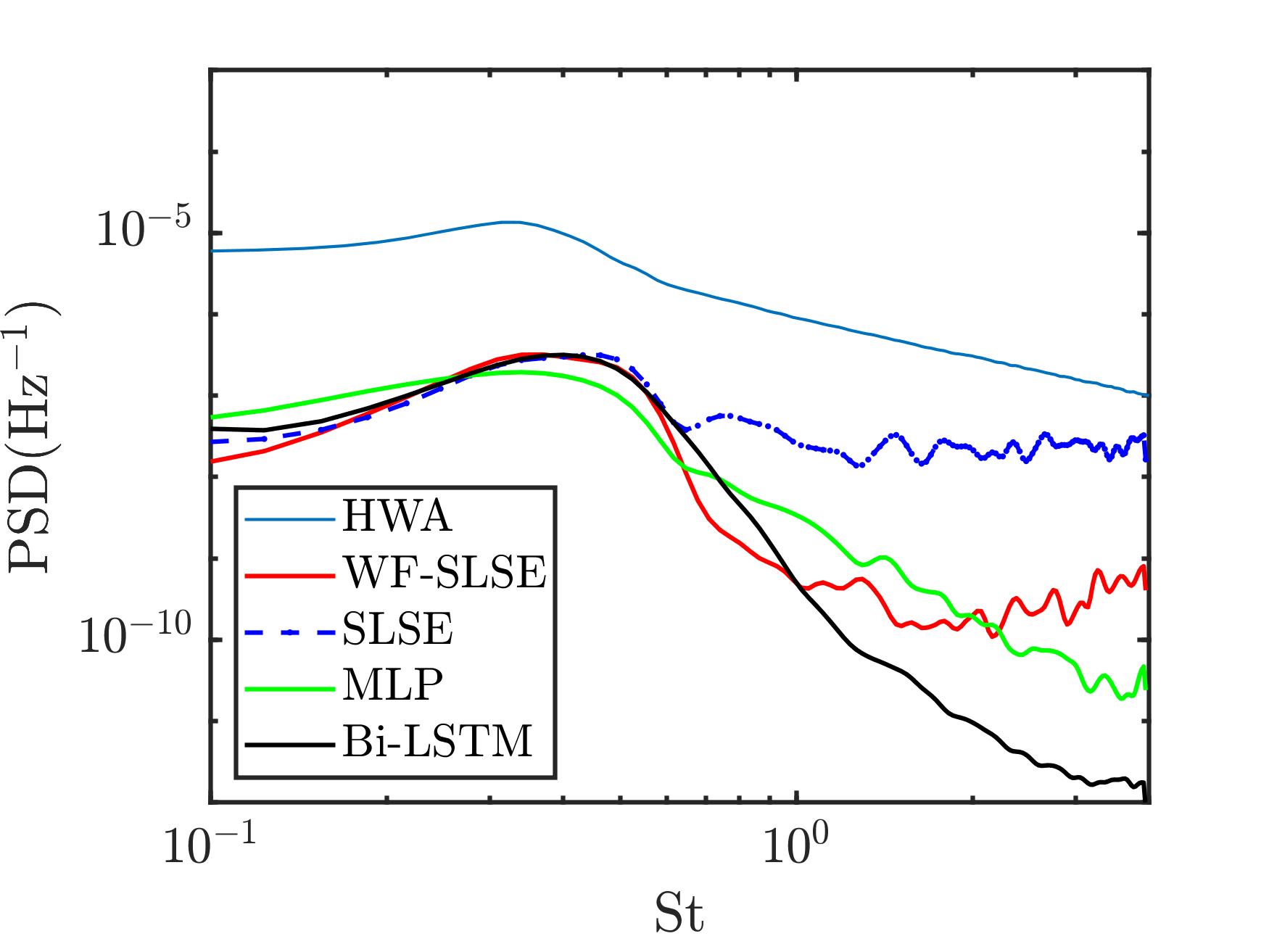}
    \caption{$x/D=5$, $r/D = 0$}
    \label{fig:5C_est}
\end{subfigure}
\begin{subfigure}{0.49\textwidth}
    \includegraphics[width=0.9\textwidth]{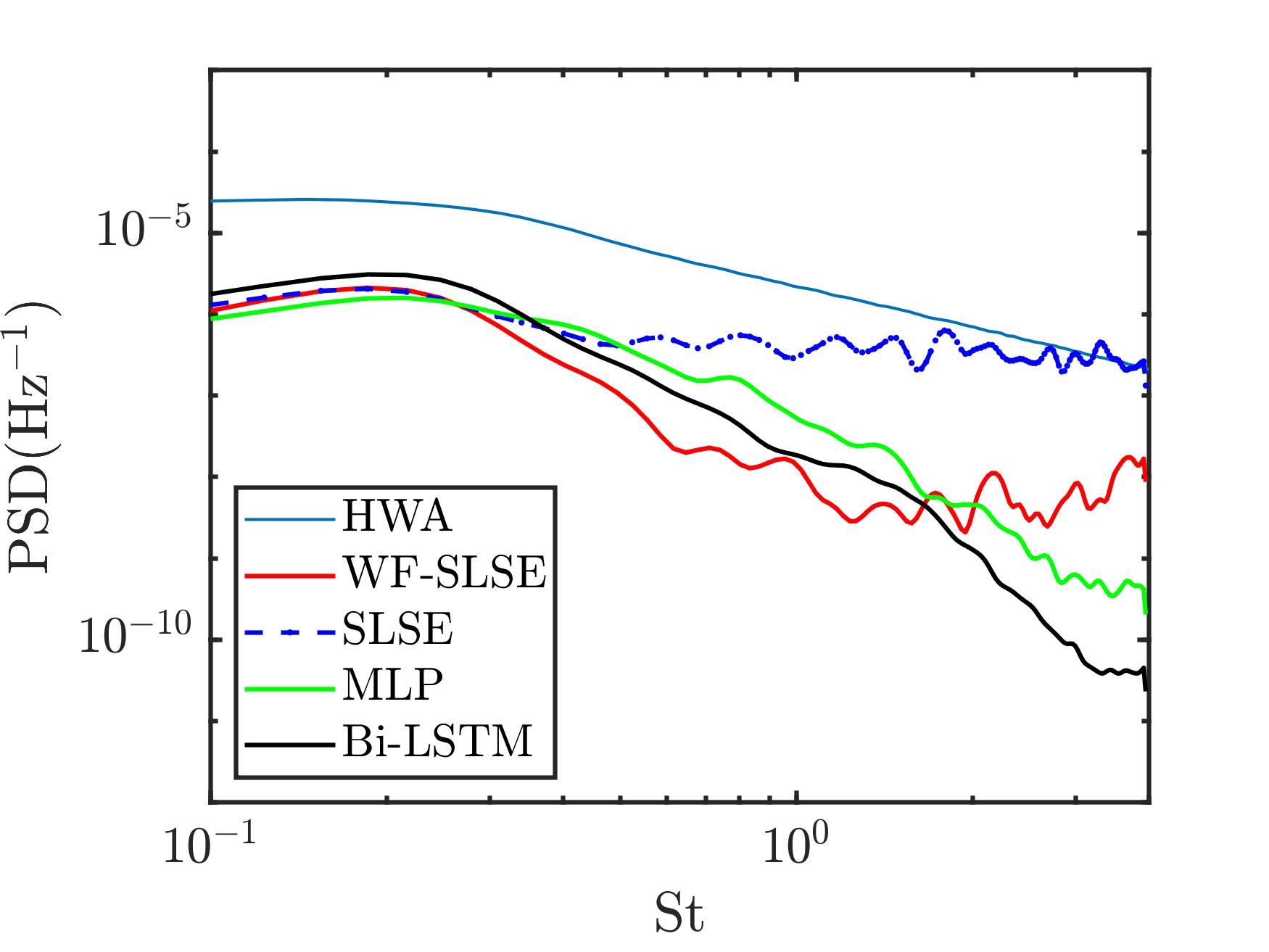}
    \caption{$x/D=5$,$r/D = 0.5$}
    \label{fig:5S_est}
\end{subfigure}\\
\begin{subfigure}{0.49\textwidth}
    \includegraphics[width=0.9\textwidth]{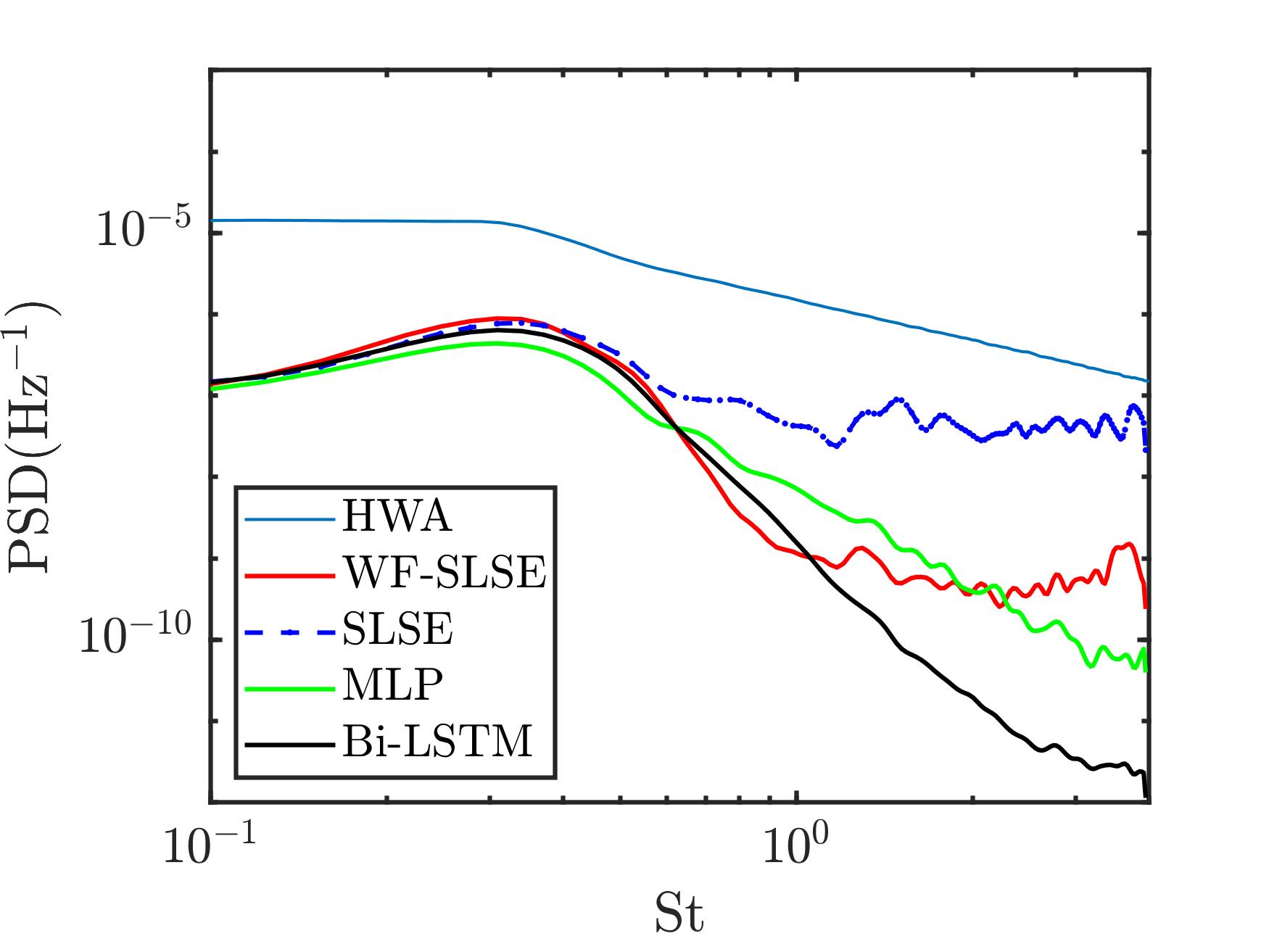}
    \caption{$x/D=6$, $r/D = 0$}
    \label{fig:6C_est}
\end{subfigure}
\begin{subfigure}{0.49\textwidth}
    \includegraphics[width=0.9\textwidth]{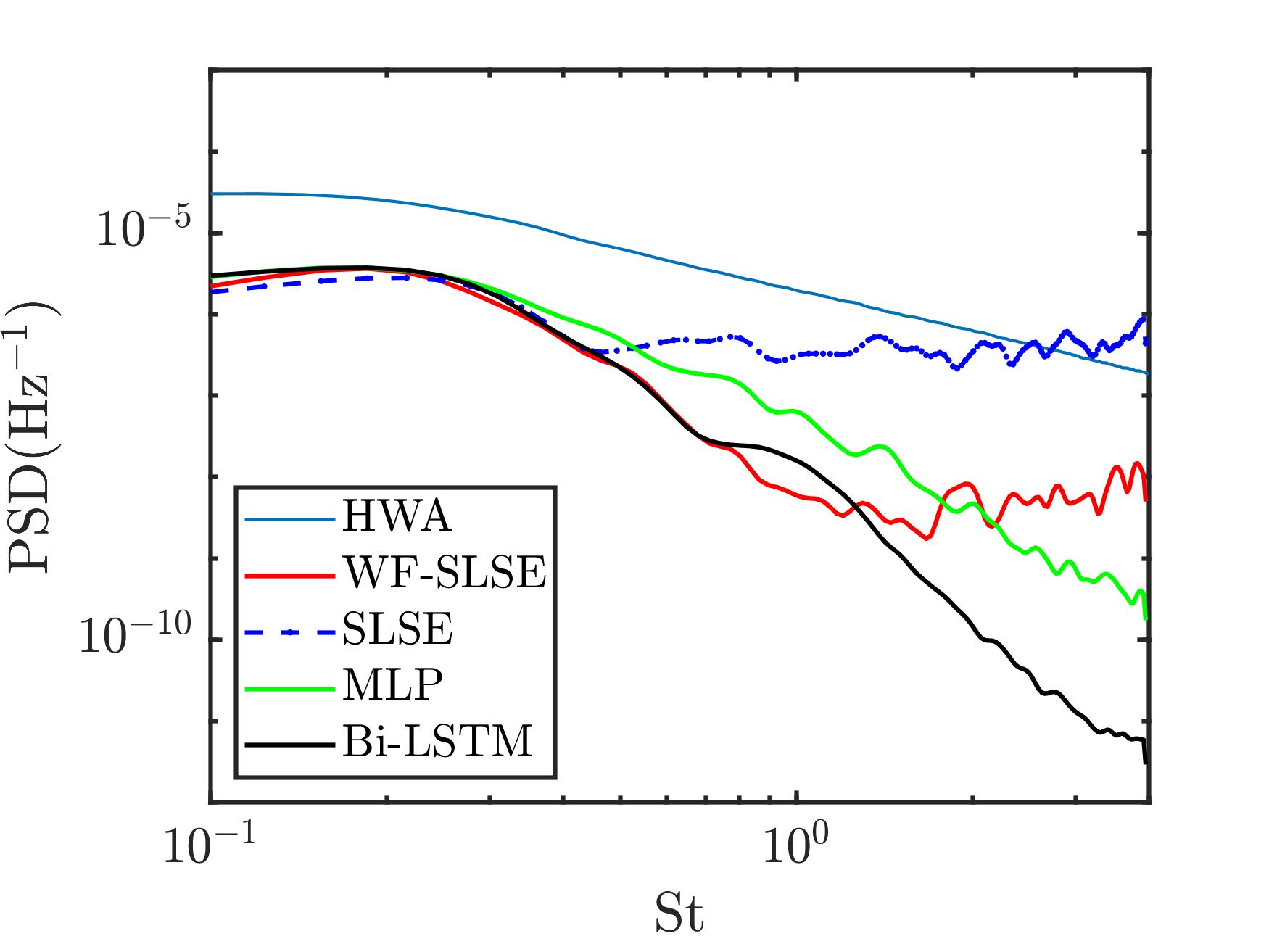}
    \caption{$x/D=6$, $r/D = 0.5$}
    \label{fig:6S_est}
\end{subfigure}
\caption{A comparison of velocity spectra from neural networks and stochastic estimations.}
\label{fig:cmp_est}    
\end{figure}

The streamwise velocity spectra estimated from machine learning architectures are compared to the ones from SLSE-POD, and results at three representative streamwise locations are shown in \Cref{fig:cmp_est}. 
Direct measurements from a hotwire measurement are also superimposed on the figures. 
Given the relatively high noise level between pressure and POD expansion coefficients, the wavelet-filtered (WF) SLSE-POD is also utilized to generate the spectral estimation from the wavelet-filtered cross-correlation.
For all cases, intensive noise can be observed in the SLSE spectra after $St = 0.5$, which is caused by the noise contained in the raw cross-correlation functions. 
Although the implementation of a wavelet-filter is able to attenuate the noise level by several orders of magnitude, the empirical choice of the filter threshold will impair the rolling-off of the spectra and the curves will stop decaying after $St = 1$. 
In contrast, the high frequency noise is effectively damped in the spectra predicted from both NN architectures. 
The bidirectional LSTM spectra always present a steeper rolling-off rate than the MLP spectra and decay rates from both networks are faster than the ones from the hotwire measurements.
Since the rolling-off of velocity spectra is contributed from the less-organized fine-scale turbulence, the POD-based NN architectures, which only preserve the linear combination of the large-scale spatial features to reproduce the flow field, are reasonable to possess a faster decay than the measured spectra.
These facts also confirm the effectiveness for NN architectures to distill useful information associated with the dominant spatial features from the pressure inputs.

From $St = 0.1$ to $0.6$ where large-scale coherent structures are dominant, four estimation techniques are observed to generate similar spectral estimations. 
The high degree of similarity between spectral estimates from neural networks and linear stochastic estimations proves that relation between pressure and the large-scale structures appearing in the low-frequency range is predominantly linear. 
Spectral densities from SLSE-POD and WF-SLSE-POD are seen to be the closest to each other, which verifies the assumption that the wavelet-filtering process will preserve the prevailing wave-packet features from the noisy cross-correlation functions. 
As for NN architectures, velocity spectra predicted from the bidirectional LSTM scheme are more consistent to the denoised WF-SLSE-POD results.
Since it has been verified that the WF-SLSE-POD is capable to preserve the most significant wave-packet features in the input-output models, the similarity between bidirectional LSTM and WF-SLSE-POD demonstrates great advantage of this neural network architecture given the fact that this method can produce high-quality spectral estimates and no filtering threshold need to be determined at the same time in an empirical manner. 
Such similarity also validates the general advantage of RNN-based neural networks to effectively handle sequential input data to extract dominant features. 

\begin{figure}[!h]
\centering
\begin{subfigure}{1\textwidth}
\includegraphics[width=0.9\linewidth,trim={18cm 5cm 18cm 3cm },clip]{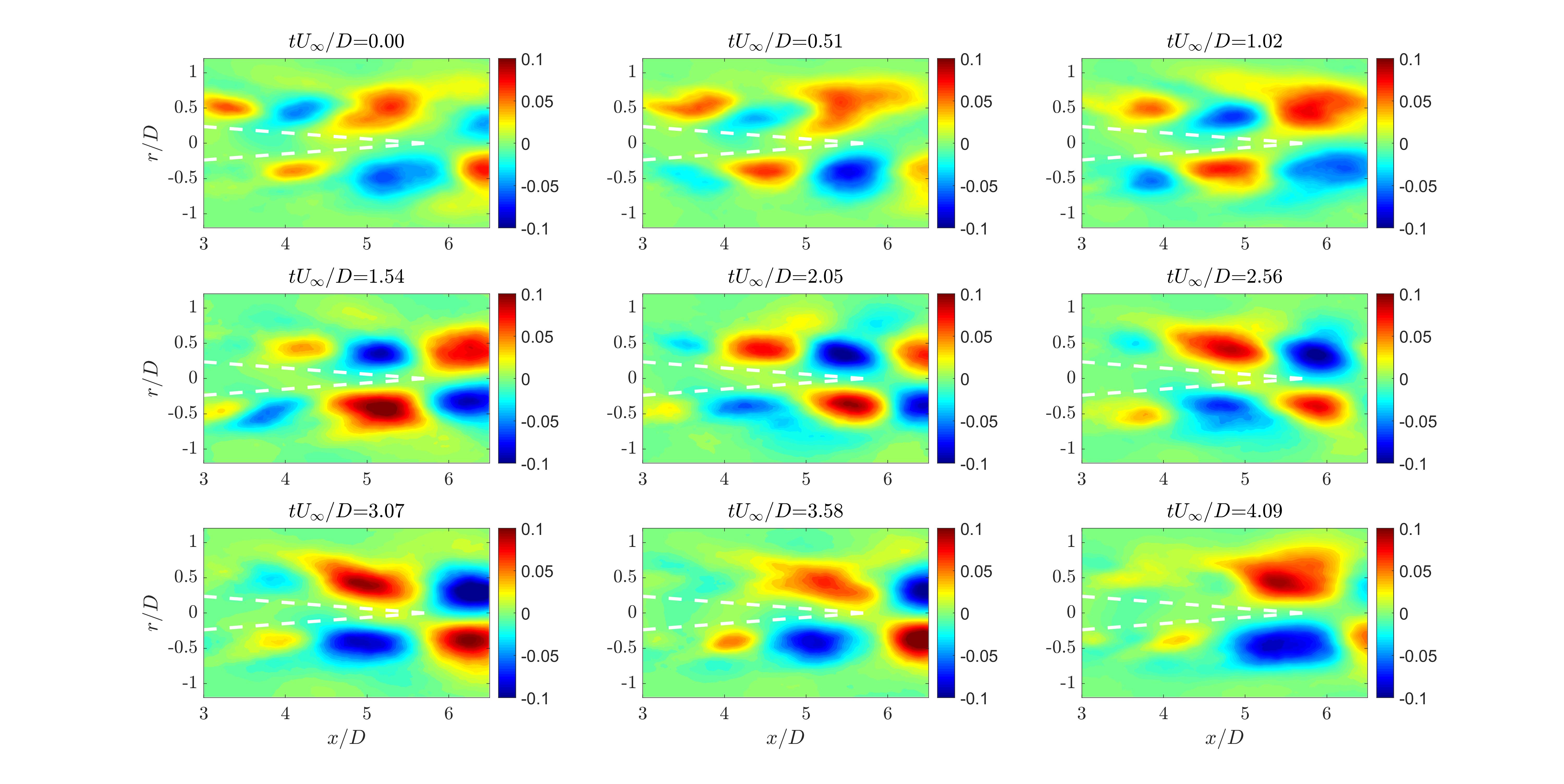}
\caption{Streamwise velocity $u_x/U_{\infty}$}
\label{fig:u_2d}
\end{subfigure}\\
\begin{subfigure}{1\textwidth}
\includegraphics[width=0.9\linewidth,trim={18cm 5cm 18cm 3cm },clip]{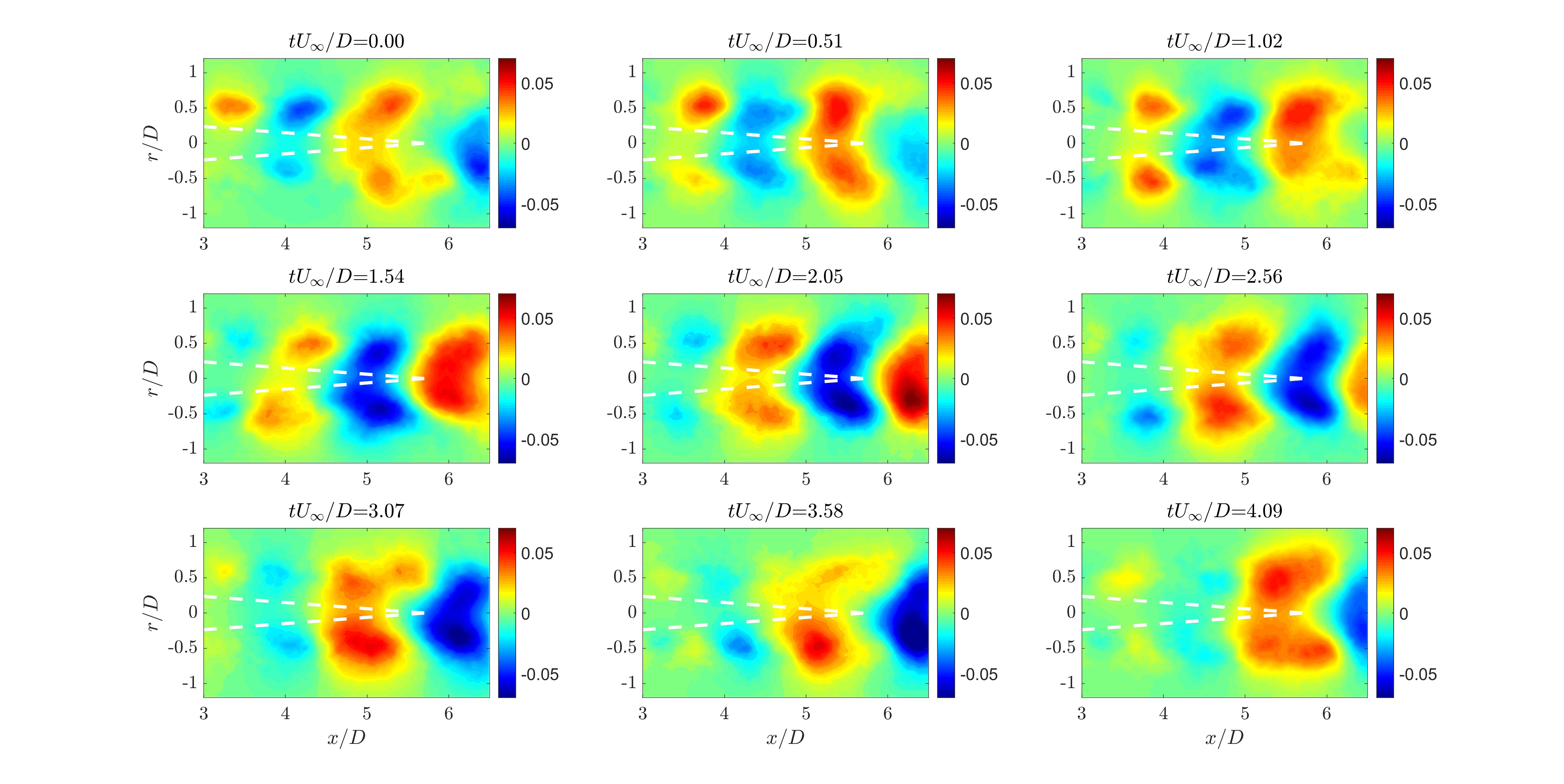}
\caption{Radial velocity $u_r/U_{\infty}$}
\label{fig:v_2d}
\end{subfigure}
\caption{Instantaneous velocity reconstruction from bidirectional LSTM. White dashed lines depict the jet potential core.}
\label{fig:2d_recon}
\end{figure}

\begin{figure}[!h]
\includegraphics[width=0.9\linewidth,trim={18cm 5cm 18cm 3cm },clip]{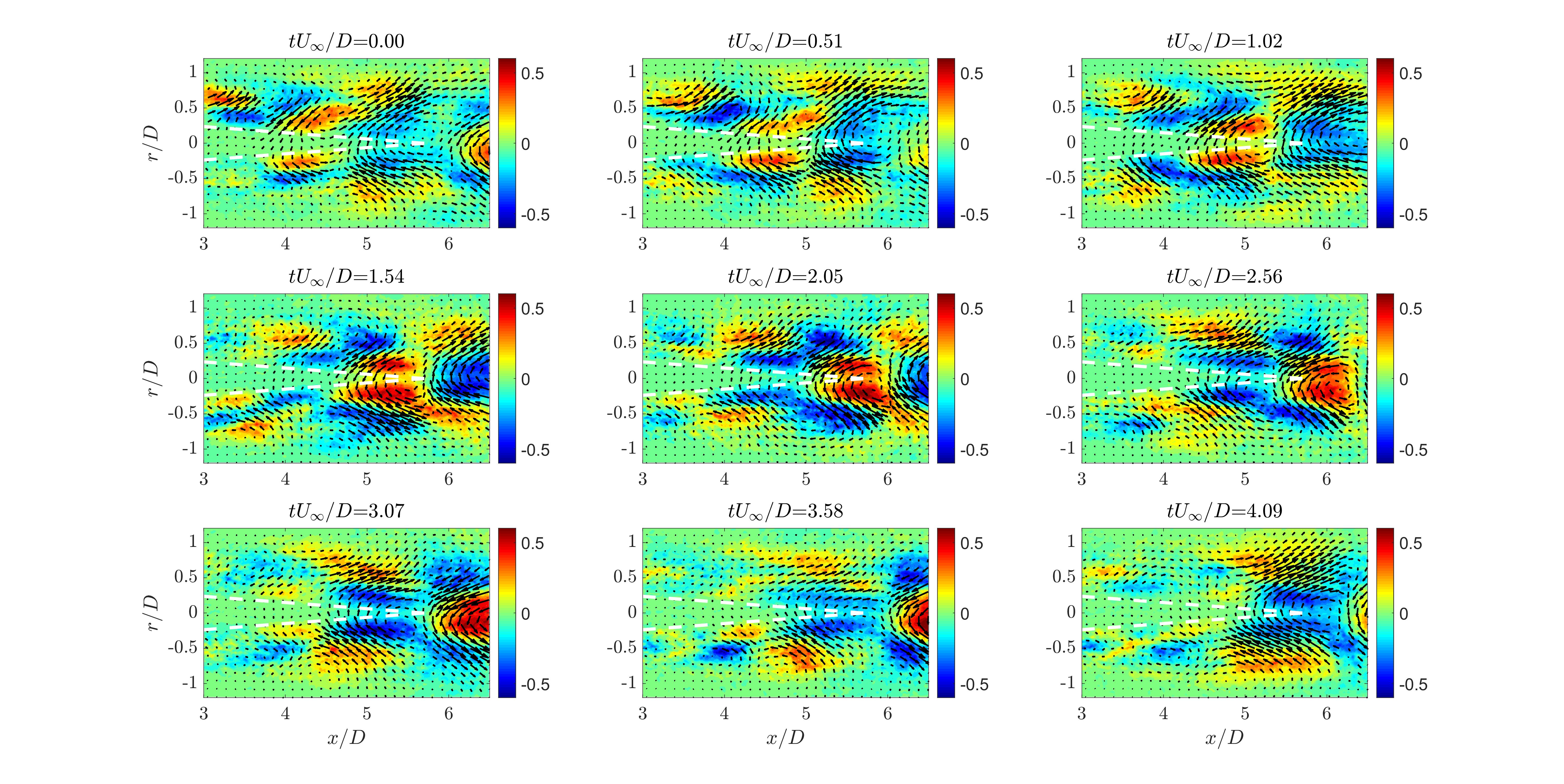}
\caption{2D vorticity ($\omega\frac{D}{U_{\infty}}$) of the reconstructed flow field from bidirectional LSTM overlaid by velocity vectors.}
\label{fig:vor_2d}
\end{figure}

When comparing to the hotwire measurements, all estimation methods will underestimate the overall spectral density in general. 
This is consistent with the fact that only the portion of velocity associated with pressure inputs can be well reflected from the estimations. 
The difference in amplitude will be gradually narrowed down when $x$ moves from $4D$ to $6D$, indicating a better relation between pressure and velocity can be constructed when the streamwise distance is confined. 
For large-scale events at $x/D = 4$, one may observe that the hump between $St = 0.2$ and $0.6$ on the jet lipline, as well as the broadband event at $St<0.3$ on the jet centerline, are both faithfully reflected from spectral estimations. 
However at $5D$ and $6D$ downstream, a discrepancy on the peak location is found on the jet centerline although the broadband events between $0.1<St<0.3$ on the jet lipline can still be fairly captured. 
For example, at $r=0$, the measured velocity spectrum peaks around $St=0.3$ at $x/D = 5$ and a flattened peak is seen to form between $0.1<St<0.3$ at $x/D = 6$. 
However, for both stochastic estimations and neural networks, velocity estimates will peak around $St = 0.4$ (see \Cref{fig:5C_est,fig:6C_est}). 
Since the centerline velocity estimation is informed by the pressure recorded on the jet liplines, the spectral estimations only represent the part of centerline velocity that is related to the pressure fluctuations on liplines. 
In other words, the peak location of spectral estimates won't necessarily match the actual peak location without knowing related information from pressure inputs. 
The addition of pressure measurements on the jet axis should be capable to resolve the shift of the peak location.

\Cref{fig:2d_recon} shows time-resolved velocity reconstruction from the bidirectional LSTM using the first 50 POD modes which corresponds to the portion of the overall velocity field that is associated with pressure measured downstream.
The jet potential core is also depicted in all instances based on the jet core length of $5.73D$. 
The core length was calculated from the mean velocity profile based on the definition in \cite{bres2018}.
\Cref{fig:u_2d,fig:v_2d} show the temporal evolution of the streamwise and radial velocity components, respectively.
Following the time sequence, the emergence, growth, and orderly convection of the large-scale structures along the streamwise direction can be clearly observed inside the jet shear layer in \Cref{fig:u_2d}, and these structures corresponds to the streamwise propagation of wave-like patterns in \Cref{fig:v_2d} where regularly distributed structures in the radial direction move oppositely to their neighbors. 
As these structures move downstream, one can clearly observe the growth and intensification of the turbulent eddies.
A comprehensive evaluation of the velocity reconstruction is presented in \Cref{fig:vor_2d} in which the 2D vorticity, $\omega=\nabla_{2D}\times u_i$ is displayed and velocity vectors are overlaid on top.
Before the end of the potential core, the dominant feature is the growth and streamwise convection of the vortical structures. 
The appearance of counter-rotating vortex pairs can also be observed inside the jet mixing layer.
At the same time, the vorticity is at an insignificant level inside the potential core which is typically accepted to be irrotational although flapping motion can be detected by the end of the jet core. 
Around the end of the jet core, vortical structures on both sides of the potential core are merged into some larger structures spanning across the jet centerline and keep travelling with the flow.
These eddies are highlighted not only by their large sizes but also by the high overall vorticity magnitude.
Since this pressure-informed estimation filters out the part of the flow that is associated with the downstream pressure, these turbulent structures emerged downstream of the jet core are highly associated with the pressure wave-packets in the flow which serves as noise sources propagating to the acoustic farfield.

\section{Conclusions} \label{sec:conclusions}
In this study, time-resolved source activities inside the jet mixing layer were estimated based on PIV and downstream in-flow pressure measurements.
Stochastic estimation methods and neural networks were proposed to model the input-output relationship between in-flow pressure fluctuations and POD expansion coefficients.
For the implementation of stochastic estimation, a wavelet-based filter was also utilized to extract the dominant the wave-packet structures from the noisy cross-correlation functions obtained from the experiments.

Two sets of experiments were performed in combination with the proposed estimation architectures. The space-time dynamics of the axisymmetric velocity components was characterized from cross-stream PIV measurements as well as downstream pressure measurements on the jet axis. 
For axisymmetric velocity components, strong mode-1 dominance in space can be observed at all streamwise locations from the azimuthal-Fourier POD. 
Cross-correlation between pressure and the first POD expansion coefficients shows significant wave-like behavior in the ﬂow at a convection speed of $0.7U_{\infty}$.
Time-resolved estimation of axisymmetric velocity from wavelet-filtered SLSE-POD reproduce the convective nature of the axisymmetric velocity components. A comparison to the measured velocity spectra from hotwire measurements demonstrates the capacity of this hybrid approach to faithfully reflect the broadband hump in the low frequency range although a discrepancy in the overall spectral amplitude is observed. 

Pressure measurements on the upper and lower jet liplines were synchronously performed with streamwise planar PIV to obtain real-time estimates of two-dimensional velocity vectors on a streamwise plane.
The most energetic spatial structures are seen to distribute within a wide range of POD modes, and the cross-correlation coefficients between lipline pressure and the dominant POD modes are at a relatively low level.
Stochastic estimations and neural networks were employed to reconstruct the 2D source behaviors using the first 50 POD modes. 
All estimation approaches utilized in this work are capable of highlighting the broadband peak at low Strouhal numbers in spite of the existence of amplitude discrepancies compared to the measured spectra. 
In the comparison of velocity spectra estimates, the unique advantage of bidirectional LSTM architecture to highlight the broadband hump as well as attenuating the high frequency noise is observed.
Real-time velocity reconstruction from bidirectional LSTM highlights the streamwise convection of the coherent structures inside the jet mixing layer as well as the formation of larger eddies downstream of the jet potential core.
These results are beneficial to enhance the understanding of the space-time dynamics of the acoustic sources in the jet flow field, and the resulting velocity field estimates could be used to calculate the pressure field via Poisson’s equation \cite{vdk2019} which allows one to evaluate the pressure wave-packets in the flow.

\section*{Acknowledgments}
The authors acknowledge the support from Wenyan Li for the development of the machine learning code.
The authors also acknowledge University of Florida Research Computing (UFRC) for providing computational resources and support that have contributed to the research results reported in this publication.
This research was funded by the National Science Foundation (NSF) under award CBET-1704768.


\bibliography{references}

\begin{thebibliography}{58}%
\makeatletter
\providecommand \@ifxundefined [1]{%
 \@ifx{#1\undefined}
}%
\providecommand \@ifnum [1]{%
 \ifnum #1\expandafter \@firstoftwo
 \else \expandafter \@secondoftwo
 \fi
}%
\providecommand \@ifx [1]{%
 \ifx #1\expandafter \@firstoftwo
 \else \expandafter \@secondoftwo
 \fi
}%
\providecommand \natexlab [1]{#1}%
\providecommand \enquote  [1]{``#1''}%
\providecommand \bibnamefont  [1]{#1}%
\providecommand \bibfnamefont [1]{#1}%
\providecommand \citenamefont [1]{#1}%
\providecommand \href@noop [0]{\@secondoftwo}%
\providecommand \href [0]{\begingroup \@sanitize@url \@href}%
\providecommand \@href[1]{\@@startlink{#1}\@@href}%
\providecommand \@@href[1]{\endgroup#1\@@endlink}%
\providecommand \@sanitize@url [0]{\catcode `\\12\catcode `\$12\catcode
  `\&12\catcode `\#12\catcode `\^12\catcode `\_12\catcode `\%12\relax}%
\providecommand \@@startlink[1]{}%
\providecommand \@@endlink[0]{}%
\providecommand \url  [0]{\begingroup\@sanitize@url \@url }%
\providecommand \@url [1]{\endgroup\@href {#1}{\urlprefix }}%
\providecommand \urlprefix  [0]{URL }%
\providecommand \Eprint [0]{\href }%
\providecommand \doibase [0]{https://doi.org/}%
\providecommand \selectlanguage [0]{\@gobble}%
\providecommand \bibinfo  [0]{\@secondoftwo}%
\providecommand \bibfield  [0]{\@secondoftwo}%
\providecommand \translation [1]{[#1]}%
\providecommand \BibitemOpen [0]{}%
\providecommand \bibitemStop [0]{}%
\providecommand \bibitemNoStop [0]{.\EOS\space}%
\providecommand \EOS [0]{\spacefactor3000\relax}%
\providecommand \BibitemShut  [1]{\csname bibitem#1\endcsname}%
\let\auto@bib@innerbib\@empty
\bibitem [{\citenamefont {Dougherty}(2002)}]{dougherty2002}%
  \BibitemOpen
  \bibfield  {author} {\bibinfo {author} {\bibfnamefont {R.~P.}\ \bibnamefont
  {Dougherty}},\ }\bibfield  {title} {\bibinfo {title} {Beamforming in acoustic
  testing},\ }in\ \href {https://doi.org/10.1007/978-3-662-05058-3_2} {\emph
  {\bibinfo {booktitle} {Aeroacoustic Measurements}}}\ (\bibinfo  {publisher}
  {Springer Berlin Heidelberg},\ \bibinfo {year} {2002})\ pp.\ \bibinfo {pages}
  {62--97}\BibitemShut {NoStop}%
\bibitem [{\citenamefont {Papamoschou}(2011)}]{papamoschou2011}%
  \BibitemOpen
  \bibfield  {author} {\bibinfo {author} {\bibfnamefont {D.}~\bibnamefont
  {Papamoschou}},\ }\bibfield  {title} {\bibinfo {title} {Imaging of
  directional distributed noise sources},\ }\href
  {https://doi.org/https://doi.org/10.1016/j.jsv.2010.11.025} {\bibfield
  {journal} {\bibinfo  {journal} {Journal of Sound and Vibration}\ }\textbf
  {\bibinfo {volume} {330}},\ \bibinfo {pages} {2265} (\bibinfo {year}
  {2011})}\BibitemShut {NoStop}%
\bibitem [{\citenamefont {Breen}\ and\ \citenamefont
  {Ahuja}(2015)}]{breen2015}%
  \BibitemOpen
  \bibfield  {author} {\bibinfo {author} {\bibfnamefont {N.~P.}\ \bibnamefont
  {Breen}}\ and\ \bibinfo {author} {\bibfnamefont {K.~K.}\ \bibnamefont
  {Ahuja}},\ }\bibfield  {title} {\bibinfo {title} {Measuring jet noise source
  locations with acoustic beamforming},\ }in\ \href
  {https://doi.org/10.2514/6.2015-0735} {\emph {\bibinfo {booktitle} {53rd
  {AIAA} Aerospace Sciences Meeting}}}\ (\bibinfo  {publisher} {American
  Institute of Aeronautics and Astronautics},\ \bibinfo {year}
  {2015})\BibitemShut {NoStop}%
\bibitem [{\citenamefont {Semeraro}\ \emph {et~al.}(2012)\citenamefont
  {Semeraro}, \citenamefont {Bellani},\ and\ \citenamefont
  {Lundell}}]{semeraro2012}%
  \BibitemOpen
  \bibfield  {author} {\bibinfo {author} {\bibfnamefont {O.}~\bibnamefont
  {Semeraro}}, \bibinfo {author} {\bibfnamefont {G.}~\bibnamefont {Bellani}},\
  and\ \bibinfo {author} {\bibfnamefont {F.}~\bibnamefont {Lundell}},\
  }\bibfield  {title} {\bibinfo {title} {Analysis of time-resolved {PIV}
  measurements of a confined turbulent jet using {POD} and koopman modes},\
  }\href {https://doi.org/10.1007/s00348-012-1354-9} {\bibfield  {journal}
  {\bibinfo  {journal} {Experiments in Fluids}\ }\textbf {\bibinfo {volume}
  {53}},\ \bibinfo {pages} {1203} (\bibinfo {year} {2012})}\BibitemShut
  {NoStop}%
\bibitem [{\citenamefont {Berger}\ \emph {et~al.}(2014)\citenamefont {Berger},
  \citenamefont {Berry}, \citenamefont {Shea}, \citenamefont {Glauser},
  \citenamefont {Jiang}, \citenamefont {Gogineni}, \citenamefont {Kaiser},
  \citenamefont {Noack},\ and\ \citenamefont {Spohn}}]{berger2014}%
  \BibitemOpen
  \bibfield  {author} {\bibinfo {author} {\bibfnamefont {Z.~P.}\ \bibnamefont
  {Berger}}, \bibinfo {author} {\bibfnamefont {M.~G.}\ \bibnamefont {Berry}},
  \bibinfo {author} {\bibfnamefont {P.~R.}\ \bibnamefont {Shea}}, \bibinfo
  {author} {\bibfnamefont {M.~N.}\ \bibnamefont {Glauser}}, \bibinfo {author}
  {\bibfnamefont {N.}~\bibnamefont {Jiang}}, \bibinfo {author} {\bibfnamefont
  {S.}~\bibnamefont {Gogineni}}, \bibinfo {author} {\bibfnamefont
  {E.}~\bibnamefont {Kaiser}}, \bibinfo {author} {\bibfnamefont {B.~R.}\
  \bibnamefont {Noack}},\ and\ \bibinfo {author} {\bibfnamefont
  {A.}~\bibnamefont {Spohn}},\ }\bibfield  {title} {\bibinfo {title} {Analysis
  of high speed jet flow physics with time-resolved {PIV}},\ }in\ \href
  {https://doi.org/10.2514/6.2014-1226} {\emph {\bibinfo {booktitle} {52nd
  Aerospace Sciences Meeting}}}\ (\bibinfo  {publisher} {American Institute of
  Aeronautics and Astronautics},\ \bibinfo {year} {2014})\BibitemShut {NoStop}%
\bibitem [{\citenamefont {Citriniti}\ and\ \citenamefont
  {George}(2000)}]{citriniti2000jfm}%
  \BibitemOpen
  \bibfield  {author} {\bibinfo {author} {\bibfnamefont {J.}~\bibnamefont
  {Citriniti}}\ and\ \bibinfo {author} {\bibfnamefont {W.~K.}\ \bibnamefont
  {George}},\ }\bibfield  {title} {\bibinfo {title} {Reconstruction of the
  global velocity field in the axisymmetric mixing layer utilizing the proper
  orthogonal decomposition},\ }\href@noop {} {\bibfield  {journal} {\bibinfo
  {journal} {Journal of Fluid Mechanics}\ }\textbf {\bibinfo {volume} {418}},\
  \bibinfo {pages} {137} (\bibinfo {year} {2000})}\BibitemShut {NoStop}%
\bibitem [{\citenamefont {IQBAL}\ and\ \citenamefont
  {THOMAS}(2007)}]{iqbal2007}%
  \BibitemOpen
  \bibfield  {author} {\bibinfo {author} {\bibfnamefont {M.~O.}\ \bibnamefont
  {IQBAL}}\ and\ \bibinfo {author} {\bibfnamefont {F.~O.}\ \bibnamefont
  {THOMAS}},\ }\bibfield  {title} {\bibinfo {title} {Coherent structure in a
  turbulent jet via a vector implementation of the proper orthogonal
  decomposition},\ }\href {https://doi.org/10.1017/s0022112006003351}
  {\bibfield  {journal} {\bibinfo  {journal} {Journal of Fluid Mechanics}\
  }\textbf {\bibinfo {volume} {571}},\ \bibinfo {pages} {281} (\bibinfo {year}
  {2007})}\BibitemShut {NoStop}%
\bibitem [{\citenamefont {Adrian}(1977)}]{adrian1977}%
  \BibitemOpen
  \bibfield  {author} {\bibinfo {author} {\bibfnamefont {R.}~\bibnamefont
  {Adrian}},\ }\bibfield  {title} {\bibinfo {title} {On the role of conditional
  averages in turbulence theory},\ }\href@noop {} {\bibfield  {journal}
  {\bibinfo  {journal} {Turbulence in Liquids}\ ,\ \bibinfo {pages} {322 }}
  (\bibinfo {year} {1977})}\BibitemShut {NoStop}%
\bibitem [{\citenamefont {Bonnet}\ \emph {et~al.}(1994)\citenamefont {Bonnet},
  \citenamefont {Cole}, \citenamefont {Delville}, \citenamefont {Glauser},\
  and\ \citenamefont {Ukeiley}}]{bonnet1994}%
  \BibitemOpen
  \bibfield  {author} {\bibinfo {author} {\bibfnamefont {J.~P.}\ \bibnamefont
  {Bonnet}}, \bibinfo {author} {\bibfnamefont {D.~R.}\ \bibnamefont {Cole}},
  \bibinfo {author} {\bibfnamefont {J.}~\bibnamefont {Delville}}, \bibinfo
  {author} {\bibfnamefont {M.~N.}\ \bibnamefont {Glauser}},\ and\ \bibinfo
  {author} {\bibfnamefont {L.~S.}\ \bibnamefont {Ukeiley}},\ }\bibfield
  {title} {\bibinfo {title} {Stochastic estimation and proper orthogonal
  decomposition: Complementary techniques for identifying structure},\ }\href
  {https://doi.org/10.1007/bf01874409} {\bibfield  {journal} {\bibinfo
  {journal} {Experiments in Fluids}\ }\textbf {\bibinfo {volume} {17}},\
  \bibinfo {pages} {307} (\bibinfo {year} {1994})}\BibitemShut {NoStop}%
\bibitem [{\citenamefont {Guezennec}(1989)}]{guezennec1989}%
  \BibitemOpen
  \bibfield  {author} {\bibinfo {author} {\bibfnamefont {Y.~G.}\ \bibnamefont
  {Guezennec}},\ }\bibfield  {title} {\bibinfo {title} {Stochastic estimation
  of coherent structures in turbulent boundary layers},\ }\href
  {https://doi.org/10.1063/1.857396} {\bibfield  {journal} {\bibinfo  {journal}
  {Physics of Fluids A: Fluid Dynamics}\ }\textbf {\bibinfo {volume} {1}},\
  \bibinfo {pages} {1054} (\bibinfo {year} {1989})}\BibitemShut {NoStop}%
\bibitem [{\citenamefont {Cole}\ \emph {et~al.}(1992)\citenamefont {Cole},
  \citenamefont {Glauser},\ and\ \citenamefont {Guezennec}}]{cole1992}%
  \BibitemOpen
  \bibfield  {author} {\bibinfo {author} {\bibfnamefont {D.~R.}\ \bibnamefont
  {Cole}}, \bibinfo {author} {\bibfnamefont {M.~N.}\ \bibnamefont {Glauser}},\
  and\ \bibinfo {author} {\bibfnamefont {Y.~G.}\ \bibnamefont {Guezennec}},\
  }\bibfield  {title} {\bibinfo {title} {An application of the stochastic
  estimation to the jet mixing layer},\ }\href@noop {} {\bibfield  {journal}
  {\bibinfo  {journal} {Physics of Fluids A: Fluid Dynamics}\ }\textbf
  {\bibinfo {volume} {4}},\ \bibinfo {pages} {192} (\bibinfo {year}
  {1992})}\BibitemShut {NoStop}%
\bibitem [{\citenamefont {Cole}\ and\ \citenamefont
  {Glauser}(1998)}]{cole1998}%
  \BibitemOpen
  \bibfield  {author} {\bibinfo {author} {\bibfnamefont {D.~R.}\ \bibnamefont
  {Cole}}\ and\ \bibinfo {author} {\bibfnamefont {M.~N.}\ \bibnamefont
  {Glauser}},\ }\bibfield  {title} {\bibinfo {title} {Applications of
  stochastic estimation in the axisymmetric sudden expansion},\ }\href
  {https://doi.org/10.1063/1.869814} {\bibfield  {journal} {\bibinfo  {journal}
  {Physics of Fluids}\ }\textbf {\bibinfo {volume} {10}},\ \bibinfo {pages}
  {2941} (\bibinfo {year} {1998})}\BibitemShut {NoStop}%
\bibitem [{\citenamefont {Murray}\ and\ \citenamefont
  {Ukeiley}(2003)}]{murray2003}%
  \BibitemOpen
  \bibfield  {author} {\bibinfo {author} {\bibfnamefont {N.~E.}\ \bibnamefont
  {Murray}}\ and\ \bibinfo {author} {\bibfnamefont {L.~S.}\ \bibnamefont
  {Ukeiley}},\ }\bibfield  {title} {\bibinfo {title} {Estimation of the
  flowfield from surface pressure measurements in an open cavity},\ }\href
  {https://doi.org/10.2514/2.2035} {\bibfield  {journal} {\bibinfo  {journal}
  {AIAA Journal}\ }\textbf {\bibinfo {volume} {41}},\ \bibinfo {pages} {969}
  (\bibinfo {year} {2003})}\BibitemShut {NoStop}%
\bibitem [{\citenamefont {Arunajatesan}\ \emph {et~al.}(2007)\citenamefont
  {Arunajatesan}, \citenamefont {Kannepalli},\ and\ \citenamefont
  {Ukeiley}}]{arunajatesan2007}%
  \BibitemOpen
  \bibfield  {author} {\bibinfo {author} {\bibfnamefont {S.}~\bibnamefont
  {Arunajatesan}}, \bibinfo {author} {\bibfnamefont {C.}~\bibnamefont
  {Kannepalli}},\ and\ \bibinfo {author} {\bibfnamefont {L.}~\bibnamefont
  {Ukeiley}},\ }\bibfield  {title} {\bibinfo {title} {Three dimensional
  stochastic estimation applied to cavity flow fields},\ }in\ \href@noop {}
  {\emph {\bibinfo {booktitle} {37th AIAA Fluid Dynamics Conference and
  Exhibit}}}\ (\bibinfo {year} {2007})\ p.\ \bibinfo {pages} {4227}\BibitemShut
  {NoStop}%
\bibitem [{\citenamefont {Tinney}\ \emph {et~al.}(2008)\citenamefont {Tinney},
  \citenamefont {Ukeiley},\ and\ \citenamefont {Glauser}}]{tinney2008}%
  \BibitemOpen
  \bibfield  {author} {\bibinfo {author} {\bibfnamefont {C.~E.}\ \bibnamefont
  {Tinney}}, \bibinfo {author} {\bibfnamefont {L.~S.}\ \bibnamefont
  {Ukeiley}},\ and\ \bibinfo {author} {\bibfnamefont {M.~N.}\ \bibnamefont
  {Glauser}},\ }\bibfield  {title} {\bibinfo {title} {Low-dimensional
  characteristics of a transonic jet. part 2. estimate and far-field
  prediction},\ }\href {https://doi.org/10.1017/S0022112008003601} {\bibfield
  {journal} {\bibinfo  {journal} {Journal of Fluid Mechanics}\ }\textbf
  {\bibinfo {volume} {615}},\ \bibinfo {pages} {53–92} (\bibinfo {year}
  {2008})}\BibitemShut {NoStop}%
\bibitem [{\citenamefont {Durgesh}\ and\ \citenamefont
  {Naughton}(2010)}]{durgesh2010}%
  \BibitemOpen
  \bibfield  {author} {\bibinfo {author} {\bibfnamefont {V.}~\bibnamefont
  {Durgesh}}\ and\ \bibinfo {author} {\bibfnamefont {J.}~\bibnamefont
  {Naughton}},\ }\bibfield  {title} {\bibinfo {title} {Multi-time-delay
  {LSE}-{POD} complementary approach applied to unsteady high-{R}eynolds-number
  near wake flow},\ }\href@noop {} {\bibfield  {journal} {\bibinfo  {journal}
  {Experiments in {f}luids}\ }\textbf {\bibinfo {volume} {49}},\ \bibinfo
  {pages} {571} (\bibinfo {year} {2010})}\BibitemShut {NoStop}%
\bibitem [{\citenamefont {Nickels}\ \emph {et~al.}(2020)\citenamefont
  {Nickels}, \citenamefont {Ukeiley}, \citenamefont {Reger},\ and\
  \citenamefont {{Cattafesta III}}}]{nickels2020}%
  \BibitemOpen
  \bibfield  {author} {\bibinfo {author} {\bibfnamefont {A.}~\bibnamefont
  {Nickels}}, \bibinfo {author} {\bibfnamefont {L.}~\bibnamefont {Ukeiley}},
  \bibinfo {author} {\bibfnamefont {R.}~\bibnamefont {Reger}},\ and\ \bibinfo
  {author} {\bibfnamefont {L.}~\bibnamefont {{Cattafesta III}}},\ }\bibfield
  {title} {\bibinfo {title} {Low-order estimation of the velocity, hydrodynamic
  pressure, and acoustic radiation for a three-dimensional turbulent wall
  jet},\ }\href
  {https://doi.org/https://doi.org/10.1016/j.expthermflusci.2020.110101}
  {\bibfield  {journal} {\bibinfo  {journal} {Experimental Thermal and Fluid
  Science}\ }\textbf {\bibinfo {volume} {116}},\ \bibinfo {pages} {110101}
  (\bibinfo {year} {2020})}\BibitemShut {NoStop}%
\bibitem [{\citenamefont {Zhang}\ \emph {et~al.}(2020)\citenamefont {Zhang},
  \citenamefont {Cattafesta},\ and\ \citenamefont {Ukeiley}}]{zhang2020}%
  \BibitemOpen
  \bibfield  {author} {\bibinfo {author} {\bibfnamefont {Y.}~\bibnamefont
  {Zhang}}, \bibinfo {author} {\bibfnamefont {L.~N.}\ \bibnamefont
  {Cattafesta}},\ and\ \bibinfo {author} {\bibfnamefont {L.}~\bibnamefont
  {Ukeiley}},\ }\bibfield  {title} {\bibinfo {title} {Spectral analysis modal
  methods (samms) using non-time-resolved piv},\ }\href@noop {} {\bibfield
  {journal} {\bibinfo  {journal} {Experiments in Fluids}\ }\textbf {\bibinfo
  {volume} {61}},\ \bibinfo {pages} {1} (\bibinfo {year} {2020})}\BibitemShut
  {NoStop}%
\bibitem [{\citenamefont {Ewing}\ and\ \citenamefont
  {Citriniti}(1999)}]{ewing1999}%
  \BibitemOpen
  \bibfield  {author} {\bibinfo {author} {\bibfnamefont {D.}~\bibnamefont
  {Ewing}}\ and\ \bibinfo {author} {\bibfnamefont {J.~H.}\ \bibnamefont
  {Citriniti}},\ }\bibfield  {title} {\bibinfo {title} {Examination of a
  lse/pod complementary technique using single and multi-time information in
  the axisymmetric shear layer},\ }in\ \href@noop {} {\emph {\bibinfo
  {booktitle} {IUTAM Symposium on Simulation and Identification of Organized
  Structures in Flows}}}\ (\bibinfo {organization} {Springer},\ \bibinfo {year}
  {1999})\ pp.\ \bibinfo {pages} {375--384}\BibitemShut {NoStop}%
\bibitem [{\citenamefont {Magstadt}\ \emph {et~al.}(2016)\citenamefont
  {Magstadt}, \citenamefont {Berry}, \citenamefont {Berger}, \citenamefont
  {Shea}, \citenamefont {Ruscher}, \citenamefont {Gogineni},\ and\
  \citenamefont {Glauser}}]{magstadt2016}%
  \BibitemOpen
  \bibfield  {author} {\bibinfo {author} {\bibfnamefont {A.~S.}\ \bibnamefont
  {Magstadt}}, \bibinfo {author} {\bibfnamefont {M.~G.}\ \bibnamefont {Berry}},
  \bibinfo {author} {\bibfnamefont {Z.~P.}\ \bibnamefont {Berger}}, \bibinfo
  {author} {\bibfnamefont {P.~R.}\ \bibnamefont {Shea}}, \bibinfo {author}
  {\bibfnamefont {C.~J.}\ \bibnamefont {Ruscher}}, \bibinfo {author}
  {\bibfnamefont {S.~P.}\ \bibnamefont {Gogineni}},\ and\ \bibinfo {author}
  {\bibfnamefont {M.~N.}\ \bibnamefont {Glauser}},\ }\bibfield  {title}
  {\bibinfo {title} {Flow structures associated with turbulent mixing noise and
  screech tones in axisymmetric jets},\ }\href
  {https://doi.org/10.1007/s10494-016-9784-8} {\bibfield  {journal} {\bibinfo
  {journal} {Flow, Turbulence and Combustion}\ }\textbf {\bibinfo {volume}
  {98}},\ \bibinfo {pages} {725} (\bibinfo {year} {2016})}\BibitemShut
  {NoStop}%
\bibitem [{\citenamefont {Picard}\ and\ \citenamefont
  {Delville}(2000)}]{picard2000}%
  \BibitemOpen
  \bibfield  {author} {\bibinfo {author} {\bibfnamefont {C.}~\bibnamefont
  {Picard}}\ and\ \bibinfo {author} {\bibfnamefont {J.}~\bibnamefont
  {Delville}},\ }\bibfield  {title} {\bibinfo {title} {Pressure velocity
  coupling in a subsonic round jet},\ }\href@noop {} {\bibfield  {journal}
  {\bibinfo  {journal} {International Journal of Heat and Fluid Flow}\ }\textbf
  {\bibinfo {volume} {21}},\ \bibinfo {pages} {359 } (\bibinfo {year}
  {2000})}\BibitemShut {NoStop}%
\bibitem [{\citenamefont {Fuchs}(1972{\natexlab{a}})}]{fuchs1972a}%
  \BibitemOpen
  \bibfield  {author} {\bibinfo {author} {\bibfnamefont {H.}~\bibnamefont
  {Fuchs}},\ }\bibfield  {title} {\bibinfo {title} {Measurement of pressure
  fluctuations within subsonic turbulent jets},\ }\href
  {https://doi.org/10.1016/0022-460x(72)90171-x} {\bibfield  {journal}
  {\bibinfo  {journal} {Journal of Sound and Vibration}\ }\textbf {\bibinfo
  {volume} {22}},\ \bibinfo {pages} {361} (\bibinfo {year}
  {1972}{\natexlab{a}})}\BibitemShut {NoStop}%
\bibitem [{\citenamefont {Fuchs}(1972{\natexlab{b}})}]{fuchs1972b}%
  \BibitemOpen
  \bibfield  {author} {\bibinfo {author} {\bibfnamefont {H.}~\bibnamefont
  {Fuchs}},\ }\bibfield  {title} {\bibinfo {title} {Space correlations of the
  fluctuating pressure in subsonic turbulent jets},\ }\href
  {https://doi.org/10.1016/0022-460x(72)90790-0} {\bibfield  {journal}
  {\bibinfo  {journal} {Journal of Sound and Vibration}\ }\textbf {\bibinfo
  {volume} {23}},\ \bibinfo {pages} {77} (\bibinfo {year}
  {1972}{\natexlab{b}})}\BibitemShut {NoStop}%
\bibitem [{\citenamefont {Jones}\ \emph {et~al.}(1979)\citenamefont {Jones},
  \citenamefont {Adrian}, \citenamefont {Nithianandan},\ and\ \citenamefont
  {Planchon}}]{jones1979}%
  \BibitemOpen
  \bibfield  {author} {\bibinfo {author} {\bibfnamefont {B.~G.}\ \bibnamefont
  {Jones}}, \bibinfo {author} {\bibfnamefont {R.~J.}\ \bibnamefont {Adrian}},
  \bibinfo {author} {\bibfnamefont {C.~K.}\ \bibnamefont {Nithianandan}},\ and\
  \bibinfo {author} {\bibfnamefont {H.~P.}\ \bibnamefont {Planchon}},\
  }\bibfield  {title} {\bibinfo {title} {Spectra of turbulent static pressure
  fluctuations in jet mixing layers},\ }\href {https://doi.org/10.2514/3.61154}
  {\bibfield  {journal} {\bibinfo  {journal} {{AIAA} Journal}\ }\textbf
  {\bibinfo {volume} {17}},\ \bibinfo {pages} {449} (\bibinfo {year}
  {1979})}\BibitemShut {NoStop}%
\bibitem [{\citenamefont {George}\ \emph {et~al.}(1984)\citenamefont {George},
  \citenamefont {Beuther},\ and\ \citenamefont {Arndt}}]{george1984}%
  \BibitemOpen
  \bibfield  {author} {\bibinfo {author} {\bibfnamefont {W.~K.}\ \bibnamefont
  {George}}, \bibinfo {author} {\bibfnamefont {P.~D.}\ \bibnamefont
  {Beuther}},\ and\ \bibinfo {author} {\bibfnamefont {R.~E.~A.}\ \bibnamefont
  {Arndt}},\ }\bibfield  {title} {\bibinfo {title} {Pressure spectra in
  turbulent free shear flows},\ }\href
  {https://doi.org/10.1017/s0022112084002299} {\bibfield  {journal} {\bibinfo
  {journal} {Journal of Fluid Mechanics}\ }\textbf {\bibinfo {volume} {148}},\
  \bibinfo {pages} {155} (\bibinfo {year} {1984})}\BibitemShut {NoStop}%
\bibitem [{\citenamefont {Li}\ and\ \citenamefont {Ukeiley}(2021)}]{li2021}%
  \BibitemOpen
  \bibfield  {author} {\bibinfo {author} {\bibfnamefont {S.}~\bibnamefont
  {Li}}\ and\ \bibinfo {author} {\bibfnamefont {L.~S.}\ \bibnamefont
  {Ukeiley}},\ }\bibfield  {title} {\bibinfo {title} {Experimental
  investigation of the fluctuating static pressure in a subsonic axisymmetric
  jet},\ }\href {https://doi.org/10.1177/1475472x211004854} {\bibfield
  {journal} {\bibinfo  {journal} {International Journal of Aeroacoustics}\
  }\textbf {\bibinfo {volume} {20}},\ \bibinfo {pages} {196} (\bibinfo {year}
  {2021})}\BibitemShut {NoStop}%
\bibitem [{\citenamefont {Jordan}\ and\ \citenamefont
  {Colonius}(2013)}]{jordan2013}%
  \BibitemOpen
  \bibfield  {author} {\bibinfo {author} {\bibfnamefont {P.}~\bibnamefont
  {Jordan}}\ and\ \bibinfo {author} {\bibfnamefont {T.}~\bibnamefont
  {Colonius}},\ }\bibfield  {title} {\bibinfo {title} {Wave packets and
  turbulent jet noise},\ }\href
  {https://doi.org/10.1146/annurev-fluid-011212-140756} {\bibfield  {journal}
  {\bibinfo  {journal} {Annual Review of Fluid Mechanics}\ }\textbf {\bibinfo
  {volume} {45}},\ \bibinfo {pages} {173} (\bibinfo {year} {2013})}\BibitemShut
  {NoStop}%
\bibitem [{\citenamefont {Cavalieri}\ \emph {et~al.}(2012)\citenamefont
  {Cavalieri}, \citenamefont {Jordan}, \citenamefont {Colonius},\ and\
  \citenamefont {Gervais}}]{cavalieri2012}%
  \BibitemOpen
  \bibfield  {author} {\bibinfo {author} {\bibfnamefont {A.~V.~G.}\
  \bibnamefont {Cavalieri}}, \bibinfo {author} {\bibfnamefont {P.}~\bibnamefont
  {Jordan}}, \bibinfo {author} {\bibfnamefont {T.}~\bibnamefont {Colonius}},\
  and\ \bibinfo {author} {\bibfnamefont {Y.}~\bibnamefont {Gervais}},\
  }\bibfield  {title} {\bibinfo {title} {Axisymmetric superdirectivity in
  subsonic jets},\ }\href {https://doi.org/10.1017/jfm.2012.247} {\bibfield
  {journal} {\bibinfo  {journal} {Journal of Fluid Mechanics}\ }\textbf
  {\bibinfo {volume} {704}},\ \bibinfo {pages} {388} (\bibinfo {year}
  {2012})}\BibitemShut {NoStop}%
\bibitem [{\citenamefont {Cavalieri}\ \emph {et~al.}(2019)\citenamefont
  {Cavalieri}, \citenamefont {Jordan},\ and\ \citenamefont
  {Lesshafft}}]{cavalieri2019}%
  \BibitemOpen
  \bibfield  {author} {\bibinfo {author} {\bibfnamefont {A.~V.~G.}\
  \bibnamefont {Cavalieri}}, \bibinfo {author} {\bibfnamefont {P.}~\bibnamefont
  {Jordan}},\ and\ \bibinfo {author} {\bibfnamefont {L.}~\bibnamefont
  {Lesshafft}},\ }\bibfield  {title} {\bibinfo {title} {Wave-packet models for
  jet dynamics and sound radiation},\ }\bibfield  {journal} {\bibinfo
  {journal} {Applied Mechanics Reviews}\ }\textbf {\bibinfo {volume} {71}},\
  \href {https://doi.org/10.1115/1.4042736} {10.1115/1.4042736} (\bibinfo
  {year} {2019})\BibitemShut {NoStop}%
\bibitem [{\citenamefont {Brunton}\ \emph {et~al.}(2020)\citenamefont
  {Brunton}, \citenamefont {Noack},\ and\ \citenamefont
  {Koumoutsakos}}]{brunton2020}%
  \BibitemOpen
  \bibfield  {author} {\bibinfo {author} {\bibfnamefont {S.~L.}\ \bibnamefont
  {Brunton}}, \bibinfo {author} {\bibfnamefont {B.~R.}\ \bibnamefont {Noack}},\
  and\ \bibinfo {author} {\bibfnamefont {P.}~\bibnamefont {Koumoutsakos}},\
  }\bibfield  {title} {\bibinfo {title} {Machine learning for fluid
  mechanics},\ }\href {https://doi.org/10.1146/annurev-fluid-010719-060214}
  {\bibfield  {journal} {\bibinfo  {journal} {Annual Review of Fluid
  Mechanics}\ }\textbf {\bibinfo {volume} {52}},\ \bibinfo {pages} {477}
  (\bibinfo {year} {2020})}\BibitemShut {NoStop}%
\bibitem [{\citenamefont {Kutz}(2017)}]{kutz2017}%
  \BibitemOpen
  \bibfield  {author} {\bibinfo {author} {\bibfnamefont {J.~N.}\ \bibnamefont
  {Kutz}},\ }\bibfield  {title} {\bibinfo {title} {Deep learning in fluid
  dynamics},\ }\href {https://doi.org/10.1017/jfm.2016.803} {\bibfield
  {journal} {\bibinfo  {journal} {Journal of Fluid Mechanics}\ }\textbf
  {\bibinfo {volume} {814}},\ \bibinfo {pages} {1} (\bibinfo {year}
  {2017})}\BibitemShut {NoStop}%
\bibitem [{\citenamefont {Tenney}\ \emph {et~al.}(2020)\citenamefont {Tenney},
  \citenamefont {Glauser}, \citenamefont {Ruscher},\ and\ \citenamefont
  {Berger}}]{tenney2020}%
  \BibitemOpen
  \bibfield  {author} {\bibinfo {author} {\bibfnamefont {A.~S.}\ \bibnamefont
  {Tenney}}, \bibinfo {author} {\bibfnamefont {M.~N.}\ \bibnamefont {Glauser}},
  \bibinfo {author} {\bibfnamefont {C.~J.}\ \bibnamefont {Ruscher}},\ and\
  \bibinfo {author} {\bibfnamefont {Z.~P.}\ \bibnamefont {Berger}},\ }\bibfield
   {title} {\bibinfo {title} {Application of artificial neural networks to
  stochastic estimation and jet noise modeling},\ }\href
  {https://doi.org/10.2514/1.j058638} {\bibfield  {journal} {\bibinfo
  {journal} {{AIAA} Journal}\ }\textbf {\bibinfo {volume} {58}},\ \bibinfo
  {pages} {647} (\bibinfo {year} {2020})}\BibitemShut {NoStop}%
\bibitem [{\citenamefont {Mohan}\ and\ \citenamefont
  {Gaitonde}(2018)}]{mohan2018}%
  \BibitemOpen
  \bibfield  {author} {\bibinfo {author} {\bibfnamefont {A.~T.}\ \bibnamefont
  {Mohan}}\ and\ \bibinfo {author} {\bibfnamefont {D.~V.}\ \bibnamefont
  {Gaitonde}},\ }\href@noop {} {\bibinfo {title} {A deep learning based
  approach to reduced order modeling for turbulent flow control using lstm
  neural networks}} (\bibinfo {year} {2018}),\ \Eprint
  {https://arxiv.org/abs/arXiv:1804.09269} {arXiv:1804.09269} \BibitemShut
  {NoStop}%
\bibitem [{\citenamefont {Jin}\ \emph {et~al.}(2020)\citenamefont {Jin},
  \citenamefont {Laima}, \citenamefont {Chen},\ and\ \citenamefont
  {Li}}]{jin2020}%
  \BibitemOpen
  \bibfield  {author} {\bibinfo {author} {\bibfnamefont {X.}~\bibnamefont
  {Jin}}, \bibinfo {author} {\bibfnamefont {S.}~\bibnamefont {Laima}}, \bibinfo
  {author} {\bibfnamefont {W.~L.}\ \bibnamefont {Chen}},\ and\ \bibinfo
  {author} {\bibfnamefont {H.}~\bibnamefont {Li}},\ }\bibfield  {title}
  {\bibinfo {title} {Time-resolved reconstruction of flow field around a
  circular cylinder by recurrent neural networks based on non-time-resolved
  particle image velocimetry measurements},\ }\bibfield  {journal} {\bibinfo
  {journal} {Experiments in Fluids}\ }\textbf {\bibinfo {volume} {61}},\ \href
  {https://doi.org/10.1007/s00348-020-2928-6} {10.1007/s00348-020-2928-6}
  (\bibinfo {year} {2020})\BibitemShut {NoStop}%
\bibitem [{\citenamefont {Deng}\ \emph {et~al.}(2019)\citenamefont {Deng},
  \citenamefont {Chen}, \citenamefont {Liu},\ and\ \citenamefont
  {Kim}}]{deng2019}%
  \BibitemOpen
  \bibfield  {author} {\bibinfo {author} {\bibfnamefont {Z.}~\bibnamefont
  {Deng}}, \bibinfo {author} {\bibfnamefont {Y.}~\bibnamefont {Chen}}, \bibinfo
  {author} {\bibfnamefont {Y.}~\bibnamefont {Liu}},\ and\ \bibinfo {author}
  {\bibfnamefont {K.~C.}\ \bibnamefont {Kim}},\ }\bibfield  {title} {\bibinfo
  {title} {Time-resolved turbulent velocity field reconstruction using a long
  short-term memory ({LSTM})-based artificial intelligence framework},\ }\href
  {https://doi.org/10.1063/1.5111558} {\bibfield  {journal} {\bibinfo
  {journal} {Physics of Fluids}\ }\textbf {\bibinfo {volume} {31}},\ \bibinfo
  {pages} {075108} (\bibinfo {year} {2019})}\BibitemShut {NoStop}%
\bibitem [{\citenamefont {Choi}\ \emph {et~al.}(2017)\citenamefont {Choi},
  \citenamefont {Fazekas}, \citenamefont {Sandler},\ and\ \citenamefont
  {Cho}}]{choi2017}%
  \BibitemOpen
  \bibfield  {author} {\bibinfo {author} {\bibfnamefont {K.}~\bibnamefont
  {Choi}}, \bibinfo {author} {\bibfnamefont {G.}~\bibnamefont {Fazekas}},
  \bibinfo {author} {\bibfnamefont {M.}~\bibnamefont {Sandler}},\ and\ \bibinfo
  {author} {\bibfnamefont {K.}~\bibnamefont {Cho}},\ }\bibfield  {title}
  {\bibinfo {title} {Convolutional recurrent neural networks for music
  classification},\ }in\ \href {https://doi.org/10.1109/icassp.2017.7952585}
  {\emph {\bibinfo {booktitle} {2017 {IEEE} International Conference on
  Acoustics, Speech and Signal Processing ({ICASSP})}}}\ (\bibinfo  {publisher}
  {{IEEE}},\ \bibinfo {year} {2017})\BibitemShut {NoStop}%
\bibitem [{\citenamefont {Selvin}\ \emph {et~al.}(2017)\citenamefont {Selvin},
  \citenamefont {Vinayakumar}, \citenamefont {Gopalakrishnan}, \citenamefont
  {Menon},\ and\ \citenamefont {Soman}}]{selvin2017}%
  \BibitemOpen
  \bibfield  {author} {\bibinfo {author} {\bibfnamefont {S.}~\bibnamefont
  {Selvin}}, \bibinfo {author} {\bibfnamefont {R.}~\bibnamefont {Vinayakumar}},
  \bibinfo {author} {\bibfnamefont {E.~A.}\ \bibnamefont {Gopalakrishnan}},
  \bibinfo {author} {\bibfnamefont {V.~K.}\ \bibnamefont {Menon}},\ and\
  \bibinfo {author} {\bibfnamefont {K.~P.}\ \bibnamefont {Soman}},\ }\bibfield
  {title} {\bibinfo {title} {Stock price prediction using {LSTM}, {RNN} and
  {CNN}-sliding window model},\ }in\ \href
  {https://doi.org/10.1109/icacci.2017.8126078} {\emph {\bibinfo {booktitle}
  {2017 International Conference on Advances in Computing, Communications and
  Informatics ({ICACCI})}}}\ (\bibinfo  {publisher} {{IEEE}},\ \bibinfo {year}
  {2017})\BibitemShut {NoStop}%
\bibitem [{\citenamefont {Mathew}\ \emph {et~al.}(2005)\citenamefont {Mathew},
  \citenamefont {Bahr}, \citenamefont {Carroll}, \citenamefont {Sheplak},\ and\
  \citenamefont {Cattafesta}}]{mathew2005}%
  \BibitemOpen
  \bibfield  {author} {\bibinfo {author} {\bibfnamefont {J.}~\bibnamefont
  {Mathew}}, \bibinfo {author} {\bibfnamefont {C.}~\bibnamefont {Bahr}},
  \bibinfo {author} {\bibfnamefont {B.}~\bibnamefont {Carroll}}, \bibinfo
  {author} {\bibfnamefont {M.}~\bibnamefont {Sheplak}},\ and\ \bibinfo {author}
  {\bibfnamefont {L.}~\bibnamefont {Cattafesta}},\ }\bibfield  {title}
  {\bibinfo {title} {Design, fabrication, and characterization of an anechoic
  wind tunnel facility},\ }in\ \href {https://doi.org/10.2514/6.2005-3052}
  {\emph {\bibinfo {booktitle} {11th {AIAA}/{CEAS} Aeroacoustics
  Conference}}},\ \bibinfo {series and number} {\bibinfo {number} {2005-3052}}\
  (\bibinfo {year} {2005})\BibitemShut {NoStop}%
\bibitem [{\citenamefont {Li}\ \emph {et~al.}(2020)\citenamefont {Li},
  \citenamefont {Ukeiley},\ and\ \citenamefont {Sheplak}}]{li2020}%
  \BibitemOpen
  \bibfield  {author} {\bibinfo {author} {\bibfnamefont {S.}~\bibnamefont
  {Li}}, \bibinfo {author} {\bibfnamefont {L.~S.}\ \bibnamefont {Ukeiley}},\
  and\ \bibinfo {author} {\bibfnamefont {M.}~\bibnamefont {Sheplak}},\
  }\bibfield  {title} {\bibinfo {title} {{PIV} measurements and reduced-order
  characterization of a mach 0.3 axisymmetric jet},\ }in\ \href
  {https://doi.org/10.2514/6.2020-2039} {\emph {\bibinfo {booktitle} {{AIAA}
  Scitech 2020 Forum}}},\ \bibinfo {series and number} {\bibinfo {number}
  {2020-2039}}\ (\bibinfo {year} {2020})\BibitemShut {NoStop}%
\bibitem [{\citenamefont {Dassen}\ \emph {et~al.}(1996)\citenamefont {Dassen},
  \citenamefont {Holthusen},\ and\ \citenamefont {Beukema}}]{dassen1996}%
  \BibitemOpen
  \bibfield  {author} {\bibinfo {author} {\bibfnamefont {T.}~\bibnamefont
  {Dassen}}, \bibinfo {author} {\bibfnamefont {H.}~\bibnamefont {Holthusen}},\
  and\ \bibinfo {author} {\bibfnamefont {M.}~\bibnamefont {Beukema}},\
  }\bibfield  {title} {\bibinfo {title} {Design and testing of a low self-noise
  aerodynamic microphone forebody},\ }in\ \href@noop {} {\emph {\bibinfo
  {booktitle} {2nd AIAA/CEAS Aeroacoustics Conference}}},\ \bibinfo {series and
  number} {\bibinfo {number} {1996-1711}}\ (\bibinfo {year} {1996})\ p.\
  \bibinfo {pages} {1711}\BibitemShut {NoStop}%
\bibitem [{\citenamefont {Soderman}\ and\ \citenamefont
  {Allen}(2002)}]{soderman2002}%
  \BibitemOpen
  \bibfield  {author} {\bibinfo {author} {\bibfnamefont {P.~T.}\ \bibnamefont
  {Soderman}}\ and\ \bibinfo {author} {\bibfnamefont {C.~S.}\ \bibnamefont
  {Allen}},\ }\bibfield  {title} {\bibinfo {title} {Microphone measurements in
  and out of airstream},\ }in\ \href
  {https://doi.org/10.1007/978-3-662-05058-3_1} {\emph {\bibinfo {booktitle}
  {Aeroacoustic Measurements}}}\ (\bibinfo  {publisher} {Springer Berlin
  Heidelberg},\ \bibinfo {year} {2002})\ pp.\ \bibinfo {pages}
  {1--61}\BibitemShut {NoStop}%
\bibitem [{\citenamefont {Wieneke}(2015)}]{wieneke2015}%
  \BibitemOpen
  \bibfield  {author} {\bibinfo {author} {\bibfnamefont {B.}~\bibnamefont
  {Wieneke}},\ }\bibfield  {title} {\bibinfo {title} {{PIV} uncertainty
  quantification from correlation statistics},\ }\href
  {https://doi.org/10.1088/0957-0233/26/7/074002} {\bibfield  {journal}
  {\bibinfo  {journal} {Measurement Science and Technology}\ }\textbf {\bibinfo
  {volume} {26}},\ \bibinfo {pages} {074002} (\bibinfo {year}
  {2015})}\BibitemShut {NoStop}%
\bibitem [{\citenamefont {Cavalieri}\ \emph {et~al.}(2013)\citenamefont
  {Cavalieri}, \citenamefont {Rodr{\'{\i}}guez}, \citenamefont {Jordan},
  \citenamefont {Colonius},\ and\ \citenamefont {Gervais}}]{cavalieri2013}%
  \BibitemOpen
  \bibfield  {author} {\bibinfo {author} {\bibfnamefont {A.~V.~G.}\
  \bibnamefont {Cavalieri}}, \bibinfo {author} {\bibfnamefont {D.}~\bibnamefont
  {Rodr{\'{\i}}guez}}, \bibinfo {author} {\bibfnamefont {P.}~\bibnamefont
  {Jordan}}, \bibinfo {author} {\bibfnamefont {T.}~\bibnamefont {Colonius}},\
  and\ \bibinfo {author} {\bibfnamefont {Y.}~\bibnamefont {Gervais}},\
  }\bibfield  {title} {\bibinfo {title} {Wavepackets in the velocity field of
  turbulent jets},\ }\href {https://doi.org/10.1017/jfm.2013.346} {\bibfield
  {journal} {\bibinfo  {journal} {Journal of Fluid Mechanics}\ }\textbf
  {\bibinfo {volume} {730}},\ \bibinfo {pages} {559} (\bibinfo {year}
  {2013})}\BibitemShut {NoStop}%
\bibitem [{\citenamefont {Camussi}\ and\ \citenamefont
  {Bogey}(2021)}]{camussi2021}%
  \BibitemOpen
  \bibfield  {author} {\bibinfo {author} {\bibfnamefont {R.}~\bibnamefont
  {Camussi}}\ and\ \bibinfo {author} {\bibfnamefont {C.}~\bibnamefont
  {Bogey}},\ }\bibfield  {title} {\bibinfo {title} {Intermittent statistics of
  the 0-mode pressure fluctuations in the near field of mach 0.9 circular jets
  at low and high reynolds numbers},\ }\href@noop {} {\bibfield  {journal}
  {\bibinfo  {journal} {Theoretical and Computational Fluid Dynamics}\ }\textbf
  {\bibinfo {volume} {35}},\ \bibinfo {pages} {229} (\bibinfo {year}
  {2021})}\BibitemShut {NoStop}%
\bibitem [{\citenamefont {Batchelor}\ and\ \citenamefont
  {Gill}(1962)}]{batchelor1962}%
  \BibitemOpen
  \bibfield  {author} {\bibinfo {author} {\bibfnamefont {G.}~\bibnamefont
  {Batchelor}}\ and\ \bibinfo {author} {\bibfnamefont {A.~E.}\ \bibnamefont
  {Gill}},\ }\bibfield  {title} {\bibinfo {title} {Analysis of the stability of
  axisymmetric jets},\ }\href@noop {} {\bibfield  {journal} {\bibinfo
  {journal} {Journal of fluid mechanics}\ }\textbf {\bibinfo {volume} {14}},\
  \bibinfo {pages} {529} (\bibinfo {year} {1962})}\BibitemShut {NoStop}%
\bibitem [{\citenamefont {Sirovich}(1987)}]{sirovich1987}%
  \BibitemOpen
  \bibfield  {author} {\bibinfo {author} {\bibfnamefont {L.}~\bibnamefont
  {Sirovich}},\ }\bibfield  {title} {\bibinfo {title} {Turbulence and the
  dynamics of coherent structures part i: coherent structures},\ }\href@noop {}
  {\bibfield  {journal} {\bibinfo  {journal} {Quarterly of Applied
  Mathematics}\ }\textbf {\bibinfo {volume} {45}},\ \bibinfo {pages} {561}
  (\bibinfo {year} {1987})}\BibitemShut {NoStop}%
\bibitem [{\citenamefont {Tinney}\ \emph {et~al.}(2006)\citenamefont {Tinney},
  \citenamefont {Coiffet}, \citenamefont {Delville}, \citenamefont {Hall},
  \citenamefont {Jordan},\ and\ \citenamefont {Glauser}}]{tinney2006}%
  \BibitemOpen
  \bibfield  {author} {\bibinfo {author} {\bibfnamefont {C.}~\bibnamefont
  {Tinney}}, \bibinfo {author} {\bibfnamefont {F.}~\bibnamefont {Coiffet}},
  \bibinfo {author} {\bibfnamefont {J.}~\bibnamefont {Delville}}, \bibinfo
  {author} {\bibfnamefont {A.}~\bibnamefont {Hall}}, \bibinfo {author}
  {\bibfnamefont {P.}~\bibnamefont {Jordan}},\ and\ \bibinfo {author}
  {\bibfnamefont {M.}~\bibnamefont {Glauser}},\ }\bibfield  {title} {\bibinfo
  {title} {On spectral linear stochastic estimation},\ }\href@noop {}
  {\bibfield  {journal} {\bibinfo  {journal} {Experiments in fluids}\ }\textbf
  {\bibinfo {volume} {41}},\ \bibinfo {pages} {763} (\bibinfo {year}
  {2006})}\BibitemShut {NoStop}%
\bibitem [{\citenamefont {Mallat}(1999)}]{mallat1999}%
  \BibitemOpen
  \bibfield  {author} {\bibinfo {author} {\bibfnamefont {S.}~\bibnamefont
  {Mallat}},\ }\href@noop {} {\emph {\bibinfo {title} {A wavelet tour of signal
  processing}}}\ (\bibinfo  {publisher} {Elsevier},\ \bibinfo {year}
  {1999})\BibitemShut {NoStop}%
\bibitem [{\citenamefont {Torrence}\ and\ \citenamefont
  {Compo}(1998)}]{torrence1998}%
  \BibitemOpen
  \bibfield  {author} {\bibinfo {author} {\bibfnamefont {C.}~\bibnamefont
  {Torrence}}\ and\ \bibinfo {author} {\bibfnamefont {G.~P.}\ \bibnamefont
  {Compo}},\ }\bibfield  {title} {\bibinfo {title} {A practical guide to
  wavelet analysis},\ }\href@noop {} {\bibfield  {journal} {\bibinfo  {journal}
  {Bulletin of the American Meteorological society}\ }\textbf {\bibinfo
  {volume} {79}},\ \bibinfo {pages} {61} (\bibinfo {year} {1998})}\BibitemShut
  {NoStop}%
\bibitem [{\citenamefont {Hochreiter}\ and\ \citenamefont
  {Schmidhuber}(1997)}]{hochreiter1997}%
  \BibitemOpen
  \bibfield  {author} {\bibinfo {author} {\bibfnamefont {S.}~\bibnamefont
  {Hochreiter}}\ and\ \bibinfo {author} {\bibfnamefont {J.}~\bibnamefont
  {Schmidhuber}},\ }\bibfield  {title} {\bibinfo {title} {Long short-term
  memory},\ }\href {https://doi.org/10.1162/neco.1997.9.8.1735} {\bibfield
  {journal} {\bibinfo  {journal} {Neural Computation}\ }\textbf {\bibinfo
  {volume} {9}},\ \bibinfo {pages} {1735} (\bibinfo {year} {1997})}\BibitemShut
  {NoStop}%
\bibitem [{\citenamefont {Aggarwal}(2018)}]{aggarwal2018}%
  \BibitemOpen
  \bibfield  {author} {\bibinfo {author} {\bibfnamefont {C.~C.}\ \bibnamefont
  {Aggarwal}},\ }\href@noop {} {\emph {\bibinfo {title} {Neural networks and
  deep learning: a textbook}}}\ (\bibinfo  {publisher} {Springer},\ \bibinfo
  {year} {2018})\BibitemShut {NoStop}%
\bibitem [{\citenamefont {Schuster}\ and\ \citenamefont
  {Paliwal}(1997)}]{schuster1997}%
  \BibitemOpen
  \bibfield  {author} {\bibinfo {author} {\bibfnamefont {M.}~\bibnamefont
  {Schuster}}\ and\ \bibinfo {author} {\bibfnamefont {K.}~\bibnamefont
  {Paliwal}},\ }\bibfield  {title} {\bibinfo {title} {Bidirectional recurrent
  neural networks},\ }\href {https://doi.org/10.1109/78.650093} {\bibfield
  {journal} {\bibinfo  {journal} {{IEEE} Transactions on Signal Processing}\
  }\textbf {\bibinfo {volume} {45}},\ \bibinfo {pages} {2673} (\bibinfo {year}
  {1997})}\BibitemShut {NoStop}%
\bibitem [{\citenamefont {Goodfellow}\ \emph {et~al.}(2016)\citenamefont
  {Goodfellow}, \citenamefont {Bengio},\ and\ \citenamefont
  {Courville}}]{goodfellow2016}%
  \BibitemOpen
  \bibfield  {author} {\bibinfo {author} {\bibfnamefont {I.}~\bibnamefont
  {Goodfellow}}, \bibinfo {author} {\bibfnamefont {Y.}~\bibnamefont {Bengio}},\
  and\ \bibinfo {author} {\bibfnamefont {A.}~\bibnamefont {Courville}},\
  }\href@noop {} {\emph {\bibinfo {title} {Deep learning}}},\ Adaptive
  computation and machine learning\ (\bibinfo  {publisher} {The MIT Press},\
  \bibinfo {address} {Cambridge, Massachusetts},\ \bibinfo {year}
  {2016})\BibitemShut {NoStop}%
\bibitem [{\citenamefont {Kingma}\ and\ \citenamefont {Ba}(2014)}]{kingma2014}%
  \BibitemOpen
  \bibfield  {author} {\bibinfo {author} {\bibfnamefont {D.~P.}\ \bibnamefont
  {Kingma}}\ and\ \bibinfo {author} {\bibfnamefont {J.}~\bibnamefont {Ba}},\
  }\href@noop {} {\bibinfo {title} {Adam: A method for stochastic
  optimization}} (\bibinfo {year} {2014}),\ \Eprint
  {https://arxiv.org/abs/arXiv:1412.6980} {arXiv:1412.6980} \BibitemShut
  {NoStop}%
\bibitem [{\citenamefont {Lau}\ \emph {et~al.}(1972)\citenamefont {Lau},
  \citenamefont {Fisher},\ and\ \citenamefont {Fuchs}}]{lau1972}%
  \BibitemOpen
  \bibfield  {author} {\bibinfo {author} {\bibfnamefont {J.}~\bibnamefont
  {Lau}}, \bibinfo {author} {\bibfnamefont {M.}~\bibnamefont {Fisher}},\ and\
  \bibinfo {author} {\bibfnamefont {H.}~\bibnamefont {Fuchs}},\ }\bibfield
  {title} {\bibinfo {title} {The intrinsic structure of turbulent jets},\
  }\href {https://doi.org/10.1016/0022-460x(72)90451-8} {\bibfield  {journal}
  {\bibinfo  {journal} {Journal of Sound and Vibration}\ }\textbf {\bibinfo
  {volume} {22}},\ \bibinfo {pages} {379} (\bibinfo {year} {1972})}\BibitemShut
  {NoStop}%
\bibitem [{\citenamefont {Breakey}(2014)}]{breakey2014}%
  \BibitemOpen
  \bibfield  {author} {\bibinfo {author} {\bibfnamefont {D.~E.}\ \bibnamefont
  {Breakey}},\ }\emph {\bibinfo {title} {Time-resolved noise source analysis in
  subsonic turbulent jets}},\ \href@noop {} {Ph.D. thesis},\ \bibinfo  {school}
  {Trinity College Dublin} (\bibinfo {year} {2014})\BibitemShut {NoStop}%
\bibitem [{\citenamefont {Br{\`{e}}s}\ \emph {et~al.}(2018)\citenamefont
  {Br{\`{e}}s}, \citenamefont {Jordan}, \citenamefont {Jaunet}, \citenamefont
  {Rallic}, \citenamefont {Cavalieri}, \citenamefont {Towne}, \citenamefont
  {Lele}, \citenamefont {Colonius},\ and\ \citenamefont {Schmidt}}]{bres2018}%
  \BibitemOpen
  \bibfield  {author} {\bibinfo {author} {\bibfnamefont {G.~A.}\ \bibnamefont
  {Br{\`{e}}s}}, \bibinfo {author} {\bibfnamefont {P.}~\bibnamefont {Jordan}},
  \bibinfo {author} {\bibfnamefont {V.}~\bibnamefont {Jaunet}}, \bibinfo
  {author} {\bibfnamefont {M.~L.}\ \bibnamefont {Rallic}}, \bibinfo {author}
  {\bibfnamefont {A.~V.~G.}\ \bibnamefont {Cavalieri}}, \bibinfo {author}
  {\bibfnamefont {A.}~\bibnamefont {Towne}}, \bibinfo {author} {\bibfnamefont
  {S.~K.}\ \bibnamefont {Lele}}, \bibinfo {author} {\bibfnamefont
  {T.}~\bibnamefont {Colonius}},\ and\ \bibinfo {author} {\bibfnamefont
  {O.~T.}\ \bibnamefont {Schmidt}},\ }\bibfield  {title} {\bibinfo {title}
  {Importance of the nozzle-exit boundary-layer state in subsonic turbulent
  jets},\ }\href {https://doi.org/10.1017/jfm.2018.476} {\bibfield  {journal}
  {\bibinfo  {journal} {Journal of Fluid Mechanics}\ }\textbf {\bibinfo
  {volume} {851}},\ \bibinfo {pages} {83} (\bibinfo {year} {2018})}\BibitemShut
  {NoStop}%
\bibitem [{\citenamefont {der Kindere}\ \emph {et~al.}(2019)\citenamefont {der
  Kindere}, \citenamefont {Laskari}, \citenamefont {Ganapathisubramani},\ and\
  \citenamefont {de~Kat}}]{vdk2019}%
  \BibitemOpen
  \bibfield  {author} {\bibinfo {author} {\bibfnamefont {J.~W.~V.}\
  \bibnamefont {der Kindere}}, \bibinfo {author} {\bibfnamefont
  {A.}~\bibnamefont {Laskari}}, \bibinfo {author} {\bibfnamefont
  {B.}~\bibnamefont {Ganapathisubramani}},\ and\ \bibinfo {author}
  {\bibfnamefont {R.}~\bibnamefont {de~Kat}},\ }\bibfield  {title} {\bibinfo
  {title} {Pressure from 2d snapshot {PIV}},\ }\bibfield  {journal} {\bibinfo
  {journal} {Experiments in Fluids}\ }\textbf {\bibinfo {volume} {60}},\ \href
  {https://doi.org/10.1007/s00348-019-2678-5} {10.1007/s00348-019-2678-5}
  (\bibinfo {year} {2019})\BibitemShut {NoStop}%
\end{thebibliography}%

\end{document}